\documentclass{aa}

\usepackage{graphicx}

\newcommand {\arcminp}{\mbox{$.\!\!^{\prime}$}}
\newcommand {\grd}{\mbox{$^{\circ}$}}
\newcommand {\grdp}{\mbox{$.\!\!^{\circ}$}}

\newcommand {\mim}{\mbox{$\mu$m}}

\newcommand {\scm}{\mbox{cm$^{-2}$}}
\newcommand {\ccm}{\mbox{cm$^{-3}$}}

\newcommand {\lsun}{\mbox{L$_{\rm \odot}$}}
\newcommand {\msun}{\mbox{M$_{\rm \odot}$}}

\newcommand {\ia}{$^{a)}$}
\newcommand {\ib}{$^{b)}$}
\newcommand {\ic}{$^{c)}$}
\newcommand {\id}{$^{d)}$}
\newcommand {\ie}{$^{e)}$}
\newcommand {\ief}{$^{f)}$}
\newcommand {\ig}{$^{g)}$}

\begin{document}

   \title{The Nuclear Bulge of the Galaxy}

   \subtitle{III. Large-Scale Physical Characteristics of Stars and Interstellar Matter}

   \author{R. Launhardt\inst{1,2} %\thanks{\emph{Present address}}
          \and
          R. Zylka\inst{3,4}
          \and
          P.G. Mezger\inst{2}
          }

   \offprints{R. Launhardt or P.G. Mezger}

   \institute{Division of Physics, Mathematics and Astronomy, 
              California Institute of Technology, 
              MS 105-24, Pasadena, CA\,91125, U.S.A., \email{rl@astro.caltech.edu}
         \and 
             Max-Planck-Institut f\"ur Radioastronomie (MPIfR),
             Auf dem H\"ugel 69, D-53121 Bonn, Germany,\\ \email{mezgerm@mpifr-bonn.mpg.de}
         \and 
             Institut f\"ur Theoretische Astrophysik (ITA), 
             Tiergartenstra\ss e 15, D-69121 Heidelberg, Germany
         \and 
             I. Phys. Institut d. Universit\"at zu K\"oln, Z\"ulpicher Str. 77, 
             D-50937 K\"oln, Germany
             }

   \date{Received August 15, 2001; accepted November 16, 2001}

   \titlerunning{Large-scale characteristics of the Nuclear Bulge}
   \authorrunning{Launhardt et al.}

%________________________________________________________________

\abstract{
We analyse IRAS and COBE DIRBE data at wavelengths between 2.2 and 
240\,\mim\ of the central 500\,pc of the Galaxy 
and derive the large-scale distribution of stars and interstellar 
matter in the Nuclear Bulge.
Models of the Galactic Disk and Bulge are developed
in order to correctly decompose the total surface brightness maps of the inner Galaxy 
and to apply proper extinction corrections. 
The Nuclear Bulge appears as a distinct, massive disk-like complex 
of stars and molecular clouds which is, on a large scale, symmetric with respect to the 
Galactic Centre. It is distinguished from the Galactic Bulge by its 
flat disk-like morphology, very high density of stars and molecular gas, 
and ongoing star formation. 
The Nuclear Bulge consists of an $R^{-2}$\ Nuclear Stellar Cluster at the centre, 
a large Nuclear Stellar Disk with radius 230$\pm$20\,pc and scale height 45$\pm$5\,pc, 
and the Nuclear Molecular Disk of same size.
\newline
The total stellar mass and luminosity of the Nuclear Bulge are 
1.4$\pm$0.6$\times 10^9$\,\msun\ and
2.5$\pm$1$\times 10^9$\,\lsun, respectively.
About 70\%\ of the luminosity is due to optical and UV radiation from 
young massive Main-Sequence stars which are most abundant in the Nuclear 
Stellar Cluster. 
For the first time, we derive a photometric mass distribution 
for the central 500\,pc of the Galaxy which is fully consistent with 
the kinematic mass distribution. 
We find that the often cited $R^{-2}$ distribution holds only for the central 
$\sim$\,30\,pc and that at larger radii the mass distribution is dominated 
by the Nuclear Stellar Disk which has a flatter density profile. \newline
The total interstellar hydrogen mass in the Nuclear Bulge is 
$M_{\rm H}$=2$\pm$0.3$\times 10^7$\,\msun, distributed  
in a warm inner disk with $R$=110$\pm$20\,pc and a massive, cold outer torus 
which contains more than 80\%\ of this mass. 
Interstellar matter in the Nuclear Bulge is very clumpy 
with $\sim$90\%\ of the mass contained in dense and massive molecular 
clouds with a volume filling factor of only a few per cent. 
This extreme clumpiness, probably caused by the tidal stability limit in the gravitational 
potential of the Nuclear Bulge, enables the strong interstellar radiation field 
to penetrate the entire Nuclear Bulge and explains the relatively low 
average extinction towards the Galactic Centre.
In addition, we find 3$\times 10^7$\,\msun\ of cold and dense material 
outside the Nuclear Bulge at positive longitudes and 1$\times 10^7$\,\msun\ 
at negative longitudes. This material is not heated by the stars in the Nuclear 
Bulge and gives rise to the observed asymmetry in the distribution of interstellar 
matter in the Central Molecular Zone. 
}
   
\maketitle

%________________________________________________________________

\section{Introduction}                                \label{intro}

\begin{table}[htb]                  
 \caption[]{Abbreviations used in this paper  \label{abbtab}}\vspace{-0mm}
  \begin{flushleft}
   \begin{tabular}[t]{ll} 
\hline \noalign{\smallskip}
Abbreviation  & Meaning                       \\
\noalign{\smallskip} \hline\noalign{\smallskip} 
~~~~BH   & Black Hole                         \\
~~~~BW   & Baade's Window                     \\
~~~~CMZ  & Central Molecular Zone             \\
~~~~FIR  & far-infrared ($\sim${\bf 30--300}\,\mim) \\
~~~~FWHM & Full Width Half Maximum            \\
~~~~GB   & Galactic Bulge                     \\
~~~~GC   & Galactic Centre                    \\
~~~~GD   & Galactic Disk                      \\
~~~~GMC  & Giant Molecular Cloud              \\
~~~~HPBW & Half Power Beam Width              \\
~~~~HWHM & Half Width Half Maximum            \\
~~~~ISM  & Interstellar Matter                \\
~~~~KLF  & K-band Luminosity Function         \\
~~~~MIR  & mid-infrared ($\sim${\bf 7--30})\,\mim)  \\
~~~~MS   & Main Sequence                      \\
~~~~{\bf NB}  & {\bf Nuclear Bulge}           \\
~~~~{\bf NMD} & {\bf Nuclear Molecular Disk}  \\
~~~~{\bf NSC} & {\bf Nuclear Stellar Cluster} \\
~~~~{\bf NSD} & {\bf Nuclear Stellar Disk}    \\
~~~~NIR  & near-infrared ($\sim${\bf 1--7}\,\mim)   \\
~~~~PAH  & Polycyclic Aromatic Hydrocarbon    \\
~~~~SED  & Spectral Energy Distribution       \\
~~~~ZL   & Zodiacal Light                     \\
\noalign{\smallskip} \hline\noalign{\smallskip} 
\end{tabular} 
\end{flushleft}
\end{table}

Our present knowledge of the Galactic Centre (GC) Region has recently been reviewed
by, e.g., Blitz et al. (1993), Genzel et al. (1994), Mezger et al. 
(1996; hereafter MDZ96), and Morris \& Serabyn (1996). 
The gas dynamics of the inner Galaxy has recently been re-investigated by 
Englmaier \& Gerhard (1999). To keep this introduction as
concise as possible, we quote here only the most relevant or recent
papers and refer otherwise to the corresponding sections of MDZ96 and references 
therein. For the same reason we use a number of abbreviations whose
meanings are explained in Table \ref{abbtab}.
For consistency, we adopt a distance to the Galactic Centre of $R_0 = 8.5$\,kpc 
throughout this paper, although recent studies suggest a somewhat lower value 
(e.g., McNamara et al. 2000: $R_0 = 7.9\pm 0.3$\,kpc; see also Reid 1993).

MDZ96 classify the centre of our Galaxy as a mildly active Seyfert nucleus.
Although the presence of a black hole of $\sim 2.6\times 10^6$\,\msun\ 
is strongly supported by recent observations (e.g., Eckart \& Genzel 1998) 
it has also become clear that most of the activity in the centre of our Galaxy 
is due to massive star formation in the central parsec (MDZ96, Sect. 5).

The mass of the central region of our Galaxy ($R \le 3$\,kpc) is dominated by the 
Galactic Bulge which consists mainly of old, evolved stars. 
Both the stellar near-infrared (NIR) surface brightness distribution and the kinematics 
of the gas suggest that the Galactic Bulge has a bar structure with its near end in the first 
galactic quadrant (i.e., at positive $l$) (e.g., Binney et al. 1991; 
Blitz \& Spergel 1991; Weiland et al. 1994).
As a consequence of the dynamics caused by the gravitational potential of this bar, 
the region around the co-rotation radius $R_{\rm CR} \sim (3.5\pm 0.5)$\,kpc is de-populated 
of gas, since Interstellar Matter (ISM) is transported efficiently inward from the 
{\it Galactic Disk Molecular Ring} 
at the bar's outer Lindblad resonance at $R_{\rm OLR} \sim (4\pm 0.5)$\,kpc. 
Inside $R < 2$\,kpc the gas settles on closed elongated ($X_1$) orbits. 
This gas is observed as a tilted disk of atomic hydrogen ($R \le 1.5$\,kpc), 
usually referred to as the {\it ``H\,I Central Disk''} (or {\it ``Nuclear Disk''})
($M_{\rm H} \sim 4\times 10^7$\,\msun; Burton \& Liszt, 1978). 
Inside the inner Lindblad resonance the $X_1$ orbits become self-intersecting, 
and shocks and angular momentum loss compress the gas into molecular form and 
drive it further inward where it finally settles on more circular, stable $X_2$\ orbits 
(e.g., Englmaier \& Gerhard 1999). 
The existence of a distinct, unusually dense molecular cloud complex in the central 
few hundred pc of our Galaxy, often referred to as the {\it ``Central Molecular Zone''}, 
(CMZ) is well-established since the early 1970s (e.g., reviews by Genzel \& Townes 1987  
and MDZ96). The gas distribution in the CMZ is highly asymmetric with most of the mass 
being located at positive longitudes and positive velocities 
(see MDZ96, Sects. 2.2.3 and 3.4).
One of the most remarkable features in the $l-v$\ plane of the CMZ is the 
{\it ``180-pc Molecular Ring''} which is hypothesized to be a shock region 
between the innermost stable $X_1$\ orbit of the bar and the more circular 
$X_2$\ orbits in the centre (Binney et al. 1991), but was also interpreted 
as an expanding molecular ring or shell (e.g., Scoville 1972; Bally et al. 1987; 
Sofue 1995b). 

The large and dense stellar complex in this region was originally thought of as the 
innermost part of the more extended Galactic Bulge, with its population of old and 
evolved stars. Various observations, e.g. ongoing 
star formation, the presence of ionizing stars, and its extraordinary high 
surface brightness suggest that this region is distinct from the 
old Galactic Bulge (as originally proposed by Serabyn \& Morris 1996) 
and may be associated with the CMZ. 
MDZ96 therefore call the innermost region $R < 300$\,pc {\it ``Nuclear Bulge''} (NB), 
and clearly distinguish it from the {\it ``Galactic Bulge''} (GB).  
Due to its relative proximity, the physical characteristics of the
NB of our Galaxy can be studied in detail.  However, the layer of interstellar
dust in the Galactic plane restricts observations of the NB
to wavelengths $\lambda \ge 2.2$\,\mim\ and the edge-on projection 
makes a derivation of the true three-dimensional morphology difficult.

We have begun a systematic investigation of the physical characteristics of 
the NB the results
of which are being published in a series of papers.  Papers\,I (Philipp
et al. 1999a) and II (Mezger et al. 1999) analysed the stellar
population of the central $\sim$30\,pc.
Based on a high-resolution $\lambda$\,2.2\,\mim\  survey, we determined the 
K-band luminosity function (KLF) of the central 30\,pc and interpreted it in terms 
of a present-day  bolometric luminosity and mass function. 
Main Sequence (MS) stars with masses $\le 1$\,\msun\ account 
for $\sim$90\% 
of the dynamical mass, but only for 6\% 
of the K-band flux density. 
MS stars with masses $\ge 1$\,\msun\ account 
for $\sim$6\% 
of the dynamical mass and a similar percentage of the integrated K-band flux 
density, but are responsible for $\sim$80\% 
of the bolometric stellar luminosity as well as the ionization of the 
observed H{\small II} regions. 
The bulk of the K-band emission comes from stars evolved from the MS such as 
giants, supergiants, and Wolf-Rayet stars. We find a deficiency of low-mass 
stars within the central 1.25\,pc and indications of a high star formation 
activity during the past $10^7-10^8$\,years.

Here, in paper III, we analyse the large-scale distribution of
stars and ISM in the NB using IRAS and 
COBE DIRBE data. The size scales addressed in this paper span tens 
to hundreds of parsecs.
The complex structure in the central few parsecs, including, e.g.,  
the circumnuclear disk, the mini-spiral, and the central cavity, 
is not addressed. 
In a succeeding paper, based on ground-based single-dish mm observations, 
we investigate the morphology and kinematics 
of ISM in the central part of the NB in more detail (Zylka et al., in prep.).

The paper is organized as follows: 
Section \ref{data} describes briefly the observational data used in this paper. 
Section \ref{datan} describes the data reduction and analysis. 
In Sect. \ref{results} we present the basic observational results, 
and in Sect. \ref{disc} we derive physical characteristics of the NB 
and the CMZ. 
Section \ref{nbmod} summarizes the results in terms of a coherent picture of the  
NB. 
%In Sect. \ref{comp} we discuss and compare this picture with Active Galactic Nuclei 
%(AGN) in Seyfert galaxies.

%__________________________________________________________________

\section{The data}                        \label{data}

This paper is mainly based on IRAS (Infrared Astronomical Satellite) and 
COBE (Cosmic Background Explorer) data covering the 
wavelength range from 2.2\,\mim\ to 240\,\mim, 
which were obtained through {\it SkyView}
%\footnote{$SkyView$\ was developed and is maintained under 
%   NASA ADP Grant NAS 5-32068 with P.I. Thomas A. McGlynn under the
%   auspices of the High Energy Astrophysics Science Archive Research
%   Center (HEASARC) at the Goddard Space Flight Center Laboratory for 
%   High Energy Astrophysics.} 
on the WorldWideWeb.
Additional data at mm and cm wavelengths are used for comparison.
Characteristics of the surveys of the Galactic Centre Region used in the following
analysis are given in Table \ref{datsum}, followed by a more detailed
description of the data. Throughout this paper we use Galactic coordinates 
($l^{\rm II},b^{\rm II}$) and all maps are centered on  
$l^{\rm II},b^{\rm II} = 0$\degr,0\degr\ 
(and not on Sgr\,A$^{\ast}$, 
which is at $l^{\rm II},b^{\rm II} = -$0\grdp054,$-$0\grdp046).

\begin{table*}[htb]                  
 \caption[]{Surveys of the GC referred to in this paper
            \label{datsum}}\vspace{-0mm}   
  \begin{flushleft}
   \begin{tabular}[t]{llclll} 
\hline \noalign{\smallskip}
\multicolumn{2}{c}{Wavelength range}  &
Angular      &
Observations &
Telescopes   &
References   \\
             &
             &
resolution   &
             &
(see References)&
             \\
\noalign{\smallskip} \hline \noalign{\smallskip}
NIR -- FIR         & 1.25\,\mim\ -- 240\,\mim & 0.7\degr   & 10 continuum bands  & COBE DIRBE           & 
                     Hauser et al. (1991) \\[0.5ex] 
MIR -- FIR         & 12\,\mim\ -- 100\,\mim   & 1\arcminp5 & 4 continuum bands   & IRAS                 & 
                     IRAS Sky Survey Atlas (ISSA) \\[0.5ex] 
Millimetre         & 800\,\mim                & 30\arcmin  & continuum map       & CSO 10.4\,m          &
                     Lis \& Carlstrom (1994)\\[0.5ex] 
                   & 1.2\,mm                  & 11\arcsec  & continuum map       & IRAM 30\,m MRT       & 
                     Zylka et al. (in prep.)\\[0.5ex] 
                   & 2.6\,mm                  & 8\arcminp8 & $^{12}$CO(1--0) map & CSMT 1.2\,m          & 
                     Bitran et al. (1997)\\
%                   & 2.6\,mm                  & 9\arcmin   & C$^{18}$O(1--0) map & CSMT                 & 
%                     Dahmen et al. (1997) \\
%                   & 2.6\,mm                  & 6\arcmin   & $^{13}$CO(1--0) map & 7\,m AT\&T Bell.Lab. & 
%                     Bally et al. (1987) \\
%                   & 3\,mm                    & 6\arcmin   & CS(2--1) map        & 7\,m AT\&T Bell.Lab. & 
%                     Bally et al. (1987) \\[0.5ex] 
Radio              & 3\,cm (10\,GHz)          & 3\arcmin   & continuum map       & Nobeyama 45\,m       & 
                     Handa et al. (1987)\\
                   & 6\,cm (5000\,MHz)        & 4\arcminp1 & continuum map       & Parkes 64\,m         & 
                     Haynes et al. (1978)\\
                   & 11\,cm (2695\,MHz)       & 4\arcminp3 & continuum map       & Effelsberg 100\,m    & 
                     Reich et al. (1990b)\\  
                   & 21\,cm (1408\,MHz)       & 9\arcminp4 & continuum map       & Effelsberg 100\,m    & 
                     Reich et al. (1990a)\\ 
\noalign{\smallskip} \hline\noalign{\smallskip}  
\end{tabular} 
\end{flushleft} 
%$^{\ast}$: Summary in Mauersberger \& Bronfman (1998)
\end{table*}

%____________________________

\subsection{IRAS data}

The IRAS data include all data distributed as
part of the IRAS Sky Survey Atlas (ISSA) and were processed by the 
Infrared Processing and Analysis Center (IPAC) to a
uniform standard with the Zodiacal Light (ZL) already subtracted.  For the 
present investigation, 
ISSA  maps in all four bands were used (i.e., $\lambda = 12$, 25, 60, and 100\,\mim).  
These maps have a mean angular resolution of 2\arcmin,
although IPAC has optimized the processing of these data for features
of size 5' or more.  
The 100\,\mim\ map is heavily saturated in the central part of the NB, 
and the 60\,\mim\ map may be partially saturated towards the Sgr\,A complex.
At each of the four wavelengths we obtained maps
of 15\grd\,$\times$\,10\grd\ with a pixel scale of 1\arcminp5, and 
40\grd\,$\times$\,40\grd\ with a pixel scale of 5\arcmin, respectively.

%____________________________

\subsection{COBE data}

The COBE Diffuse Infrared Background Experiment (DIRBE) was
planned primarily to investigate the cosmic infrared background
radiation in 10 filter bands centered at
1.25, 2.2, 3.5, 4.9, 25, 60, 100, 140, and 240\,\mim\ (for details see
the COBE DIRBE Explanatory Supplement 1997; Hauser et al. 1998;
Kelsall et al. 1998).  
The COBE DIRBE maps
cover the entire sky and provide an estimate
of the infrared intensity at each pixel and wavelength band based on
an interpolation of the observations made at various times at solar
elongations close to 90\grd.  To first order, these maps depict the
sky as if it were observed through a temporally constant
interplanetary dust foreground, thus enabling straightforward modeling
and subtraction of the ZL.  The DIRBE instrument has a
42\arcmin\,$\times$\,42\arcmin\ instantaneous field of view.
For this study, maps in all 10 bands were obtained. The maps have an
angular resolution of $\sim$0.7\grd\ and a pixel scale of $\sim$0.35\grd\ 
(21\arcmin).

%____________________________

\subsection{Complementary millimetre and radio continuum data}

Millimetre continuum maps at 11\arcsec\ resolution 
of the central region of the NB were obtained 
with the IRAM 30-m telescope during several observing runs between 
1995 and 1999 (see Table \ref{datsum}). 
These maps are only used for comparison with the COBE and 
IRAS data and will be presented and discussed 
in detail in Paper IV by Zylka et al. (in preparation).

In addition, we obtained radio continuum maps of the GC region 
at 3, 6, 11, and 21\,cm (Fig. \ref{radiomapsplo})  
from the MPIfR Survey archive. These data are already published 
elsewhere (see Table \ref{datsum}) and 
the 11\,cm and 21\,cm maps are already decomposed into source and background 
components. From the total intensity 3\,cm and 6\,cm maps, we subtracted the 
background emission to obtain the source contribution. 
Due to the different beam sizes, observing modes, and data reduction procedures, 
the four data sets may not recover extended emission in the same way. 
Therefore, the derived flux densities given in Table \ref{nbpar} and shown in 
Fig. \ref{gcnb_sed} may contain systematic uncertainties and are not suitable 
to derive physical characteristics of the gas from flux density ratios. 
We use these maps only to compare the morphology of the ionized gas with 
that of the stars and dust in the NB.

%%%%%%%%%%%%%%%%%%%%%%%%%%%%%%%%%%%%%%%%%%%%%%%%%%%%%%%%%%%%%%%%%%%%%%%%%%%%%%

\section{Data analysis}                         \label{datan}

%___________________________

\subsection{Sources of radiation and extinction}    \label{radext}

\begin{figure*}[htb]
\includegraphics[width=0.99\textwidth] {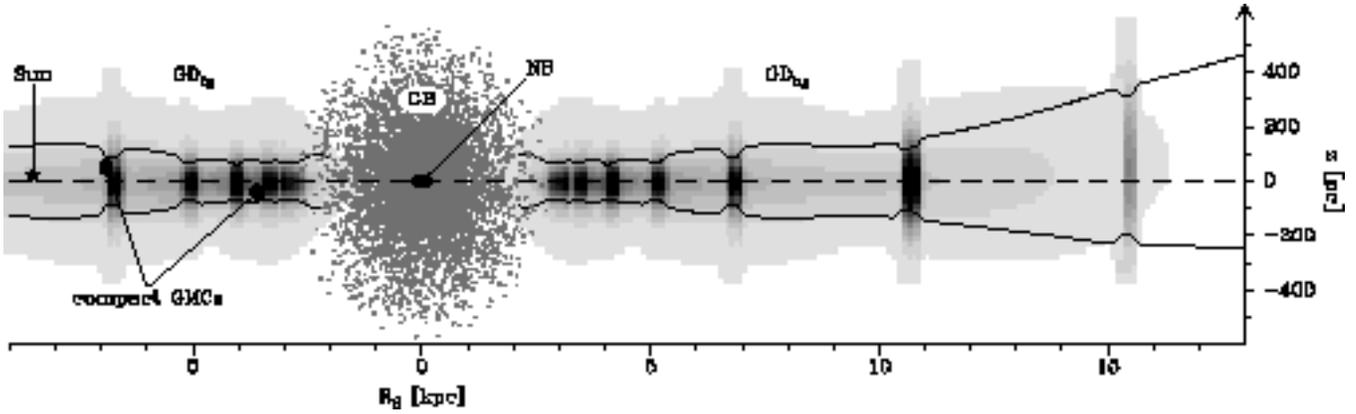}
%{galsketch_paper.ps} (galsketch.graphic)
\caption{\label{galsketch} 
 Schematic presentation of the different galactic features which contribute to 
 emission and absorption in the direction of the Galactic Centre Region. 
 The grey-scale image shows the average hydrogen density in the Galactic Disk (GD) 
 (0.1 to 1 cm$^{-3}$; slice through the disk model described in Appendix \ref{apgalmod}).
 Narrow dark features correspond to spiral arms which have a 3 to 30 times enhanced 
 average density. Subscripts 
 'fg' and 'bg' refer to the foreground and background parts of the GD with respect to the GC.  
 The dashed line denotes the Galactic mid-plane and solid lines mark the density 
 FWHM thickness of the gaseous disk (H$_2$\,+\,H{\small I}\,+\,H{\small II}). 
 The interplanetary dust ring of the solar system (Zodiacal Light) is not shown.  
 Note that the $z$\ scale is stretched with respect to the $R_{\rm G}$\ scale. 
 The Galactic Bulge (GB) is not drawn in its full extent in latitude. 
 The Nuclear Bulge (NB) is shown on scale.}
\end{figure*}

\begin{table*}[htb]                  
 \caption[]{Sources which dominate in different 
            wavelength regions the emission and absorption in and towards the NB 
            \label{radmech}}
  \begin{flushleft}
   \begin{tabular}[t]{llll} 
\hline \noalign{\smallskip}
Wavelength region    &
$\lambda$ range      &
Emission due to      &
Absorption due to    \\
\noalign{\smallskip} \hline\noalign{\smallskip}  
NIR       & 1\,\mim\,--\,7\,\mim   & stars                     & dust       \\
MIR       & 7\,\mim\,--\,30\,\mim  & hot dust (VSGs), PAHs     & dust       \\
Submm/FIR & 30\,\mim\,--\,3\,mm    & ``classical'' dust grains & dust, negligible for $\lambda \ge 100$\,\mim \\
Radio     & 3\,mm\,--\,$\sim$6\,cm & free-free                 & ---        \\
          & $\geq$\,$\sim$6\,cm    & synchrotron               & free-free        \\
\noalign{\smallskip} \hline\noalign{\smallskip} 
\end{tabular} 
\end{flushleft}
\end{table*}

\begin{figure}[htb]
\includegraphics[width=0.47\textwidth] {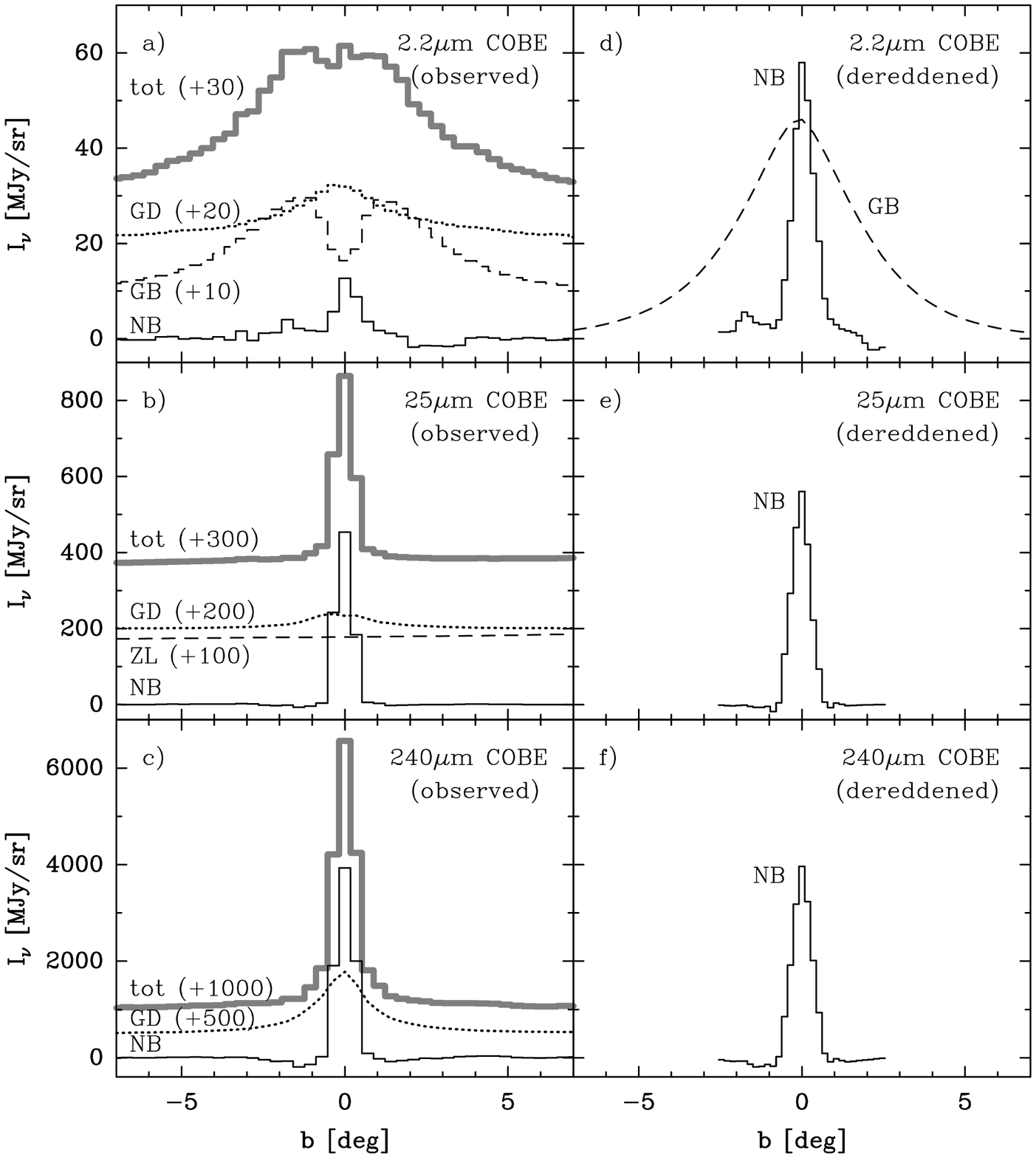}
%{gc_rawcuts_paper.ps} (gc_rawcuts_paper_plo.graphic)
\caption{\label{rawcuts} 
 Latitude profiles of the observed total surface brightness at $b = 0$\degr\ 
 (left panels, thick grey lines) at 
 {\bf a)} $\lambda$\,2.2\,\mim, 
 {\bf b)} $\lambda$\,25\,\mim, and 
 {\bf c)} $\lambda$\,240\,\mim\ as 
 observed by COBE DIRBE. 
 Black solid, dashed, and dotted lines show the 
 contributions of NB, GB, ZL, and GD, respectively, to the observed surface brightness. 
 Right panels d) through f) show the final dereddened latitude profiles of the NB 
 (solid lines). In addition, the dereddened 2.2\,\mim\ surface brightness profile 
 of the Galactic Bulge is shown as dashed line in panel d).} 
\end{figure}

In order to derive surface brightness maps of the GC region from 
the data, the total surface brightness maps were decomposed into the 
individual contributing sources of radiation and extinction which were, 
in turn, modeled and interpolated over the GC region 
before they were subtracted from the total surface brightness maps. 
Figure \ref{galsketch} illustrates the different Galactic features
which contribute to emission and absorption in the direction of 
the GC and Table \ref{radmech} summarizes the different sources dominating 
emission and absorption in different wavelength regions.  
Figure \ref{rawcuts} illustrates that, particularly at NIR 
wavelengths, the resulting maps of the NB depend crucially on proper decomposition, 
extinction corrections, and modeling of the different galactic features. 
Therefore, this Section gives a detailed description of the data analysis process.

The major emission sources at NIR wavelengths are stars in 
the Galactic Disk (GD) and GB. In addition to these features, the ZL 
from interplanetary dust in the solar system contributes to the extended 
near- and mid-infrared (MIR) emission. 
Extinction by dust in the GD and in the 
GC region itself prevents observations of the stellar component 
of the NB at wavelengths shorter than 2\,\mim, and must be taken into 
account for wavelengths as long as $\lambda \sim 100$\,\mim\ 
(see Table \ref{radmech}). Most of the extinction in the GD arises from the 
Disk Molecular Ring at galactocentric radius $R_{\rm G} \sim 4-5$\,kpc. 
The far sides of the GB and GD also suffer from extinction by 
dust in the GC region.
In addition to these large-scale absorption features, some compact Giant Molecular 
Clouds (GMCs) along the line of sight appear as discrete patches of high extinction. 
These clouds could not be modeled individually, but had to be masked out before the 
models were fitted. 
Since GD and GB have nearly exponential emission profiles, 
proper modeling was essential to obtain reliable surface brightness interpolations 
over the GC region. 

At wavelengths $\lambda \ge 7$\,\mim, dust emission (i.e., mainly 
re-radiation of absorbed stellar emission) from the GD and GC 
region dominates over stellar emission. 
The MIR emission is generally dominated by UV-excited, non-transiently-heated 
very small grains and Polycyclic Aromatic Hydrocarbons (PAH's) 
with a possible contribution by hot dust in circumstellar shells and photospheric 
emission from cold luminous supergiants.  
The bulk of the dust ($> 99$\% by mass) 
in the GD and NB is relatively cold ($\sim$\,15--30\,K). 
Therefore, dust emission as tracer of the mass of ISM is best observed at FIR 
and submm wavelengths. 
In the range 3\,mm\,$\le \lambda \le$\,6\,cm, free-free emission from gas 
ionized by early-type MS stars dominates the radio emission of the NB. 
Synchrotron radiation from relativistic electrons becomes the strongest radio 
component for $\lambda \ge 6$\,cm, but free-free absorption has to be taken into 
account already at cm wavelengths, depending on the free-free emission measure. 

The total surface brightness $I_{\rm tot}$\ observed in the direction of the GC 
can be written as:
\begin{eqnarray}                               \label{eqitot}
I_{\rm tot} & = & I_{\rm ZL} + I_{\rm GD}^{\rm fg} + \nonumber \\ 
            &   & + e^{-\tau_{\rm GD}^{\rm fg}} (I_{\rm GB}^{\rm fg} + 
  I_{\rm NB} + e^{-\tau_{\rm CMZ}}(I_{\rm GB}^{\rm bg} + I_{\rm GD}^{\rm bg})) 
\end{eqnarray}
$I_{\rm xx}$\ are total ``observed'' intensities, i.e., 
integrated along the line-of-sight through the corresponding components, 
but not corrected for extinction by dust inside these components.
For the meaning of the subscripts we refer to Table \ref{abbtab}. 
In addition we use the superscripts ``fg'' = foreground and ``bg'' = background 
(with respect to the GC as seen from the Sun).
Surface brightness $I$\ and optical depth $\tau$\ are functions of wavelength $\lambda$, 
galactic longitude $l$, and latitude $b$. 

%____________________________

\subsection{Goal and strategy}                  \label{datangen}

\begin{figure}[htb]
\includegraphics[width=0.50\textwidth] {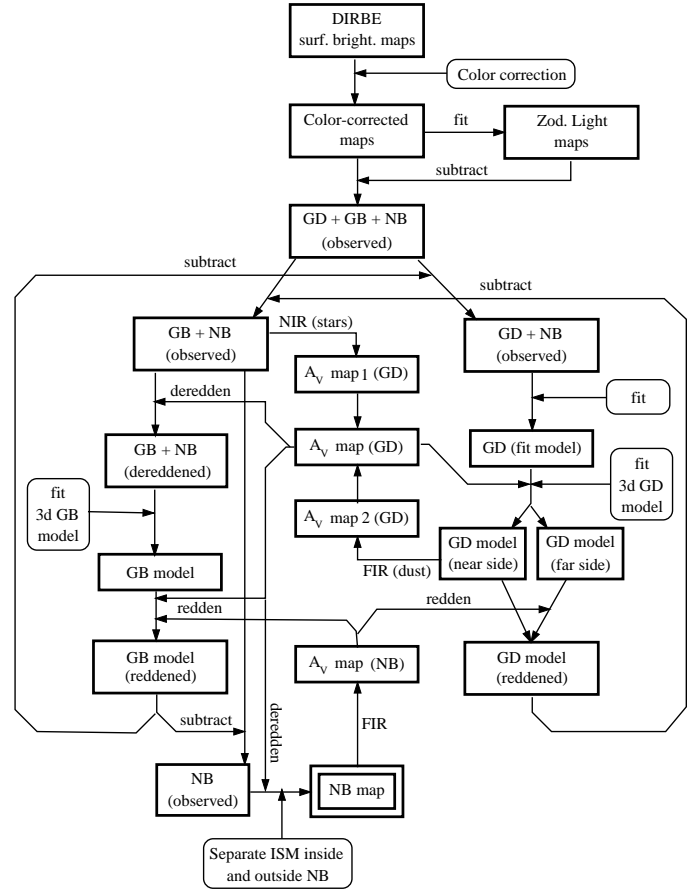}
% {nb3_fluxd_xfig_new.eps} (/scr/rl/PAPERS/NB3/nb_red_sketch_new.xfig)
\caption{\label{datanflux} 
 Flow chart of the COBE data reduction steps}
\end{figure}

The main goal of the data reduction was to derive 
surface brightness maps of the NB at different wavelengths as would be 
seen with no other emitting or absorbing matter present in the Galaxy. 
The other data mentioned in Sect. \ref{data} were reduced by its respective 
authors and used as published (references in Table \ref{datsum}).

For reasons mentioned in Sect \ref{radext}, surface brightness and 
extinction maps of the different galactic features 
had to be derived and modeled partially in an iterative way before they could 
be subtracted from the total surface brightness maps. 
The correct order of applying the different corrections 
to the data is described by Eq. (\ref{eqitot}). 
Figure \ref{datanflux} displays schematically the data flow with its 
recursive iterations.
The general data reduction steps described in the following subsections are:
\begin{enumerate}
\item Photometric corrections for the IRAS and COBE filter transmission 
      characteristics (Sect. \ref{photcorr}).
\item Model and subtract the ZL (Sect. \ref{zlcorr}).
\item Model the dust distribution in the GD 
      (Sect. \ref{datandisk} and Appendix \ref{apgalmod}).
\item Model the NIR surface brightness of the GB (Sect. \ref{datangb}).
\item Derive extinction maps due to dust in the GD and in the GC region (Sect. \ref{datanext}).
\item Fit, correct for extinction, and subtract the contributions by stars and dust 
      in the GD (Sect. \ref{datandisk}). 
\item Correct for extinction and subtract contributions by stars in the GB (Sect. \ref{datangb}). 
\item Correct the final NIR maps of the NB for extinction by dust inside the NB 
      (Sect. \ref{datanextnb}).
\item Decompose ISM in the CMZ into ISM located inside and outside the NB 
      (Sect. \ref{disismmorph}).
\end{enumerate}

%____________________________

\subsection{Photometric corrections}         \label{photcorr}

Colour correction factors for COBE (Appendix A of the 
COBE/DIRBE Explanatory Supplement) and IRAS maps (Wheelock et al. 1994) 
were derived for the peak
flux density ratios toward the NB, but were applied to the entire
images. The maximum correction had to be applied to the COBE 25\,\mim\ 
band (23\%). For all other bands, correction factors are smaller 
then 5\% of the peak flux densities.
In addition, the COBE DIRBE to IRAS photometric transformation was applied
to the IRAS ISSA maps (COBE/DIRBE Explanatory Supplement; only gain
factors were used) in order to account for systematic photometric
uncertainties in the IRAS calibration (particularly in the 60 and
100\,\mim\ bands).

%____________________________

\subsection{Zodiacal Light subtraction}   \label{zlcorr}

ZL emission was already subtracted from the IRAS data 
issued by IPAC.
In the {\it COBE} maps we modeled the ZL outside the Galactic 
Disk (between $|b|\,=\,11\ldots 14$\degr), interpolated linearly over the 
region of the disk, and subtracted the resulting maps from the COBE maps 
in the wavelength bands between 2.2\,\mim\ and 100\,\mim.  

%____________________________

\subsection{Extinction corrections}   \label{datanext}

Two major extinction-causing features have to be considered in the process of 
decomposing the observed surface brightness maps into individual galactic 
components (see Fig. \ref{galsketch} and Eq. (\ref{eqitot})):
\begin{enumerate}
\item Dust located at the near side of the GD absorbs light from the 
      GB and NB, as well as from the far side of the GD.
\item Dust in the CMZ absorbs light from the far sides of GB and GD, respectively. 
\end{enumerate}
The dust opacity spectrum used throughout this paper is described in Appendix \ref{apdust} 
(see Table \ref{optab}). 

%__________________

\subsubsection{Extinction by dust in the Galactic Disk}   \label{datanextgd}

The foreground extinction due to dust in the GD can be derived in two ways: 
\begin{enumerate}
\item From the observed reddening of the stellar NIR emission from the GB, and 
\item from the optically thin FIR dust continuum emission of the GD.
\end{enumerate} 
Neither of the two methods is straightforward. The first method requires the knowledge 
of the effective colour temperature of the GB and is very sensitive to how 
the GD emission is subtracted and where the extinction map is assumed not to be 
strongly affected by bright star-forming regions or very opaque individual GMCs. 
The second method requires a three-dimensional model of the dust distribution in 
the GD because the observed FIR continuum emission arises from lines of sight 
through the entire disk while extinction towards the GC is caused by 
dust on the near side only.

{\it Method 1:} 
Under the assumption that the intrinsic NIR colour temperature (i.e., the 
intrinsic surface brightness ratio $I_1/I_2$) of the GB is uniform (see Sect. \ref{resgb}) and 
can be derived at high latitudes where extinction by dust in the GD is negligible, 
a foreground extinction map can be derived from the observed NIR 
surface brightness ratio $I^{\prime}_1/I^{\prime}_2$:
\begin{equation}                                 \label{colav}
A_{\rm v} = 1.086\, \frac{\kappa_{\rm V}}{\kappa_2-\kappa_1} 
            \left({\rm ln}\left(\frac{I^{\prime}_1}{I^{\prime}_2}\right) - {\rm ln}\left(\frac{I_1}{I_2}\right)\right)
\end{equation}
Here, $\kappa_{\rm i}$\ are the dust opacities at the corresponding wavelengths 
(see Table \ref{optab}).
Due to the problems mentioned above, only an average latitude extinction profile 
could be derived rather than a complete extinction map. The intrinsic NIR surface 
brightness ratios $I_1/I_2$\ of the GB 
for the $\lambda$\,2.2, 3.5, and 4.8\,\mim\ DIRBE bands were  derived at $|b| > 3.5$\grd\ from the 
disk-subtracted (but not extinction-corrected) maps (see Sect. \ref{datandisk}). 
They correspond to an 
average effective temperature of stars in the GB of $\sim$4000\,K (see Sect. \ref{resgb}). 
This method yields a visual extinction between the Sun and the front side of the GB/NB 
of $A_{\rm V}\sim 14$\,mag. 
The corresponding average latitude extinction profile is shown in Fig. \ref{avlat}.

{\it Method 2:} 
In order to derive the extinction by dust on the near side of the GD from 
the optically thin FIR emission arising from lines of sight 
through the entire disk, we developed a three-dimensional model of 
the dust distribution in the GD, which is described in Appendix \ref{apgalmod}.
Our best model yields $A_{\rm V}\sim 15$\,mag towards the front side of the NB 
and the latitude $A_{\rm V}$\ profile at $l = 0$\grd\ shown in Fig. \ref{avlat}.

These two methods are completely independent. They yield similar, but systematically 
slightly different latitude extinction profiles.
Since both methods have their weak points, we finally used an extinction map 
which was obtained by scaling the extinction map derived by method 2 to the 
average of the two extinction profiles at $l = 0$\grd\ which is also shown in 
Fig. \ref{avlat}. 

%__________________

\subsubsection{Extinction by dust in the Galactic Center Region}   \label{datanextnb}

Since radiation from the far sides of GB and GD suffers from extinction 
by dust in the GC region, the modeled or interpolated surface brightness 
maps of these features have to be corrected for this extinction before they can be 
subtracted from the raw maps. 
Assuming a homogeneous and axisymmetric distribution of dust in the NB, the results from 
Sect. \ref{resgbext} (see Fig. \ref{avlat}) suggest a visual extinction of 
$\sim 30$\,mag through the entire NB. 
In contrast, we derive $A_{\rm V}{\rm (NB)}\sim 200$\,mag 
from the FIR dust continuum emission, assuming a 
homogeneous dust distribution (Sect. \ref{disismprop}) . 
Uncertainties in our dust model may contribute to, but cannot fully explain this discrepancy. 
Rather we argue in Sect. \ref{disismprop}, that the ISM in the CMZ 
is extremely clumpy with most of the mass being concentrated in 
small, ultra-opaque molecular clouds which cover $\sim$\,10\% 
of the area of the NB and block the stellar radiation nearly completely. 
In Sect. \ref{resnbsed} we derive $A_{\rm V}{\rm (NB)} \sim 20$\,mag for the average diffuse 
extinction through the NB. 
Therefore, the far side parts of GB and GD were corrected with an 
extinction map which was derived by scaling the total hydrogen column density map 
of the CMZ (Fig. \ref{nb_nh_inout}b) to 
$A_{\rm V} = 20$\,mag plus a 10\% 
blockage factor within the $N_{\rm H} = 2\times 10^{22}$\,\scm\ contour.  
The assumption that the degree of clumpiness 
of ISM is the same inside and outside the NB may not hold true completely, 
but is certainly a better approximation of the physical conditions than assuming 
homogeneous matter distribution.

To derive the size and morphology of the stellar NB, the NIR emission from the NB itself 
was, in addition, corrected for extinction by dust inside the NB. 
In Sect. \ref{disismmorph} we decompose the total hydrogen column density map of the CMZ 
into material located inside and outside the NB. 
An intrinsic extinction map of the NB was then derived by scaling its $N_{\rm H}$\ 
map (Fig. \ref{nb_nh_inout}d) in the same way as above 
(i.e., $A_{\rm V}$(NB)\,=\,20\,mag plus 10\% 
blockage). 
In correcting the NIR surface brightness of the NB for intrinsic extinction, we assumed that stars 
and diffuse dust within the NB are homogeneously mixed. 
Intrinsic ($I_{\nu}^\star$) and observed surface brightness ($I_{\nu}^{\prime}$) 
are then related by:
\begin{equation}                \label{eqextnbmix}
I_{\nu}^{\prime}(\lambda) = 0.9\: (1-e^{-\tau_{\rm NB}}) \frac{I_{\nu}^\star(\lambda)}{\tau_{\rm NB}}\quad .
%\frac{1-e^{-\tau_{\rm NB}}}{\tau_{\rm NB}}\, I_{\nu}^0(\lambda)\quad .
\end{equation}
%

%____________________________

\subsection{Galactic Disk subtraction}   \label{datandisk}

The ``observed'' surface brightness distribution of the GD 
at all wavelengths was 
derived by interpolating the total observed surface brightness between selected 
regions in the longitude intervals 3.8\degr$<|\,l\,|<$\,20\degr\ (i.e.
within the inner tangent to the Disk Molecular Ring, but outside the NB).
The regions used for deriving the average disk emission profile were
determined by comparing NIR and FIR maps in order to avoid 
areas which deviate strongly from the average disk profile, like
heavily FIR-emitting or NIR-absorbing GMCs, bright 
star-forming regions, or gaps in the dust disk. 
At NIR wavelengths (DIRBE bands 1 to 4), GD and GB are superimposed 
at longitudes $|l| \le 20$\degr\ and the surface brightness of the GD 
cannot be derived independently. Therefore, it was first modeled further 
out ($|l| > 20$\degr) and subtracted from the total emission before modeling and 
subtracting iteratively the surface brightness distributions of GB 
(see Sect. \ref{datangb}) and inner GD. 
Before the GD emission was finally subtracted from the total surface 
brightness maps, it was decomposed into contributions from the near and far 
sides (considering extinction by dust in the near side of the GD) 
and corrected for extinction in the GC region as described in Sect. \ref{datanext}.

%____________________________

\subsection{Galactic Bulge subtraction}     \label{datangb}

To derive NIR surface brightness maps of the NB, all contributions from  
the GB must be subtracted. 
Here, we assume that the GB and NB are morphologically 
different features and that the central surface brightness profile of the GB 
can be derived from its outer parts. Since the surface brightness profile 
of the GB has a centrally peaked exponential (or similar) shape and is 
not symmetric, proper extinction correction and interpolation over the central 
region is crucial. When properly corrected for foreground extinction, the 
often-referred peanut shape of the surface brightness distribution nearly disappears 
and a box-shaped asymmetric ellipsoid remains 
(Fig. \ref{gbint3dprof}a). 
Since there is striking evidence that the GB is a triaxial 
stellar bar with its near end at positive longitudes (e.g., Binney et al. 1991; 
Blitz \& Spergel 1991; Weiland et al. 1994), its surface brightness distribution 
cannot be modeled by a simple symmetric function. 
Therefore, we followed for the most part the approach of Dwek et al. (1995), 
Stanek et al. (1997), and Freudenreich (1998), and modeled the GB 
as a triaxial ellipsoidal bar with a centrally peaked volume emissivity $\rho$. 
We have tried sech$^2$, Gaussian, and exponential volume emissivity profiles and found 
the exponential models, represented by: 
\begin{equation}                      \label{eqigbprof}
\rho \propto {\rm exp}\,(-R_{\rm s}) 
\end{equation}
to give the best fits.
The shape of the bar was chosen to be a ``generalized ellipsoid'' 
(Althanassoula et al. 1990; Freudenreich 1998) represented by the 
effective radius $R_{\rm s}$:
\begin{equation}                      \label{eqigbrs1}
R_{\rm s}^{C_{\parallel}} = R_{\perp}^{C_{\parallel}} + 
                            \left(\frac{|Z^{\prime}|}{a_z}\right)^{C_{\parallel}}\ \ \ \ \ \ \ \ 
{\rm with} 
\end{equation}
\begin{equation}                      \label{eqigbrs2}
R_{\perp}^{C_{\perp}} = \left(\frac{|X^{\prime}|}{a_x}\right)^{C_{\perp}} +
                           \left(\frac{|Y^{\prime}|}{a_y}\right)^{C_{\perp}}
\end{equation}
where $a_x$, $a_y$, and $a_z$\ are the scale lengths and $C_{\perp}$\ and 
$C_{\parallel}$\ the face-on and edge-on shape parameters. 
The orientation of the bar was characterized 
by an in-plane tilt $\phi_x$\ (angle between bar's major axis and the line 
Sun -- Galactic Centre) 
and an out-of-plane tilt $\theta_z$. 
To derive the surface brightness, the models were integrated as seen from the 
position of the Sun which was adopted to be $Z_0 = 16$\,pc 
above the galactic mid-plane (e.g., Freudenreich 1998).

In the GD-subtracted COBE NIR maps of the GB corrected for 
extinction by dust in the GD (Sect. \ref{datanext}), both the region of 
the CMZ and of individual opaque GMCs were masked out before the 
models were fitted to the observed surface brightness distribution. 
Parameters of our best-fit GB model are given in Table \ref{gbpartab}. 
The Spectral Energy Distribution (SED) of the GB is derived and analysed in 
Sect. \ref{resgb} (Fig. \ref{gbsed}).
Figure \ref{gbint3dprof} compares the model 
with the observed surface brightness distribution. 

The modeled surface brightness maps of the GB were then corrected for 
extinction by dust in the GC region (i.e., the CMZ):
\begin{equation}                      \label{eqgbext}
I^{\prime}_{\rm GB} \simeq I_{\rm GB}\,\frac{(1+e^{-\tau_{\rm CMZ}})}{2} 
 \end{equation}
(see Sect. \ref{datanextnb} for $\tau_{\rm CMZ}$) and subtracted from the 
COBE maps.

%______________________________________________________________

\section{Basic Parameters derived from the COBE DIRBE and IRAS ISSA maps}  \label{results}

%____________________________

\subsection{Extinction towards the Galactic Centre Region}                \label{resgbext}

\begin{figure}[htb]
\includegraphics[width=0.40\textwidth] {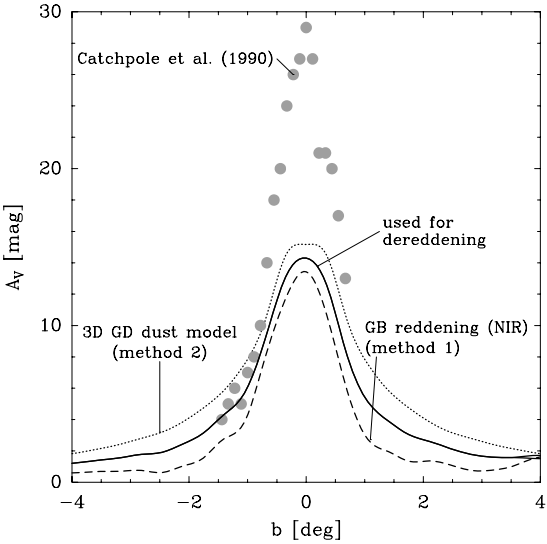}
% {avlat_paper.ps} (avlat.greg)
\caption{\label{avlat} 
 Visual extinction towards the front-side of Galactic and Nuclear Bulge, 
 due to dust in the Galactic Disk. 
 {\it Dotted curve:} Extinction profile derived from 3D model of the 
 dust distribution in the GD (Sect. \ref{datanextgd}, Method 2). 
 {\it Dashed curve:} Extinction profile derived from reddening of the 
 GB (stellar) NIR emission (Sect. \ref{datanextgd}, Method 1). 
 {\it Solid curve:} Average of the two former profiles, 
 which we use to correct the NIR maps of the NB. 
 {\it Grey dots:} Extinction profile derived by Catchpole et al. (1990) from 
 colour-magnitude diagrams of stars in the direction of the GC.
 This profile relates to the Centre of the Galaxy, i.e. it includes extinction 
 by dust inside the NB.}
\end{figure}

In Fig. \ref{avlat}, the latitude profiles of extinction between the 
Sun and the front sides of GB and NB at $l$\,=\,0\grd, as derived in 
Sect. \ref{datanext}, are shown. The average extinction profile, which we finally 
used, can be approximated by the sum of two sech$^2$\ functions:
\begin{equation}                      \label{eqavprof}
A_{\rm V}(b) \simeq \sum_{i=1}^{2} A_{{\rm V} i}(0)\,{\rm sech}^2\left(\frac{b}{\sigma_{b i}}\right)
\end{equation}
with $A_{{\rm V} 1}(0)\,\simeq$\, 11\,mag, $\sigma_{b 1} \simeq 0.78$\grd, 
$A_{{\rm V} 2}(0) \simeq 3.5$\,mag, and $\sigma_{b 2} \simeq 3.3$\grd. 
The total mid-plane extinction towards the front-sides of GB and NB 
is $A_{\rm V} = 14.5\pm2$\,mag. 
Compared to the well-established value $A_{\rm V}$(Sun--GC)\,$\simeq$\,30\,mag 
(e.g., Lebofsky \& Rieke 1987; Catchpole et al. 1990), 
our result suggests that $\sim$50\% 
of the absorbing dust column between the 
Sun and the GC is located in the GD and the other 50\%  
is located in the GC region. 
Figure \ref{avlat} also shows the extinction profile derived by Catchpole et al. (1990) 
from colour-magnitude diagrams of stars in the direction of the GC. 
This profile represents the extinction to stars in the immediate vicinity of 
the GC and accounts for extinction by dust in the GD {\it and} 
in the GC region.

%____________________________

\subsection{Galactic Bulge}                                           \label{resgb}

\begin{figure}[htb]
\includegraphics[width=0.45\textwidth]{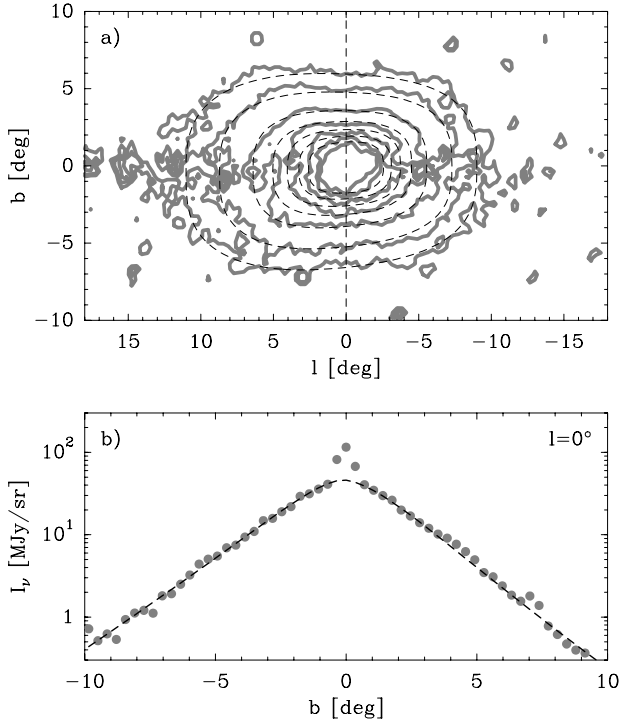}
% {gb_int3d_paper.ps} (gb_int3d_plot_paper.graphic)
\caption{\label{gbint3dprof} 
 2.2\,\mim\ surface brightness distribution of the Galactic and Nuclear Bulge. 
 In order to achieve the best signal-to-noise ratio, a weighted average of the 
 2.2, 3.5, and 4.9\,\mim\ maps, scaled to the 2.2\,\mim\ surface brightness, 
 is shown.
 Contributions from the Galactic Disk are subtracted and the emission is 
 dereddened for extinction by foreground dust.~~
 {\bf a)} Contour maps of the observed dereddened (thick grey lines) and modeled 
          (dashed lines) NIR surface brightness distribution. Levels are at 5, 10, 20, ..., 60\%\ 
          of modeled peak surface brightness.~~
 {\bf b)} Latitude profile at $l = 0$\grd\ of the observed dereddened (grey dots) 
     and modeled (dashed line) NIR surface brightness distribution of the GB.}
\end{figure}

\begin{figure}[htb]
\includegraphics[width=0.49\textwidth]{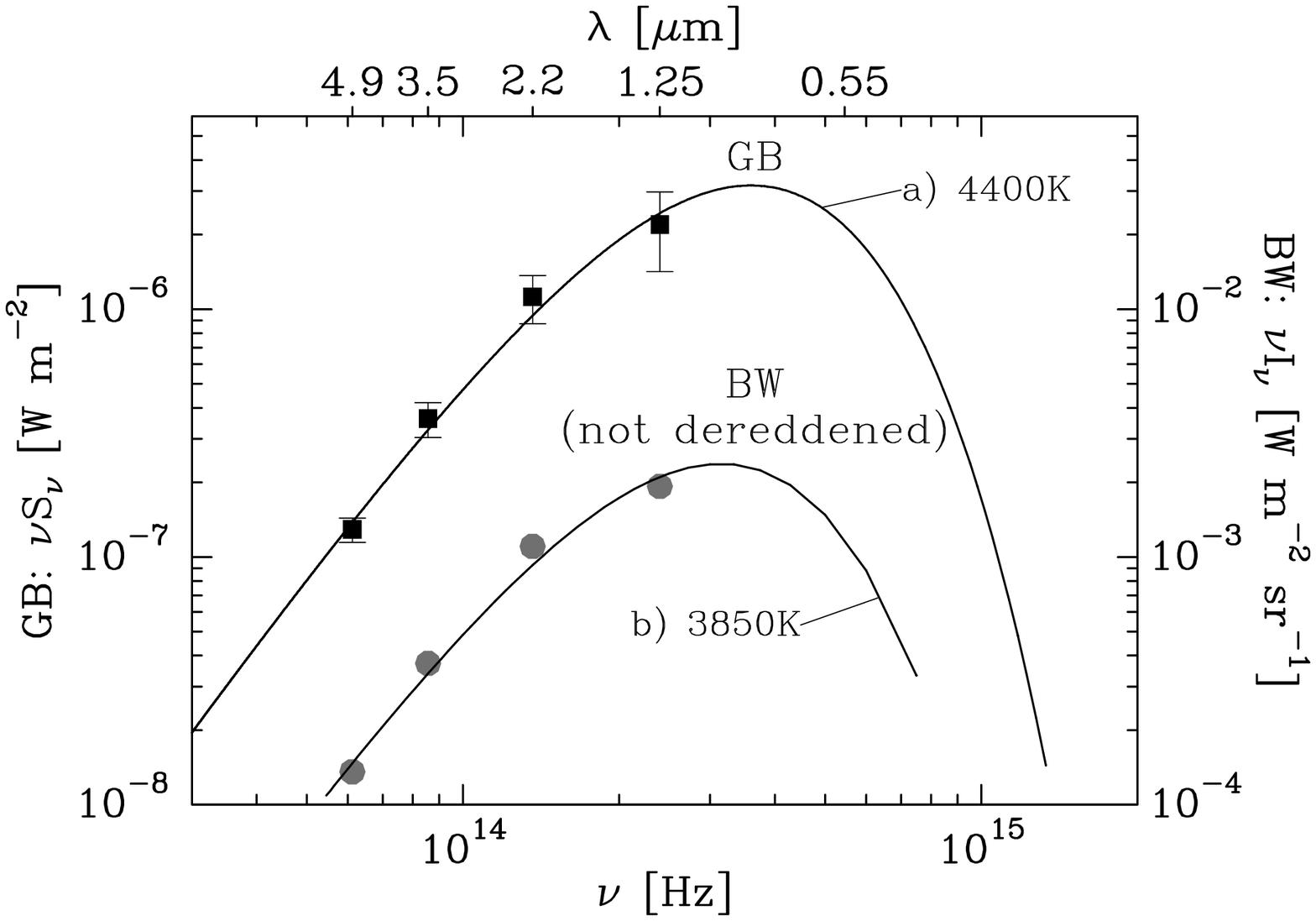}
% {gbsed.ps} (gb_sed.greg)
\caption{\label{gbsed}
 Spectral Energy Distribution of the Galactic Bulge. \newline 
 Black squares refer to the integrated flux density of the model fit (left scale). 
 Error bars refer to the uncertainties of the extinction corrections   
 ($\sigma(A_{\rm V}) \sim \pm 2$\,mag). 
 Curve a) shows the best black-body fit to these points weighted with the 
 uncertainties ($T_{\rm eff} \sim 4400$\,K). 
 Filled circles refer to the surface brightness at Baade's Window (BW) in the 
 disk-subtracted, not extinction-corrected COBE maps (right scale). 
 Curve b) shows the best black-body fit to these points ($T_{\rm eff} \sim 3850$\,K.)}
\end{figure}

\begin{table}[htb]                  
 \caption[]{Galactic Bulge model parameter values\\ 
            (see Eqs. (\ref{eqigbprof}) to (\ref{eqigbrs2}))}               \label{gbpartab}
  \begin{flushleft}
   \begin{tabular}[t]{ll} 
\hline \noalign{\smallskip}
Parameter & Value \\
\noalign{\smallskip} \hline \noalign{\smallskip}
Sun dist. from Galactic Center $R_0$\ia      & 8.5\,kpc  \\
Sun dist. from Galactic Plane $Z_0$\ia       & 16\,pc    \\
Bar $X$\ scale length $a_x$\                 & 1.1\,kpc  \\
Bar $Y$\ scale length $a_y$\                 & 0.36\,kpc \\ 
Bar $Z$\ scale length $a_z$\                 & 0.22\,kpc \\
Bar face-on shape parameter $C_{\perp}$\     & 1.6       \\
Bar edge-on shape parameter $C_{\parallel}$\ & 3.2       \\
Bar in-plane tilt angle $\phi_x$\            & 15\grd    \\
Bar out-of-plane tilt angle $\theta_z$\      & -1\grd    \\
$S_{\nu}(1.25\,\mim)$\                       & (0.92$\pm$0.32)\,MJy \\
$S_{\nu}(2.2\,\mim)$\                        & (0.82$\pm$0.18)\,MJy \\
$S_{\nu}(3.5\,\mim)$\                        & (0.42$\pm$0.07)\,MJy \\
$S_{\nu}(4.9\,\mim)$\                        & (0.21$\pm$0.02)\,MJy \\
Average effective temperature $T_{\rm eff}$\ & (4400$\pm$400)\,K      \\
Total Luminosity $L_{\rm bol}$\ib            & (1.0$\pm$0.3)$\times$10$^{10}$\,\lsun \\
\noalign{\smallskip} \hline\noalign{\smallskip} 
\end{tabular} 
\end{flushleft}
\begin{list}{}{}
\item[$^{a)}$] Fixed input parameter
\item[$^{b)}$] Integration of the SED (Fig. \ref{gbsed})
\end{list}
\end{table}

The parameters of our 'best-fit' triaxial bar model of the GB 
(Sect. \ref{datangb} and Eqs. (\ref{eqigbprof}) to (\ref{eqigbrs2}))
are summarized in Table \ref{gbpartab}, and the result is shown in 
Fig. \ref{gbint3dprof}. 
Our model yields similar axis ratios,  
($a_x$\,:\,$a_y$\,:\,$a_z$)\,$\sim$\,(3\,:\,1\,:\,0.6), and projected surface 
brightness profiles as the best-fit models of Dwek et al. (1995) and 
Freudenreich (1998). 
The tilt angle of the bar's major axis against the line of sight 
in our model is $\phi_x \sim$15\degr. 
Other studies favor a somewhat larger tilt angle of 16-25\degr\ 
(e.g., Englmaier \& Gerhard 1999).
Note that our goal was to obtain a good fit to the projected surface brightness 
distribution of the GB in order to subtract it from the COBE maps, rather than 
to derive a detailed model of the three-dimensional morphology of the bar. 
We did not explore the whole parameter space and we do not claim that our 
model is unique.

In Figure \ref{gbsed}, the integrated SED of the GB is shown. 
The NIR flux densities were derived by integrating the 
models at 2.2, 3.5, and 4.9\,\mim\  within $l = \pm 20$\grd\ and $b = \pm 10$\grd\ and 
are listed in Table \ref{gbpartab}. 
A weighted least-square black-body fit to these points yields an average effective 
temperature of the Bulge stars of $T_{\rm eff}({\rm GB}) = 4400\pm 400$\,K. 
Since we found no evidence for a colour gradient in the GB, 
a lower limit to the average effective stellar temperature can be derived 
from the uncorrected NIR surface brightness ratios at high latitudes 
where the extinction is low (e.g., Baade's Window: 
$l,b\,\sim$\,1\grd,-3.9\grd, $A_{\rm K} \sim 0.13$\,mag).
The best black-body fit to these points yields $\sim 3850$\,K 
which is in good agreement with our estimate of $T_{\rm eff}$(GB).
We tried different bar models and fitting routines; the derived total 
flux densities were nearly independent of the particular model.

Our result agrees well with the volume emissivity ratios derived by 
Freudenreich (1998) for his models S and E. 
The total luminosity of the GB derived from our SED fit to the NIR 
flux densities is $L_{\rm GB} = 1.0\pm 0.3\times 10^{10}$\,\lsun. 
Since there is no evidence for ongoing star formation and the presence of 
hot massive stars in the GB, which would contribute considerable luminosity 
at shorter wavelengths, this value represents the bolometric 
luminosity of the GB. 
Our estimate is intermediate to the GB luminosities of 
$5.3\pm 1.6\times 10^9$\,\lsun\ derived by Dwek et al. (1995) 
from COBE NIR observations and 
the 2.2\,\mim\ luminosity function of Bulge stars and of 
$\sim 2\times 10^{10}$\,\lsun\ derived by Maihara et al. (1978) 
from their 2.4\,\mim\ observations of the GC region. 
Dwek et al. (1995) derive a total stellar mass of the GB of 
$M_{\rm GB}\sim 1.3\pm 0.5\times 10^{10}$\,\msun\ 
and a mass-to-luminosity ratio of $\sim 2$\,\msun/\lsun. 
The central mass and luminosity volume densities of the GB are 
$\rho_{\rm M}\sim 8\pm2$\,\msun\,pc$^{-3}$\ and 
$\rho_{\rm L}\sim 4\pm1$\,\lsun\,pc$^{-3}$, 
respectively. 
The numbers depend only weekly on the value of the total luminosity, which 
is more sensitive to the exact integration area and how the outer GB is modeled.

%%%%%%%%%%%%%%%%%%%%%%%%%%%%%%%%%

%____________________________

\subsection{Nuclear Bulge and Central Molecular Zone}               \label{resnb}
%____________________________

\subsubsection{NIR through Radio images of the Nuclear Bulge}       \label{resnbmaps}

\begin{figure}[htbp]
\includegraphics[width=0.49\textwidth]{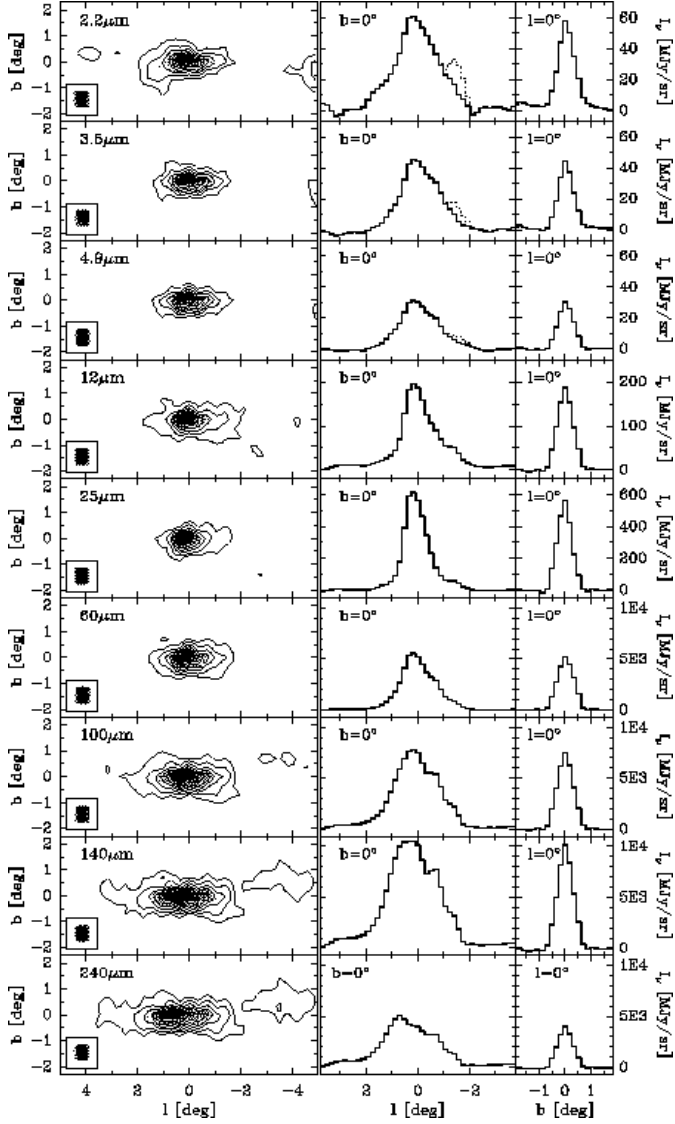}
% {nb_cobemaps_paper.ps} (gc_cobe_maps_paper_plo.graphic)
\caption{\label{cobemapsplo}
 Surface brightness maps of the Nuclear 
 Bulge at 9 wavelengths between 2.2\,\mim\ and 240\,\mim\ as seen by COBE. 
 ZL, emission from GD and GB, 
 and a point-like NIR source at $l,b \sim -1.4$\degr,0\degr\
 (dotted curve in the longitude profiles; see text) have been 
 subtracted. The data are corrected for foreground extinction by dust 
 in the GD, but not for 
 extinction by dust inside the NB. Lowest contour levels are at 10\% 
 of the maximum in the NIR maps and 5\% 
 in all other maps.
 Small boxes in the lower left of the maps show the DIRBE beam 
 (HPBW 0.7\degr) at the corresponding wavelength bands.
 The middle and right panels show the corresponding longitude 
 and latitude profiles at $b$\,=\,0\grd\ 
 and $l$\,=\,0\grd, respectively.}
\end{figure}

\begin{figure}[htbp]
\includegraphics[width=0.49\textwidth]{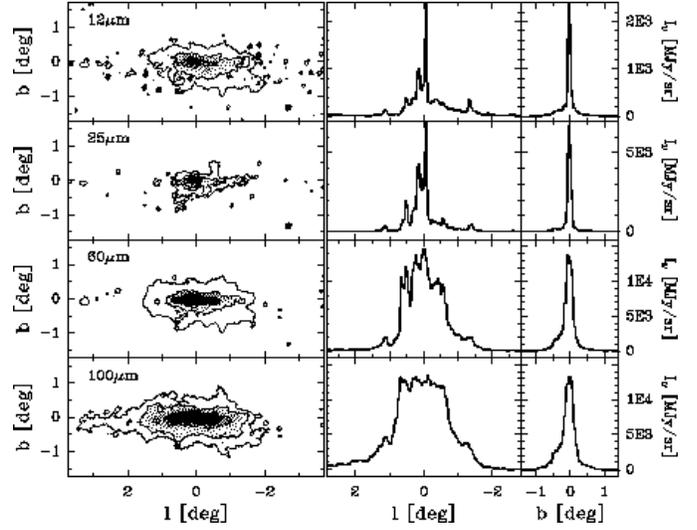}
% {nb_irasmaps_paper_new.ps} (gc_iras_fir_paper_plo.graphic)
\caption{\label{irasmapsplo}
 Thermal dust emission from the Nuclear Bulge as seen by IRAS (ISSA maps, 
 angular resolution $\sim$2\arcmin). 
 ZL and GD are subtracted and the 
 data are corrected for foreground extinction by dust in the GD.
 Middle and right panels show the corresponding longitude 
 and latitude profiles as in Fig. \ref{cobemapsplo}.}
\end{figure}

\begin{figure}[htbp]
\includegraphics[width=0.49\textwidth]{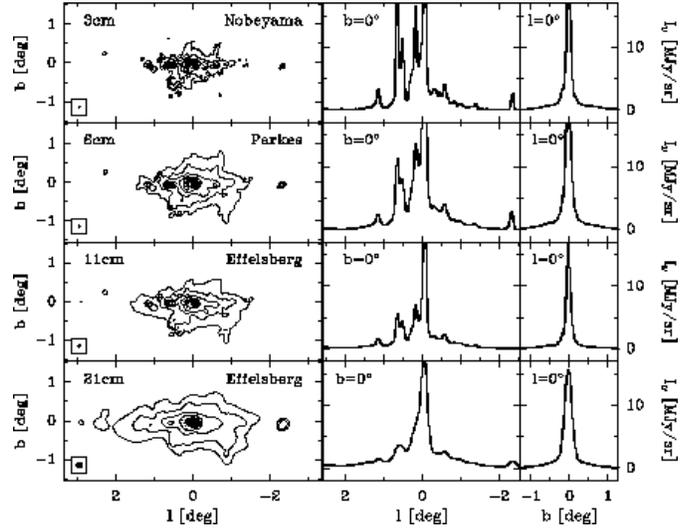}
% {nb_radiomaps_paper.ps} (gc_radio_paper_plo.graphic)
\caption{\label{radiomapsplo}
 Radio continuum emission from the Nuclear Bulge. These data were obtained 
 from different data bases (see Table \ref{datsum} for references). 
 FWHM beam sizes are shown as black circles in the lower left corners of the maps.
 Middle and right panels show the corresponding longitude 
 and latitude profiles as in Fig. \ref{cobemapsplo}.}
\end{figure}

Images and surface brightness profiles of the NB 
ranging from $\lambda$\,2.2\,\mim\ to 240\,\mim\ 
derived from the COBE DIRBE and IRAS ISSA maps 
and processed as described in Sect. \ref{datan} 
are shown in Figs. \ref{cobemapsplo} and \ref{irasmapsplo}. 
Note that these images are not corrected for extinction by dust 
inside the NB and that the FIR maps contain emission from dust in the NB 
{\it and} in the outer CMZ (see Sect. \ref{disismmorph}).
For comparison, we also show radio continuum maps of the NB which were 
obtained from different data bases (Fig. \ref{radiomapsplo}; see Table 
\ref{datsum} for references).
At NIR wavelengths, the NB emits stellar continuum radiation, mainly from 
red giants and supergiants.
The NB also emits strong MIR/FIR/submm dust emission together with free-free 
and synchrotron radio emission. In addition, the molecular gas exhibits strong 
line emission (see Fig. \ref{nb_nh_inout}a). 
Pertinent fit and other parameters are given in Table \ref{nbpar}. 

The COBE NIR images show a relatively compact source elongated $\sim$2--3\degr\ 
in longitude and unresolved in latitude (upper three panels in 
Fig. \ref{cobemapsplo}). 
The 4.9\,\mim\ image, which is least affected by extinction, shows this elongated 
structure most clearly. 
The 2.2\,\mim\ image exhibits some low-level extended emission, which may be 
due to uncertainties in the extinction correction and de-composition process 
and will, therefore, not be discussed here. 
At $\lambda$\,1.25\,\mim\ (DIRBE band 1), the extinction towards the NB and 
the uncertainties 
in the data reduction are too large to produce a meaningful 
result; therefore we don't show nor use the 1.25\,\mim\ image here.
Visible especially at 2.2\,\mim\ is a source at $l \sim -1.4$\degr\ 
which has a ``bluer'' SED than that of the average NB. 
This source, referred to as IRAS\,17393$-$3004, is the 
brightest NIR source seen in the direction of the NB. It has been 
investigated by Philipp et al. (1999b) and is classified as an 
M4 supergiant surrounded by a dust shell and located at a distance 
of $\le$\,4.7\,kpc from the sun.
Therefore, we modeled this source by a Gaussian and subtracted it from the 
final NIR images (dotted curve in the 
longitude profiles in Fig. \ref{cobemapsplo}). 
The surface brightness longitude profiles of the resulting images are 
asymmetric and peak at $l \sim +0.2$\degr. Although this is marginal 
compared to the 0.7\degr\ DIRBE beam, this asymmetry is seen at all 
wavelengths and only at $\lambda \ge 140$\,\mim\ does the peak shift 
towards larger $l$. 

The COBE MIR images are less extended in longitude than the NIR and FIR 
images, indicating that hot dust emission arises mainly from the inner part 
of the NB. The 12\,\mim\ image shows an extended, low-level halo. 
Since the strong PAH feature at 11.3\,\mim\ lies in this band, this may indicate 
that UV-excited PAHs in and around the NB are more extended than ``normal'' 
warm and hot dust. The 12\,\mim\ halo is also seen 
in the IRAS image which has a much higher angular resolution 
(Fig. \ref{irasmapsplo}). The IRAS 12 and 25\,\mim\ images, too, indicate that the 
bulk of the MIR emission arises from a very compact region which includes 
the {\it Sgr\,A Radio Complex} and the Radio {\it Arc} and {\it Bridge} 
(MDZ96; Sofue 1994; Reich 1994).  

The COBE FIR images show the same general morphology as the NIR images, i.e., 
the emission is extended in $l$\ with an asymmetry towards positive $l$, 
and is basically unresolved in $b$. 
While at negative longitudes the emission extends out to $l \sim -1.7$\degr\ 
at all wavelengths, its extent towards positive longitudes increases 
with increasing wavelength up to nearly 4\degr\ at 240\,\mim. This indicates 
a large excess of cold dust at positive $l$\ which extends much further out then 
the stellar NIR emission and the emission of warm dust. 
The excess of cold dust emission at $l > +1.7$\degr\ can also be seen in the 
100\,\mim\ IRAS image (Fig. \ref{irasmapsplo}).
The COBE 140 and 240\,\mim\ 
images likewise show some extended low-level emission from cold dust outside the NB 
at $-2$\degr\,$> l > -5$\degr. 
At much higher angular resolution than the COBE maps, the IRAS
60 and 100\,\mim\ images show a very distinct and narrow ridge of emission 
between $l \sim \pm0.7$\degr\ (Fig. \ref{irasmapsplo}), which is unresolved 
(in latitude) in the COBE maps (Fig. \ref{cobemapsplo}). 
The morphology of the extended low-level FIR emission in the IRAS maps 
compares well to that of the COBE images. 

Although the morphology of the radio continuum maps is mainly determined 
by emission from distinct compact sources and source complexes, their overall 
shape is similar to that of the MIR IRAS maps, thus indicating that 
free-free emission from ionized gas (as observed at 6\,cm) and 
IR emission from hot dust have a similar distribution. 
The 11 and 21\,cm images are dominated by non-thermal 
synchrotron emission. 

%____________________________

\subsubsection{The Spectral Energy Distribution of the Nuclear Bulge}  \label{resnbsed}

\begin{table*}[htb]                  
 \caption[]{Flux densities and source sizes derived from IR and radio images of the Nuclear Bulge 
            \label{nbpar}}\vspace{-0mm}
  \begin{flushleft}
   \begin{tabular}[t]{rcccllll} 
\hline \noalign{\smallskip}
$\lambda_{\rm c}$~ &
$\nu_{\rm c}$~     &
Telescope          &
Beam               &
~~~$I_{\nu}$\ia    &
~$S_{\nu}^{\rm tot}$&
\multicolumn{2}{c}{FWHM size (observed)}\\
                   &
                   &
                   &
(HPBW)             &
at (0\degr,0\degr) &
                   &
~$\Delta l\,\ (l_{\rm max},\,l_{\rm min})$ &
~$\Delta b\,\ (b_{\rm min},\,l_{\rm max})$ \\
{[$\mu$m]}         &
[Hz]               &
                   &
[arcmin]           &
[MJy/sr]           &
~[Jy]              &
~~~~~~[degr]       &
~~~~~~[degr]       \\
\noalign{\smallskip} \hline \noalign{\smallskip}
2.2    & 1.36E14 & COBE       & 42   & 5.6E1\,(1.2E2)\ib & 2.1E4\,(4.0E4)\ib    & 1.65 (0.90,\,$-$0.75) & 0.65 ($-$0.35,\,0.30) \\
3.5    & 8.57E13 & COBE       & 42   & 4.4E1\,(6.5E1)\ib & 1.6E4\,(2.2E4)\ib    & 1.65 (0.85,\,$-$0.80) & 0.73 ($-$0.41,\,0.32) \\
4.9    & 6.12E13 & COBE       & 42   & 3.0E1\,(3.8E1)\ib & 1.1E4\,(1.3E4)\ib    & 1.58 (0.75,\,$-$0.83) & 0.70 ($-$0.40,\,0.30) \\[0.7mm]
12     & 2.50E13 & COBE       & 42   & 1.6E2             & 5.6E4                & 1.24 (0.62,\,$-$0.62) & 0.74 ($-$0.37,\,0.37) \\
12     & 2.50E13 & IRAS       &  2   & 2.0E2\ic          & 1.2E5                & 0.30 (0.19,\,$-$0.11)\id & 0.17 ($-$0.08,\,0.09)\id \\
25     & 1.20E13 & COBE       & 42   & 5.3E2             & 1.4E5                & 1.01 (0.60,\,$-$0.41) & 0.64 ($-$0.37,\,0.27) \\
25     & 1.20E13 & IRAS       &  2   & 6.0E2\ic          & 2.0E5                & 0.30 (0.19,\,$-$0.11)\id & 0.14 ($-$0.07,\,0.07)\id \\
60     & 5.00E12 & COBE       & 42   & 5.1E3             & 2.0E6                & 1.44 (0.70,\,$-$0.74) & 0.77 ($-$0.43,\,0.34) \\
60     & 5.00E12 & IRAS       &  2   & 4.2E3\ic          & 2.1E6                & 1.30 (0.67,\,$-$0.63) & 0.27 ($-$0.17,\,0.10) \\
100    & 3.00E12 & COBE       & 42   & 7.4E3             & 4.1E6                & 1.94 (1.02,\,$-$0.92) & 0.80 ($-$0.47,\,0.33) \\
100    & 3.00E12 & IRAS       &  2   & 5.6E3\ic\ie       & 3.9E6\ie             & 1.57 (0.86,\,$-$0.71) & 0.32 ($-$0.19,\,0.13) \\
140    & 2.14E12 & COBE       & 42   & 1.0E4             & 6.2E6                & 2.22 (1.27,\,$-$0.95) & 0.78 ($-$0.46,\,0.32) \\
240    & 1.25E12 & COBE       & 42   & 5.0E3             & 2.8E6                & 2.40 (1.45,\,$-$0.95) & 0.74 ($-$0.44,\,0.30) \\[0.7mm]
800    & 3.75E11 & JCMT       & 0.5  & 1.1E2\ic          & ---\ief              & ---\ief~~~~~~~~~~~~~~~& ---\ief~~~~~~~~~~~~~~~\\
1200   & 2.50E11 & MRT        & 0.18 & 3.1E1\ic          & ---\ief              & ---\ief~~~~~~~~~~~~~~~& ---\ief~~~~~~~~~~~~~~~\\[0.7mm]
3\,cm  & 1.00E10 & Nobeyama   & 3.0  & 1.9E1\ic          & 1.8E3                & ---\id~~~~~~~~~~~~~~~ & ---\id~~~~~~~~~~~~~~~ \\
6\,cm  & 5.00E09 & Parkes     & 4.1  & 2.0E1\ic          & 2.5E3                & ---\id~~~~~~~~~~~~~~~ & ---\id~~~~~~~~~~~~~~~ \\
11\,cm & 2.70E09 & Effelsberg & 4.3  & 1.0E1\ic          & 2.0E3                & ---\id~~~~~~~~~~~~~~~ & ---\id~~~~~~~~~~~~~~~ \\
21\,cm & 1.41E09 & Effelsberg & 9.4  & 1.6E1\ic          & 1.8E3                & ---\id~~~~~~~~~~~~~~~ & ---\id~~~~~~~~~~~~~~~ \\
\noalign{\smallskip} \hline\noalign{\smallskip} 
\end{tabular} 
\end{flushleft}
\begin{list}{}{}
\item[$^{a)}$] Beam-averaged surface brightness
\item[$^{b)}$] First value: corrected for foreground extinction, (second value): additionally corrected 
      for extinction by dust inside the NB (see Sect. \ref{datanextnb})
\item[$^{c)}$] Smoothed to the COBE DIRBE resolution (0.7\degr)
\item[$^{d)}$] Emission is dominated by strong point sources. Size derived after subtraction of point sources
\item[$^{e)}$] The IRAS 100\,\mim\ emission is saturated in the central region of the NB
\item[$^{f)}$] The submm continuum maps cover only the inner $\sim$\,1.4\degr\,$\times$\,0.4\degr.
\end{list}
\end{table*}

\begin{figure}[htb]
\includegraphics[width=0.45\textwidth]{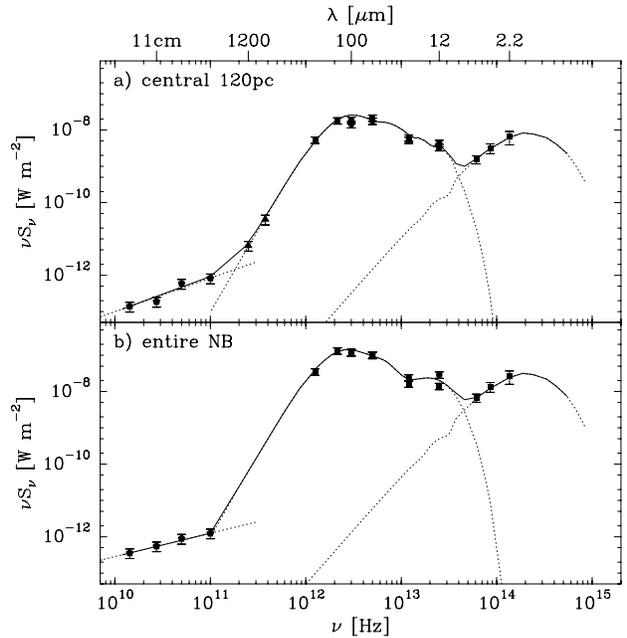}
% {gcnb_sed_av20b10_paper_new.ps} (gcnb_sed_av20b10_paper_new.graphic)
\caption{\label{gcnb_sed}
  Spectral energy distribution of a) the central 120\,pc and b) the entire NB.
  The different symbols represent 
  COBE (squares), IRAS (circles), ground-based submm ( triangles), and
  radio (pentagons) data. The IRAS and ground-based radio 
  data were smoothed to the COBE resolution.
  The solid lines show the best model fit to the observed dust and stellar 
  continuum data (corrected for foreground extinction only).    
  The dotted curves show the underlying dust and stellar continuum.
  The contribution of radio continuum (free-free and synchrotron radiation) emission 
  is also marked by dotted lines. Fit parameters are given in Table \ref{gcnb_sed_tab}. 
  Sect. \ref{resnbsed} describes how the SED fits were derived. 
  Note that these SEDs do not account for the contribution of hot massive stars 
  at $\lambda<2$\,\mim.}
\end{figure}

\begin{table*}[htb]                  
 \caption[]{Characteristics of dust and stars in three concentric regions of the NB derived from the SED
            \label{gcnb_sed_tab}}
  \begin{flushleft}
   \begin{tabular}[t]{lcrcccrclcrcl} 
\hline \noalign{\smallskip}
Region (diameter):   & \multicolumn{4}{c}{Central 1.25\,pc\,\ia} & 
              \multicolumn{4}{c}{Central 120\,pc\,\ib}  & 
              \multicolumn{4}{c}{Entire NB\,\ic}        \\
 && $T$~ & $M_{\rm H}$ & $L$      && $T$~ & $M_{\rm H}$ & ~~$L$    && $T$~ & $M_{\rm H}$ & ~~$L$    \\
 && [K] & [\msun]     & [\lsun]  && [K] & [\msun]     & ~[\lsun] && [K] & [\msun]     & ~[\lsun] \\
\noalign{\smallskip} \hline \noalign{\smallskip}
Cold dust                      &&     40  & 3.3E$+$2 & 1.7E$+$5      &&   24.7 & 3.6E$+$6 & 5.6E$+$7 &&   21.2 & 4.7E$+$7 & 2.7E$+$8 \\
Warm dust                      &&    150  & 5.8E$+$0 & 3.3E$+$6      &&   59   & 1.8E$+$4 & 3.6E$+$7 &&   49   & 4.2E$+$5 & 2.1E$+$8 \\
Hot dust                       &&    350  & 5.0E$-$3 & 2.1E$+$5      &&   200  & 1.1E$+$1 & 1.1E$+$7 &&   220  & 1.6E$+$2 & 7.4E$+$7 \\
Dust (total)                   &&     42  & 3.3E$+$2 & 3.7E$+$6      &&   24.9 & 3.6E$+$6 & 1.0E$+$8 &&   21.4 & 4.7E$+$7 & 5.5E$+$8 \\
Cool stars(obs)\,\id\ie        &&    ---  & ---      & ---           &&   ---  & ---      & 3.2E$+$7 &&   ---  & ---      & 1.2E$+$8 \\
Dust + cool stars (obs)\,\id\ie&&    ---  & ---      & 6.8E$+$6      &&   ---  & ---      & 1.3E$+$8 &&   ---  & ---      & 6.7E$+$8 \\
Cool stars(intrinsic)\,\id\ief &&  4\,000 & ---      & 3.1E$+$6\,\ig && 4\,600 & ---      & 1.9E$+$8 && 4\,400 & ---  & 6.8E$+$8 \\[0.5ex]
%Hot stars(intrinsic)           && 25\,000 & ---      & 9.2E$+$7\,\ig && ---\ih & ---      & 1.4E$+$9 && ---\ih & ---      & 8.2E$+$8 \\
%All stars                      &&         &          & 1.2E$+$8\,\ig && ---    & ---      & 1.6E$+$9 && ---    & ---      & 1.5E$+$9 \\ [0.5ex]
\noalign{\smallskip} \hline\noalign{\smallskip} 
  \end{tabular} 
 \end{flushleft}
\begin{list}{}{}
\item[$^{a)}$] See Paper I, Table 1b
\item[$^{b)}$] Corresponds to one 0\grdp7 COBE DIRBE pixel
\item[$^{c)}$] Includes dust in the immediate environment of the NB, but not the entire CMZ 
      ($-$2\grdp5\,$\le\,l\,\le$\,3\grdp8)
\item[$^{d)}$] Stellar NIR luminosity $L_{\rm NIR}$\ (see definition below), does not account for hot (i.e., young massive) stars
\item[$^{e)}$] Corrected for foreground extinction, but not for extinction by dust inside the NB
\item[$^{f)}$] Additionally corrected for extinction by dust inside the NB (see Sect. \ref{datanextnb})
\item[$^{g)}$] Values from Paper\,I, Table\,1. Note that $A_{\rm V} = 31$\,mag was used as total foreground extinction
\item[$^{h)}$] Not derived since no data points shortward of 2.2\,\mim\ available
\end{list}
\end{table*}

To compile the radio through NIR SEDs 
of the central 120\,pc and of the entire NB, the corresponding central 
surface brightnesses and integrated flux densities were derived at all wavelengths 
from the final maps presented in Figs. \ref{cobemapsplo} through \ref{radiomapsplo}. 
In these maps, all contributions but the NB were removed and the remaining emission was 
corrected for extinction by foreground dust, but not by dust inside the NB. 
The total flux densities $S_{\nu}^{\rm tot}(\lambda)$\ were integrated 
within a polygon which completely includes the 3\% 
contour of the column density map (Fig. \ref{nb_nh_inout}b) and which approximately 
covers the range \mbox{$-$2.5\grd}\,$\le\,l\,\le$\,3.8\grd, and 
$-$1.3\degr\,$\le\,b\,\le$\,1.1\degr. 
This area partially includes dust in the CMZ located outside the actual NB 
(see Sect. \ref{disismmorph}). 
In addition, we derived surface brightnesses from the two submm 
continuum maps listed in Table \ref{datsum}. The IRAS, submm, and radio continuum 
maps were convolved with the DIRBE beam to obtain comparable surface brightnesses.
The central 120\,pc refer to the square-shaped 0.7\degr\ DIRBE 
beam (see Fig. \ref{cobemapsplo}) with an equivalent circular 
aperture of 0.395\degr, which corresponds to $R_{\rm equiv.}\sim 60$\,pc. 
The resulting surface brightnesses and flux densities are compiled in 
Table \ref{nbpar} and were used to construct the corresponding SEDs 
shown in Fig. \ref{gcnb_sed}. 
The error bars correspond to the usually adopted
calibration uncertainties  of $\sim$\,20\% 
for the COBE and IRAS 12 to 240\,\mim\ data and $\sim$\,30\% 
for the submm data.  The uncertainties of the COBE NIR data were estimated from the 
variations when using different GB and extinction models (40\% 
at 2.2\,\mim, 30\% 
at 3.5 and 4.9\,\mim). 

To fit the observed SEDs and derive physical parameters of stars and 
dust in the NB, we use a simple (not self-consistent) 
radiative transfer model consisting of three 
single-temperature dust components and one stellar 
component. All four components are homogeneously mixed. 
Hot massive MS stars are not considered since 
their contribution to the K-band flux density is marginal 
(except for the central few pc; see Sects. \ref{disstars} through \ref{disstarslum}) 
and we don't have data points shortward of 2.2\,\mim. 
The cold and warm dust components represent large ``classical'' grains with their 
temperature distribution.
Hot dust emission represents very small stochastically heated grains with a possible
contribution by PAHs as well as dust in hot circumstellar shells. 
The NIR part of the SED was calculated using the model for internal 
extinction in the NB described in Sect. \ref{datanextnb}.
The final SEDs are $\chi^2$\, fits to the observed flux densities with the 
three dust temperatures and masses and the effective stellar NIR 
temperature $T_{\rm NIR}$\ being free parameters. 
Here, $T_{\rm NIR}$\ is the effective black-body temperature derived from a fit to 
the completely extinction-corrected NIR flux densities (2.2 to 4.9\,\mim). 
Although the 2.2\,\mim\ flux density may be slightly affected by emission from 
hot stars for which the SEDs shown in Fig. \ref{gcnb_sed} do not account, 
$T_{\rm NIR}$\ should approximately represent the average 
effective stellar temperature of cool evolved stars, 
which dominate the NIR luminosity of the NB (see Sect. \ref{disstars}). 
The contribution of hot stars in different regions of the NB will be further discussed in 
Sects.  \ref{disstars} through \ref{disstarslum}.

While the fit parameters for the dust emission sections of the SEDs are nearly 
independent of the particular extinction model of the NB, the results for the stellar 
(NIR) section depend strongly on the assumed distribution of extinction within the NB. 
Since the populations of evolved stars in the Nuclear and Galactic Bulge are found to 
be similar (see. Sect. \ref{disstars}), we assumed 
$T_{\rm NIR}({\rm NB}) \sim T_{\rm NIR}({\rm GB})$. 
This assumption could be fulfilled by setting 
$A_{\rm V}({\rm NB})\sim 20$\,mag (plus 10\% 
blockage; see Sects. \ref{datanextnb} and \ref{disismprop}). 
Higher values for the diffuse average extinction through the NB lead to 
unreasonably high temperatures and intrinsic luminosities 
(e.g., $A_{\rm V}({\rm NB}) = 30$\,mag yields $T_{\rm NIR} \sim 8000$\,K and 
$L_{\ast,\rm NIR}({\rm NB}) \sim 3.5\times 10^9$\,\lsun). 
The implication of this low diffuse extinction will be discussed in Sect. \ref{disismprop}.
Our model yields a total intrinsic stellar 
NIR luminosity\footnote{$L_{\rm NIR}$\ is the luminosity derived from a 
black-body fit to the extinction-corrected 2.2 to 4.9\,\mim\ flux densities,  
assuming that the 2.2\,\mim\ flux is not contaminated by emission from 
hot stars.} 
of $L_{\ast,\rm NIR}({\rm NB}) \sim 7\times 10^8$\,\lsun, 
which does not account for the contribution from hot stars.

The corresponding SED fits to the central 120\,pc and the entire NB are shown in 
Fig. \ref{gcnb_sed} and the results of the spectral decomposition and the derived masses 
and luminosities are given in Table \ref{gcnb_sed_tab}. 
For the NB we used a relative metallicity $Z/Z_{\odot}=2$\ 
(Appendix \ref{apgalmod}, for the dust model used see Appendix \ref{apdust}).
Note that the dust emission in this model, and hence the derived hydrogen masses, account 
for all interstellar matter in the central $\sim$\,6\degr. 
The uncertainties and the implications of the parameter fit to the SED 
will be discussed in Sect. \ref{disc}. 
The corresponding SED for the central 1.25\,pc and fit parameters of its spectral 
decomposition given in Table \ref{gcnb_sed_tab} are taken from Paper\,I. 
Note that in Paper\,I $A_{\rm V} = 31$\,mag was used as total foreground extinction, 
assuming that all contributing stars are actually located in the central parsec 
(no line-of-sight spread).
In Sect. \ref{disc} we make a further attempt to decompose the emission from the NB 
into contributions from different sub-components.

%____________________________

\subsubsection{Column density and mass of Interstellar Matter in 
               the Central Molecular Zone}  \label{resnbnh}

A hydrogen column density map of the CMZ (i.e., all molecular material 
in the central kpc) can be derived from the optically thin 240\,\mim\ 
dust emission (surface brightness $I_{\nu}$) if the dust 
temperature $T_{\rm d}$\ is known: 
\begin{equation}
N_{\rm H} = \frac{I_{\nu}(\lambda)}{B_{\nu}(\lambda,T_{\rm d})}\,
            \frac{1}{m_{\rm H}\,\kappa(\lambda)}\, 
            \frac{M_{\rm H}}{M_{\rm d}}\ .
\end{equation}
Here, $B_{\nu}(\lambda,T_{\rm d})$\ denotes the Planck function, 
$T_{\rm d}$\ the 
mass-averaged dust temperature\footnote{Mass-averaged dust temperature: 
average temperature of all dust components  
(mixed or separated along the line-of-sight) weighted with their respective masses. 
It is not necessarily identical with the 
effective colour temperature derived from flux density ratios.}, 
$m_{\rm H}$\ the mass of an Hydrogen atom,
$\kappa$\ the mass absorption coefficient (dust opacity), 
and $M_{\rm H}/M_{\rm d}$\ the 
hydrogen-to-dust mass ratio (for the dust model used see Appendix \ref{apdust}). 

Since the dust grains in the NB have a broad temperature distribution, 
a simple black-body colour temperature derived from two FIR maps 
is not a good representation of the true dust temperature. 
Instead, we used an empirical colour temperature map derived from the 
COBE 60-to-140\,\mim\ flux ratio map of the NB. 
The 140 and 60\,\mim\ fluxes are dominated by emission from cold (15-30\,K) 
and warm (40-100\,K) dust (``classical'' grains), respectively. 
Hot, stochastically 
heated very small grains mainly contribute at 12 and 25\,\mim.
The 60-to-140\,\mim\ surface brightness ratio was scaled to the mass-averaged 
dust temperatures derived from the SED model fits to the central 120\,pc and the entire NB 
(Fig. \ref{gcnb_sed} and Table \ref{gcnb_sed_tab}). 
We then assumed proportionality between $T_{\rm d}$\ and $S_{60}/S_{140}$\ 
in the temperature range 15$\ldots$30\,K. 
The uncertainty of this method is smaller than $\pm1$\,K, which results in a 30\% 
uncertainty in $N_{\rm H}$ at 16\,K and 10\% 
at 25\,K. 
Where the surface brightness in the 60 and 140\,\mim\ COBE maps was below a certain cut-off 
level and no reliable dust temperature could be derived, the temperature was set to 16\,K, 
the lowest value derived in the map.

The resulting dust temperature profile is shown in Fig. \ref{nb_lm}b and will be 
discussed in Sect. \ref{disismmorph}. 
Mass-averaged dust temperatures in the CMZ are in the range 16 to 25\,K. 
The corresponding dust temperature map was then used to convert 
the COBE 240\,\mim\ map (Fig. \ref{cobemapsplo}) into a 
hydrogen column density map (Fig. \ref{nb_nh_inout}b).
This map shows a convincing similarity to the 
$^{12}$CO(1--0) map of Bitran et al. (1997; see Fig. \ref{nb_nh_inout}a),  
indicating that both optically thin FIR dust emission and 
$^{12}$CO emission trace, at least on large scales, the same ISM. 
Beam-averaged column densities range from 1 to 
15\,$\times$\,10$^{22}$\,\scm\ with 
$N_{\rm H}$(0\degr,0\degr)\,$\sim 7\times 10^{22}$\,\scm\ 
(Fig. \ref{nb_lm}a).  
The asymmetry in the dust temperature and column density profiles 
and the relation of ISM to the stellar mass distribution in the NB 
will be discussed in Sect. \ref{disismmorph}.

Integrating the hydrogen column density map (Fig. \ref{nb_nh_inout}b) yields a 
total hydrogen mass of $M_{\rm H} \sim 6.8\times 10^7$\,\msun\ in the 
central 8\degr\ ($\simeq$\,1\,kpc) of the Galaxy. The area \mbox{$-$2.5\grd}\,$\le\,l\,\le$\,3.8\grd\ 
contains $M_{\rm H} \sim 5.3\times 10^7$\,\msun\
which is in good agreement with the mass of 4.2\,$\times 10^7$\,\msun\ 
derived from the SED model fit to the integrated flux densities (Sect. \ref{resnbsed} 
and Table \ref{gcnb_sed_tab}). 
In Sect. \ref{disismmorph} we decompose the $N_{\rm H}$\ map into components inside and outside 
the NB and derive the corresponding hydrogen masses and column densities.

%%%%%%%%%%%%%%%%%%%%%%%%%%%%%%%%%%%%%%%%%%%%%%%%%%%%%%%%%%%%%%%%%%%%%%%%%%%%%%%%%%
%%%%%%%%%%%%%%%%%%%%%%%%%%%%%%%%%%%%%%%%%%%%%%%%%%%%%%%%%%%%%%%%%%%%%%%%%%%%%%%%%%

\section{Physical characteristics of the Nuclear Bulge}                    \label{disc}

Based on the observational results presented in the previous Section, we 
derive and further discuss in this section physical characteristics of stars 
and ISM in the NB and its immediate environment. 

%________________________________________________________________

\subsection{The stellar population of the Nuclear Bulge}                        \label{disstars}

A comparison of the model K-band luminosity function (KLF) of the central 
30\,pc with the KLF observed in Baade's Window indicates that 
the overall populations of low and intermediate-mass MS stars 
in Nuclear and Galactic Bulge are similar, but that the central 
30\,pc have an overabundance of K-luminous giants. 
These stars, which are interpreted as remnants of 
high star-formation activity some $10^7$\ to $10^8$\ yrs ago 
(e.g., Genzel et al. 1994), are more concentrated towards the centre than 
low-mass MS stars (Paper I, Fig. 6). 
Unique for the NB is a population of massive, young MS stars 
and supergiants which are also strongly concentrated towards 
the centre (Paper II; Genzel et al. 1996).
The contribution of these stars to both dynamic mass and NIR luminosity is 
negligible ($\sim 7$\%\ and $\sim 6$\%, respectively), 
but they dominate the bolometric luminosity in the central 
30\,pc ($\sim 80$\%) and are responsible for the ionization of the 
thin intercloud medium (see Sect. \ref{disismprop}). 
The main contributor to the stellar mass are low-mass MS stars ($>90$\%), 
which account, however, for only $\sim 6$\%\ of the NIR luminosity.
The contribution of evolved stars to the stellar mass 
is negligible, but they dominate the NIR luminosity ($\sim 88$\%; 
for this and the above quantitative estimates see Paper II). 
Therefore, an estimate of the total stellar mass from the NIR luminosity is 
only possible when certain assumptions about the mixing of low-mass MS stars 
and evolved stars are made (see Sects. \ref{disstarslum} and \ref{disstarsmass}). 

%_______________________________________________

\subsection{Morphology of the stellar Nuclear Bulge and the 
            large-scale distribution of stars}                              \label{disstarsmorph}

\begin{figure}[htb]
\includegraphics[width=0.48\textwidth]{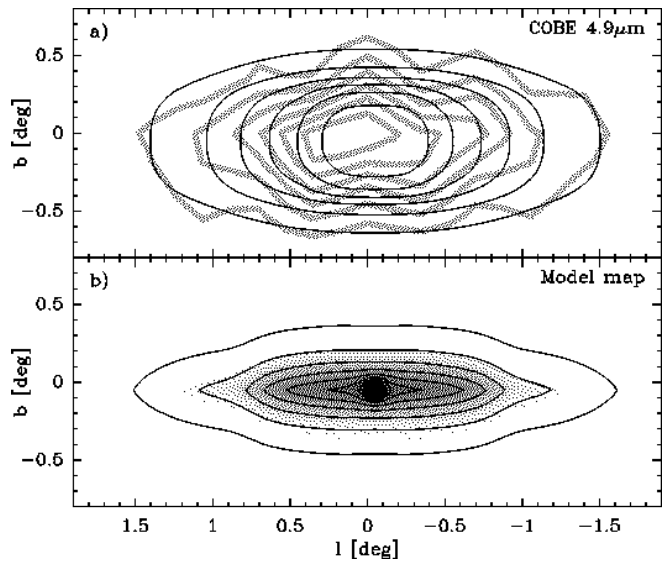}
% {gcnb_nir_deconv_paper_newa_map.ps} (gcnb_nir_deconv_paper_newa_map.graphic)
\caption{\label{nirsizemap}
 {\bf a)} Contour map of the final (observed) 4.9\,\mim\ map of the NB, corrected for 
  extinction by dust in front of and inside the NB  
 (thick grey contours, 10 to 85 in steps of 15\% of the maximum surface brightness). 
 Overlayed are the surface brightness contours of the best-fit model (shown in panel b) convolved 
 with the DIRBE beam (thin black contours, same levels).~~
 {\bf b)} Surface brightness map of the 'best fit' NIR model of the NB (see Sect. \ref{disstarsmorph}).}
\end{figure}

\begin{figure}[htb]
\includegraphics[width=0.49\textwidth]{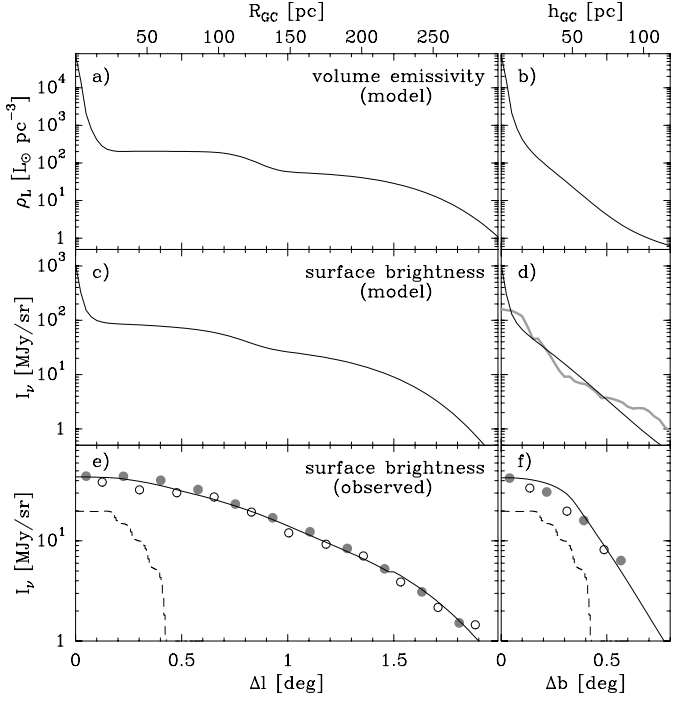}
% {gcnb_nir_deconv_paper_newa.ps} (gcnb_nir_deconv_paper_newa.graphic)
\caption{\label{nirsize}
 Longitude and latitude profiles of the stellar NIR emission from the NB. 
 Longitude and latitude offsets are measured from the position of Sgr\,A$^{\star}$.~~ 
 {\bf a)} and {\bf b)} NIR volume emissivity profiles of the NB model representing the 
  distribution of cool stars.~~
 {\bf c)} and {\bf d)} Integrated 4.9\,\mim\ surface brightness distribution of the 
  volume emissivity model. For comparison, the thick grey line shows the IRAS 60\,\mim\ 
  latitude profile (90\arcsec\ resolution), scaled to match the 4.9\,\mim\ brightness.~~ 
 {\bf e)} and {\bf f)} Observed 4.9\,\mim\ (COBE) surface brightness profiles, corrected for 
  extinction by dust in front of and inside the NB. 
  Data points from positive and negative longitude and latitude offsets are plotted as 
  filled and open circles, respectively. 
  The average profiles were used to fit the models. The scatter of the data points 
  indicates the longitudinal asymmetry, which we consider as the uncertainty of 
  the model fit. The dashed profiles show the COBE beam. Solid 
  curves show the profiles of the beam-convolved model map.}
\end{figure}

\begin{figure}[htb]
\includegraphics[width=0.50\textwidth]{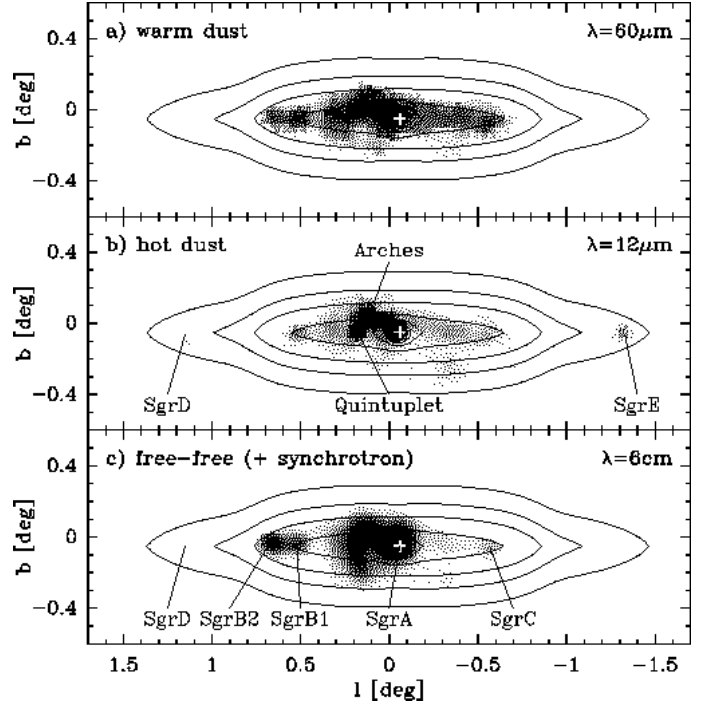}
% {gcnb_massivestars_new.ps} (gcnb_massivestars.graphic)
\caption{\label{nb_ms}
 Signatures of the distribution of massive stars in the Nuclear Bulge.
 {\bf a)} Distribution of warm dust (50--100\,K, 60\,\mim, IRAS).~~
 {\bf b)} Distribution of hot dust (200-500\,K, 12\,\mim, IRAS).~~
 {\bf c)} Distribution of free-free (and non-thermal synchrotron) emission (6\,cm, Parkes).~~
 The plot scales are logarithmic and cut-off levels are set to suppress extended low-level 
 emission. The most prominent Galactic centre features are labeled. 
 Overlayed are contours of the NIR surface brightness of the {\it Stellar Nuclear disk} model 
 (see Fig. \ref{nirsize}).}
\end{figure}

\begin{table*}[htb]                  
 \caption[]{Characteristics of Nuclear Stellar Cluster and Nuclear Stellar Disk
            \label{nscdtab}}
  \begin{flushleft}
   \begin{tabular}[t]{lllll} 
\hline \noalign{\smallskip}
 & & ~~~$M_{\ast}$ & ~~~$L_{\rm NIR}$ & ~~~$L_{\rm tot}$$^{(i)}$ \\
\noalign{\smallskip} \hline \noalign{\smallskip}
{\bf NSC:} & $\rho_0$\ (Eq. (\ref{eq_nsc}))$^{(a)}$ & 
   $3.3\times 10^6$\,\msun/pc$^{-3}$        &
   $6.6\times 10^6$\,\lsun/pc$^{-3}$        &
   $2.0\times 10^8$\,\lsun/pc$^{-3}$        \\
           & central 1pc$^3$$^{(b)}$              &
   $7.0\times 10^5$\,\msun$^{(d)}$          &
   $1.4\times 10^6$\,\lsun                  &
   $4.2\times 10^7$\,\lsun                  \\
           & central 1pc$^2$$^{(c)}$              &
   $1.5\times 10^6$\,\msun                  &
   $3.1\times 10^6$\,\lsun$^{(e)}$          &
   $9.2\times 10^7$\,\lsun$^{(e)}$          \\
           & entire NSC                           &
   $3\pm 1.5\times 10^7$\,\msun           &
   $6\pm 3\times 10^7$\,\lsun             &
   $1.8\pm 0.9\times 10^9$\,\lsun         \\
\noalign{\smallskip} \hline\noalign{\smallskip} 
{\bf NSD:} & inner $R\le120$\,pc                  &
   $8\pm 2\times 10^8$\,\msun$^{(f)}$     &
   $4\pm 1\times 10^8$\,\lsun$^{(g)}$     &
   $4\pm 1\times 10^8$\,\lsun \\
           & entire NSD                           &
   $1.4\pm 0.6\times 10^9$\,\msun$^{(f)}$ &
   $6.5\pm 2\times 10^8$\,\lsun$^{(g)}$   &
   $6.5\pm 2\times 10^8$\,\lsun           \\
\noalign{\smallskip} \hline\noalign{\smallskip} 
{\bf total NB:} &                                 &
   $1.4\pm 0.6\times 10^9$\,\msun         &
   $7\pm 2\times 10^8$\,\lsun$^{(h)}$     &
   $2.5\pm 1\times 10^9$\,\lsun           \\
\noalign{\smallskip} \hline\noalign{\smallskip} 
  \end{tabular} 
 \end{flushleft}
\begin{list}{}{}
\item[$^{(a)}$] With $n=2$\ for $R\le6$\,pc, $n=3$\ for $R>6$\,pc, and an outer
       cut-off radius $R_{\rm out}=200$\,pc
\item[$^{(b)}$] Integrated in a sphere with $R=0.625$\,pc (volume $V$\,=\,1\,pc$^3$)
\item[$^{(c)}$] Integrated in a zylinder with $R=0.625$\,pc throughout the
       entire NSC (pencil beam with area $A$\,=\,1\,pc$^2$)
\item[$^{(d)}$] Genzel et al. (1997); fixed input parameter
\item[$^{(e)}$] Paper\,I, Table 1b; fixed input parameter
\item[$^{(f)}$] Derived from $L_{\rm NIR}$\ with $M_{\star}$/$L_{\rm NIR} = 2$\,\msun/\lsun\ 
       (see. Sect. \ref{disstarsmass})
\item[$^{(g)}$] Derived from a two-component volume emissivity model fit of the form 
       of Eq. (\ref{eqnbnirmod}) to the extinction-corrected 4.9\,\mim\ 
       COBE map of the NB (Fig. \ref{nirsize}).  
\item[$^{(h)}$] Table \ref{gcnb_sed_tab}, line 7; fixed input parameter 
\item[$^{(i)}$] $L_{\rm tot}$\ includes the contribution from hot massive MS stars (see Sect. \ref{disstarslum}).
\end{list}
\end{table*}

The size and morphology of the stellar NB can best be derived from the 4.9\,\mim\ 
image where stellar emission dominates and extinction effects, i.e., 
uncertainties introduced in the data reduction 
and extinction correction process, are relatively small. 
Figure \ref{nirsizemap}a shows the final 4.9\,\mim\ COBE image of the NB 
corrected for foreground extinction {\it and} for extinction by dust 
inside the NB (see Sect. \ref{datanextnb}). 
Corresponding longitude and latitude profiles are shown in Figs. \ref{nirsize}e and f. 
Despite some sub-beam-scale asymmetry at the centre, the overall NIR emission from the NB 
is symmetric with respect to the position of Sgr\,A$^{\star}$.
% (l,b\,$\sim$\,-0.06\degr,-0.05\degr).  
Since the coarse angular resolution of the COBE data prohibits interpretation of asymmetries 
on such a small scale, we assume that the large-scale distribution of stars in the NB 
is symmetric with respect to the GC.

To derive the physical size of the stellar NB, models of the volume emissivity 
distribution were constructed, the resulting surface brightness distribution calculated, 
convolved with the COBE beam, and then compared to 
the extinction-corrected 4.9\,\mim\ map. 

It was found that the central few parsec, 
which are not resolved by COBE, are dominated by the luminous 
Nuclear Stellar Cluster with an $R^{-2\pm0.3}$\ power-law density profile 
(sometimes called the {\it ``central $R^{-2}$\ star cluster''}; 
e.g., Becklin \& Neugebauer 1968; Serabyn \& Morris 1995), 
and that a large-scale elliptical $R^{-2}$\ mass distribution 
around the GC is approximately consistent with the dynamical mass contained 
in the central 300\,pc (see Haller et al. 1996). 
However, such a distribution cannot explain the observed distribution of light on 
scales larger than 30\,pc, which is highly elliptical 
and cannot be described by a single power law. 
The observed NIR surface brightness distribution can much better be reproduced 
by a large-scale exponential disk with the 
{\it Nuclear Stellar Cluster} (NSC) at its centre. 
In the following, we call this disk the {\it Nuclear Stellar Disk} (NSD) and describe 
our two-component model.

%__________________________________________

{\bf 1.} {\it Nuclear Stellar Cluster:}
The radial K-band surface brightness distribution $\rho^{\rm L}(R)$\ 
in the central $R_{\rm GC} = 0.625$\,pc 
(Paper\,I, Fig.\,6, here corrected for extinction) 
can be best fitted by a volume density distribution of the form: 
\begin{equation}                                      \label{eq_nsc}
\rho^{\rm L}(R) = \frac{\rho_0^{\rm L}}{1+\left(\frac{R}{R_0}\right)^n}
\end{equation}
with $n=2.0$ and a core radius $R_0=0.22$\,pc. 
The observed K-band flux density and NIR luminosity due to cool stars 
in this area of $A=1$\,pc$^2$\ are $S_{\rm K}\sim 350$\,Jy and 
$L_{\rm cool\star}\sim 3.1\times10^6$\,\lsun, respectively (Paper\,I, Table\,1). 
With these values we derive $\rho_0^{\rm L}\sim 6.6\times 10^6$\,\lsun\,pc$^{-3}$\ 
for Eq. (\ref{eq_nsc}).
To fit both the observed large-scale surface brightness distribution of the NB and 
the dynamical mass in the inner $R_{\rm GC}\le 20$\,pc (see Sect. \ref{disstarsmass} 
and Fig. \ref{nb_photmass}), 
a steeper power-law dependence of the stellar density in this cluster 
is required outside $R_{\rm GC}\sim$5-10\,pc. We obtained a good fit with $n\sim 3$\ 
beyond 6\,pc, but could not put strong constraints on where and how this transition 
in the slope exactly occurs. 
With these parameters we derive a total NIR luminosity of the NSC of 
$L_{\rm NIR}=6\pm 3\times 10^7$\,\lsun. The relatively large uncertainty reflects the range of possible 
cut-off radii and outer slopes.
The luminosity contribution from high-mass MS stars and the mass distribution in the NSC will be 
discussed in Sects. \ref{disstarslum} and \ref{disstarsmass}. 
Parameters of the NSC are summarized in Table \ref{nscdtab}.

%__________________________________________

{\bf 2.} {\it Nuclear Stellar Disk:}
Although other flat configurations may be conceivable for the NSD (e.g., a symmetric bar), 
we adopt a disk as the most likely configuration. 
The COBE NIR data are inconsistent with the disk having a single power-law emissivity profile. 
Instead, we obtain a good fit with an exponential-to-a-power law radial 
dependence of the volume emissivity:
\begin{equation}   \label{eqnbnirmod}
\rho^{\rm L}(R) \propto {\rm exp}\left({\rm ln}\frac{1}{2}\,\left|\frac{R}{\sigma/2}\right|^n\right)    
\end{equation}
with $\sigma$\ the FWHM radius. 
Although the sum of two components with $\sigma_1/2 = 120$\,pc and 
$\sigma_2/2 = 220$\,pc, and $n=5\pm1$\ fits the data best, the presence of two distinct 
components may not have a physical meaning, but could 
reflect extinction effects inside the NB not accounted for in our model 
(see Sect. \ref{disismmorph}). 
The radial dependence of the volume emissivity 
in this disk can also be approximated by a range of different power laws: 
flat ($\propto R^{-0.1}$) at $R < 120$\,pc, 
$\propto R^{-3.5}$\ for 120\,pc\,$< R <$\,220\,pc, 
and a steep drop-off ($\propto R^{-10}$) at $R > 220$\,pc. 
Since the NB is unresolved in latitude, only an upper limit for its FWHM 
scale height of 60\,pc (0\grdp40) could be derived from the COBE data. 
In Sect. \ref{disismmorph} 
we derive a scale height of warm dust, which is closely related to the stellar NB, of 
45$\pm$5\,pc. 
Adopting this scale height and $n=1.4\pm0.2$\ (see Sect. \ref{disismmorph}), 
Eq. (\ref{eqnbnirmod}) 
yields a good fit to the latitude profile of COBE NIR data (Fig. \ref{nirsize}f). 
The total NIR luminosity and average midplane volume emissivity
of the NSD then amount to $L_{\rm NIR}=6.5\pm 2\times 10^8$\,\lsun\ and 
$\rho_{\star} \sim 100 $\,\lsun\,pc$^{-3}$, respectively. 
Parameters of the NSD are summarized in Table \ref{nscdtab}.

The corresponding total NIR surface brightness distribution, i.e., 
the sum of the contributions from NSC and NSD, 
is shown in Figs. \ref{nirsizemap}b and \ref{nirsize}c and d. 
For comparison, we also show in Fig. \ref{nirsize}d the observed latitude profile 
of the IRAS 60\,\mim\ emission from the NB (90\arcsec\ resolution). 
Finally, Figs. \ref{nirsizemap}a and \ref{nirsize}e and f compare, after convolution with 
the COBE beam, the modeled and observed NIR surface brightness profiles. Since we assumed 
symmetry with respect to the GC, the model was optimized to fit the average surface brightness 
at positive and negative latitude and longitude offsets, respectively. 
The observed and modeled latitude profiles do not match perfectly 
(Fig. \ref{nirsize}f). However, the small scale height of the NB together with the 
uncertainties in the beam shape, which may have slightly varied during the observations, 
do not allow to put better constraints on the vertical density profile of the NB. 

Although we cannot directly observe the emission from massive hot MS stars in the NB, 
their distribution can be derived from the emission by warm and hot 
dust and from free-free emission from H{\small II} regions. 
Figure \ref{nb_ms} compares the NIR morphology 
of the NB (cool stars, NIR model) with the distribution of the 
above signatures of massive stars. 
It becomes clear that massive stars are neither evenly distributed throughout the NB 
nor are they present in the centre only. 
They are rather concentrated in several individual clusters, 
the brightest of which is centered at Sgr\,A$^{\star}$\ and represents the 
centre of the the NSC. 
This confirms our earlier finding that young massive stars are strongly concentrated towards the 
centre (Paper I), but also shows that these stars are not exclusively formed in the central few 
pc of the NB. Our models account for this by extrapolating the (smooth) 
radial distribution of the NSC to large radii (see Sect. \ref{disstarslum}). 
A comprehensive overview over the distribution of H{\small II} regions and non-thermal 
features in the GC Region can be found in LaRosa et al. (2000).

%___________________________________________________

\subsection{The stellar luminosity of the Nuclear Bulge}                          \label{disstarslum}

The total stellar luminosity of the NB depends on the contributions 
by high-mass stars whose optical/UV emission cannot directly be observed due to 
the high extinction. It has been shown in Paper II that this contribution depends 
critically on the Present-Day Mass Function which in turn is determined by the 
star formation history of the NB (continuous star formation, star bursts, etc.). 
In view of these uncertainties we use the integrated NIR SED as the only observable 
quantity to derive the total luminosity of cool stars in the NB and discuss 
possible corrections to this value due to contributions by high-mass stars. 

The total luminosity of cool stars in the NB can be derived by 
additionally correcting the observed stellar NIR emission 
(corrected for foreground extinction) for extinction by dust inside the NB. 
Using the dust distribution and extinction model described in 
Sect. \ref{datanextnb}, we derive 
$L_{\rm NIR}({\rm NB})\sim L_{\rm cool\ast}({\rm NB})\sim 6.8\times 10^8$\,\lsun\ 
(Table \ref{gcnb_sed_tab}, line 7). 

Assuming isotropic extinction and emission from the NB and assuming further that 
stellar emission is the only dust-heating source and that all stellar emission shortward of 
$\lambda =$\,1\,\mim\ is completely absorbed by dust and re-radiated as thermal continuum 
emission, the total stellar luminosity of the NB 
(i.e., $L_{\rm cool\ast}+L_{\rm hot\ast}$) should be equal to the 
observed total luminosity (i.e., $L_{\rm NIR,obs} + L_{\rm dust}$) 
when corrected for all foreground extinction. 
This value amounts to $L_{\ast}({\rm NB}) = 6.7\times 10^8$\,\lsun\ 
(Table \ref{gcnb_sed_tab}, line 6, last column), very similar to the intrinsic 
luminosity of cool stars derived above. 
However, to conclude that hot massive MS stars do not contribute to the 
bolometric luminosity of the NB is apparently wrong.   
Due to the highly inhomogeneous distribution of ISM in the NB, 
particularly the extreme clumpiness and very small scale height  
(see Sect. \ref{disismmorph}), only a small fraction of the short-wavelength 
stellar emission is actually absorbed and re-radiated 
(i.e., $L_{\rm dust}\ll L_{\rm hot\ast}$)
and the total luminosity may be considerably underestimated. 

Therefore, we use the intrinsic NIR luminosities of the NSC and NSD derived from our 
models (Sect. \ref{disstarsmorph}), assume that they represent the total luminosity 
due to cool stars, and try to supplement these values for the contribution by hot 
massive stars. 
In Paper\,I (Table 1b) we derive for the central 1.25\,pc a luminosity ratio of 
hot (i.e., young massive) stars to cool stars (i.e. low-mass MS stars plus Giants 
and Supergiants) of $\sim$30. 
According to our analysis of the KLF of the central 30\,pc in Paper II, 
the ratio $L_{\rm hot\ast}/L_{\rm cool\ast}$\ drops to 4--5 when averaged over 
30\,pc.
This behaviour can be approximately reproduced by 
assuming that $L_{\rm hot\ast}/L_{\rm cool\ast} = 30$\ holds for the entire 
NSC and that $L_{\rm NIR}=L_{\rm tot}$\ holds for the entire NSD, i.e., no 
massive MS stars are associated with our model NSD. 
Note that NSC and NSD are two idealized model components which overlap spatially. 
The extrapolation of the NSC out to $R=230$\,pc actually provides 
$L_{\ast}\sim 5\times 10^8$\,\lsun\ due to massive MS stars outside $R>30$\,pc, 
the region which is dynamically dominated by the NSD (see Sect. \ref{disstarsmass}). 
This is approximately consistent with the number of massive star clusters and giant 
H{\small II} regions observed in the NSD (Fig. \ref{nb_ms}) if each of these regions 
emits $5-10\times 10^7$\,\lsun\ (e.g., Figer et al. 1999).
The NIR luminosities of NSC and NSD (Sect. \ref{disstarsmorph} and Table \ref{nscdtab}) 
convert then into a total stellar luminosity of the NB of 
$L_{\ast}({\rm NB})= 2.5\pm 1\times 10^9$\,\lsun. 
The large uncertainty is mainly due to the unknown contribution by massive stars 
outside the central 30\,pc.
We adopt this value as the currently best representation of the total stellar luminosity of the NB. 
It is somewhat larger than the earlier, independent estimate of 
$L({\rm NB}) \sim 1.4\times 10^9$\,\lsun\ by Cox \& Laureijs (1988; see MDZ96, Table 5), 
which, however, did not account for the contribution by hot massive stars.
The stellar luminosities for different regions of the NB as derived from this model 
are summarized in Table \ref{nbsumtab}.
Table \ref{nbcomptab} summarizes the notations used for different physical components 
and spatial regions and their interrelation. Note that different physical components like, 
e.g., NSC and NSD, do spatially overlap, and that the characteristics of different spatial 
regions listed in Table \ref{nbsumtab} represent the corresponding average of the different 
model components in that region.

%______________________________________________________________

\subsection{The stellar mass of the Nuclear Bulge}                      \label{disstarsmass}

\begin{figure}[htb]
\includegraphics[width=0.48\textwidth]{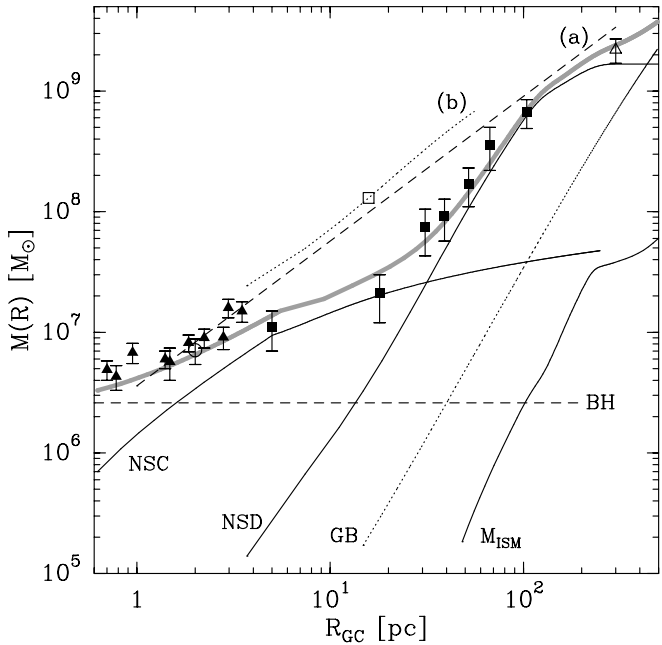}
% {gcnb_nir_photmass_new.ps} (gcnb_nir_photmass.graphic)
\caption{\label{nb_photmass}
 Enclosed mass in spheres of radius $R_{\rm GC}$\ for the inner 
 500\,pc of the Galaxy. 
 Thin solid lines show the mass distribution of 
 NSC and NSD as derived from our photometric models 
 (Sect. \ref{disstarsmorph}).
 The dotted line is the photometric mass distribution of the GB. 
 The dashed horizontal line is for the central Black Hole (Eckart \& Genzel 1998). 
 The rightmost solid line (M$_{\rm ISM}$) is for the total intestellar mass in the NB and the outer CMZ 
 (Sect. \ref{disismmorph}).
 The thick grey line represents the total enclosed mass, i.e., the sum of all these components. 
 Data points with error bars refer to the following dynamical mass estimates: 
 filled triangles: McGinn et al. (1989); 
 open circle: Rieke \& Rieke (1988); 
 filled squares: Lindqvist et al. (1992);
 open triangle: Liszt \& Burton (1978) and Burton \& Liszt (1978) 
  (adapted from Haller et al. 1996). 
 The dashed curve (a) shows, for comparison, the power-law approximation 
 to the enclosed mass by Sanders \& Lowinger (1972) (Eq. (2.1) in MDZ96).
 The dotted curve (b) shows the photometric mass distribution of the NB (without GB) 
 according to our model when integrated through the entire NB 
 in cylinders of radius $R_{\rm GC}$\ (i.e., the 
 ``observed'' mass in a given area).
 The open square shows the total ``observed'' stellar mass of the central 
 $R=15$\,pc derived from the KLF of the K-band mosaic (Paper II).} 
\end{figure}

In the following we use $M_{\ast}/L_{\rm NIR}$\ ratios observed in different regions 
to determine $M_{\ast}/L_{\rm NIR} = f(R)$, then derive individual $M_{\ast}/L_{\rm NIR}$\ 
ratios for NSC and NSD, and use these ratios to convert the NIR luminosity distributions 
of NSC and NSD into a photometric mass distribution of the NB. 
Note that only average $M_{\ast}/L_{\rm NIR}$\ ratios are known for certain areas 
and that this ratio depends critically on the ratio of mass-dominating low-mass MS 
stars to luminosity-dominating evolved stars, 
which is not constant throughout the NB (see Sect. \ref{disstars}). 
Here, we assume that NSC and NSD have different, but intrinsically constant 
$M_{\ast}/L_{\rm NIR}$\ ratios and calibrate their values to reproduce 
the effective observed $M_{\ast}/L_{\rm NIR}$\ ratios in different areas 
where both components overlap.

The total mass in the central 1.25\,pc or 30\arcsec, $3.3\times 10^6$\,\msun,  
according to Genzel et al. (1997), is dominated by the central Black Hole 
($M_{\rm BH} \sim 2.6\times 10^6$\,\msun), which leaves a total stellar mass 
of $M_{\ast} \sim 7\times 10^5$\,\msun\ in the central pc$^3$. 
Assuming that the stellar mass distribution within the NSC has the same radial dependence 
as the NIR luminosity, the central mass density in Eq. (\ref{eq_nsc}) becomes 
$\rho_0^{\rm M}\sim 3.3\times10^6$\,\msun\,pc$^{-3}$.
This yields $M_{\star}/L_{\rm NIR}\sim 0.5$\ which we assume to hold for the entire NSC. 
The total stellar mass of the NSC becomes then 
$M_{\ast}{\rm (NSC)} = 3\pm1.5\times 10^7$\,\msun. 

In Paper\,I we derive for the central 30\,pc $M_{\ast}/L_{\rm NIR}\sim 1$. 
The increase of $M_{\ast}/L_{\rm NIR}$\ when averaged over larger areas is 
consistent with the notion that the density of low-luminosity giants and 
low and intermediate-mass MS stars (spectral type O9 or later), which dominate the total 
stellar mass, decreases more slowly with increasing radius than that of 
K-luminous giants, supergiants, and early O stars (Paper I, Fig. 6).
An effective $M_{\ast}/L_{\rm NIR}$\ ratio of 1 for the central 30\,pc can be 
reproduced by adopting $M_{\ast}/L_{\rm NIR}= 2$\ for the NSD, the same 
value as derived for the GB (see Sect. \ref{resgb}). 
This is consistent with the finding that the KLFs for low and intermediate-mass 
MS stars in Nuclear and Galactic Bulge are very similar (Paper II). 
Individual high-mass star-forming regions in the outer NB, which supposedly have 
a lower $M_{\ast}/L_{\rm NIR}$\ ratio, are accounted for by extrapolating the NSC 
out to $R=230$\,pc (see Sect. \ref{disstarslum}).
Assuming a 30\%\ uncertainty for the $M_{\ast}/L_{\rm NIR}$\ ratio, 
the NIR luminosity of the NSD (Table \ref{nscdtab}) 
converts into a photometric stellar mass of $1.4\pm0.6\times 10^9$\,\msun. 
The masses of the two model components NSC and NSD and of different 
spatial regions of the NB are given in Tables \ref{nscdtab} and \ref{nbsumtab}, 
respectively.

Figure \ref{nb_photmass} shows the mass distribution in spheres of 
radius $R_{\rm GC}$\ around the GC derived from our volume emissivity model 
of the NB (NSC + NSD) together with the above discussed $M_{\ast}/L_{\rm NIR}$\ ratios. 
Also shown are the mass of the central Black Hole (Eckart \& Genzel 1998), 
the mass distribution in the GB (see Sect. \ref{resgb}), and the distribution 
of ISM in the NB and the outer CMZ (see Sect. \ref{disismmorph}, here corrected for
He and metals).
The mass distribution is dominated by 
the central Black Hole at $R_{\rm GC}<2$\,pc, 
the NSC at 2\,pc\,$\le R_{\rm GC} \le 30$\,pc,
the NSD at 30\,pc\,$\le R_{\rm GC} \le 400$\,pc,
and the GB at $R_{\rm GC}>400$\,pc. 
This photometrically derived mass distribution in the central 500\,pc with the 
Black hole included is compared to different dynamical mass estimates 
(for references see caption Fig. \ref{nb_photmass}). 
Both mass distributions agree very well. 
The presence of the NSD, which was inferred from the large-scale NIR surface 
brightness distribution of the NB, provides a logical explanation for the ``kink'' 
in the dynamical mass distribution at $R_{\rm GC}\sim 20$\,pc and the steep 
rise between $R_{\rm GC}\sim 20$\,pc and 100\,pc. 
The power-law approximation by Sanders \& Lowinger (1972) 
(Fig. \ref{nb_photmass} curve a; cf. Eq. (2.1) in MDZ96) also gives a reasonable 
estimate of the total enclosed mass in the inner 500\,pc. However, it is inconsistent 
with the observed highly 
elliptical NIR surface brightness distribution of the NB and, in detail, does not 
fit the dynamical mass distribution well. Note that these authors assumed a constant 
$M_{\ast}/L_{\rm NIR}$\ ratio for the entire GC region. 
Also shown in Fig. \ref{nb_photmass} is the stellar mass distribution 
contained in corresponding circular areas of radius $R_{\rm GC}$\ when 
integrated throughout the entire NB (curve b).
It reproduces correctly the integrated stellar mass in the central 30\,pc 
derived from an analysis of the KLF of the K-band mosaic 
($M_{\ast}(30{\rm pc})\sim 1.2\times 10^8$\,\msun, Paper II).
Curve (b) cannot be directly compared to dynamical masses which refer 
to spherical volumes of radius $R_{\rm GC}$. 

With $M_{\ast}{\rm (NB)}=M_{\ast}{\rm (NSC)}+M_{\ast}{\rm (NSD)}=(1.4\pm 0.6)\times 10^9$\,\msun, 
the NB accounts for $\sim$\,1-2\% 
of the total stellar mass in the GD and corresponds to $\sim$\,10\% 
of the mass of the GB (see Sect. \ref{resgb}).
The average mass density of stars in the NB (outside the central $\sim 20$\,pc) 
is then $\rho^{\rm M}_{\star} \sim$\,(200$\pm$50)\,\msun\,pc$^{-3}$, 
which is more than 40 times higher than the expected central mass density in the 
GB when its exponential volume emissivity/density profile (Eq. (\ref{eqigbprof})) 
is interpolated into the centre.

%________________________________________________________________

\subsection{Morphology of Interstellar Matter in the 
            Galactic Centre Region and its relation to the Nuclear Bulge}            \label{disismmorph}

\begin{figure}[htb]
\includegraphics[width=0.45\textwidth]{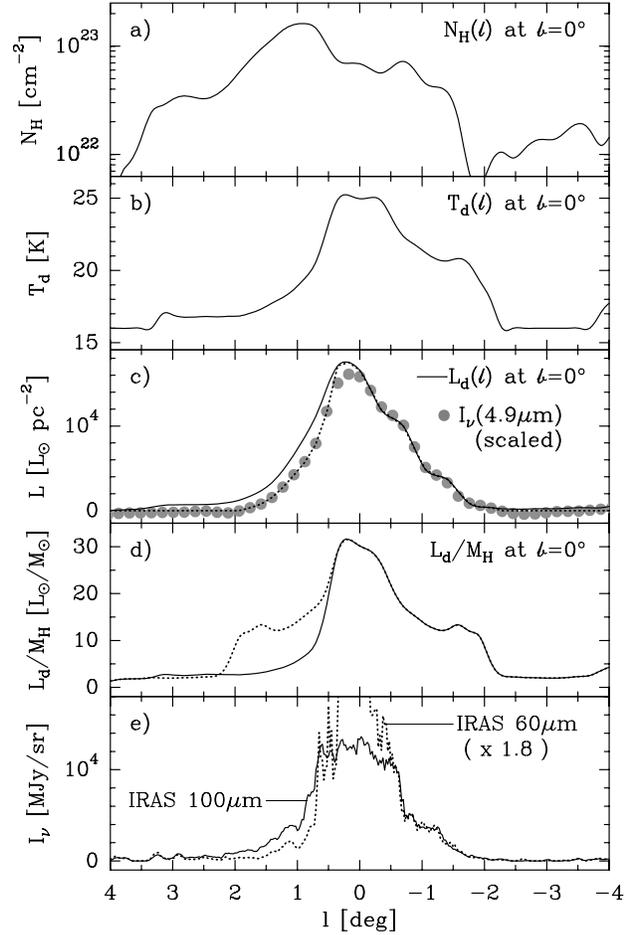}
% {nb_lm_paper.ps} (nb_lm_profiles_paper_new.graphic)
\caption{\label{nb_lm}
 {\bf a)} Longitude profile of hydrogen column density in the CMZ 
          (see Fig. \ref{nb_nh_inout}b).~ 
 {\bf b)} Mass-averaged dust temperature in the CMZ.~ 
 {\bf c)} Total dust luminosity per pc$^2$\ (solid line). 
          Grey dots show the (dereddened) 4.9\,\mim\ surface brightness which 
          represents the stellar luminosity (scaled to match $L_{\rm d}$).
          Dotted line at positive $l$: assumed intrinsic dust luminosity of the NB 
          (see Sect. \ref{disismmorph}).~ 
 {\bf d)} Specific dust luminosity ($L_{\rm d}/M_{\rm H}$, solid line). 
          Dotted line:  assumed intrinsic specific dust luminosity of the NB 
          (see Sect. \ref{disismmorph}).~ 
 {\bf e)} IRAS 60 and 100\,\mim\ surface brightness profiles, representing warm dust.}
\end{figure}

\begin{figure}[htb]
\includegraphics[width=0.44\textwidth]{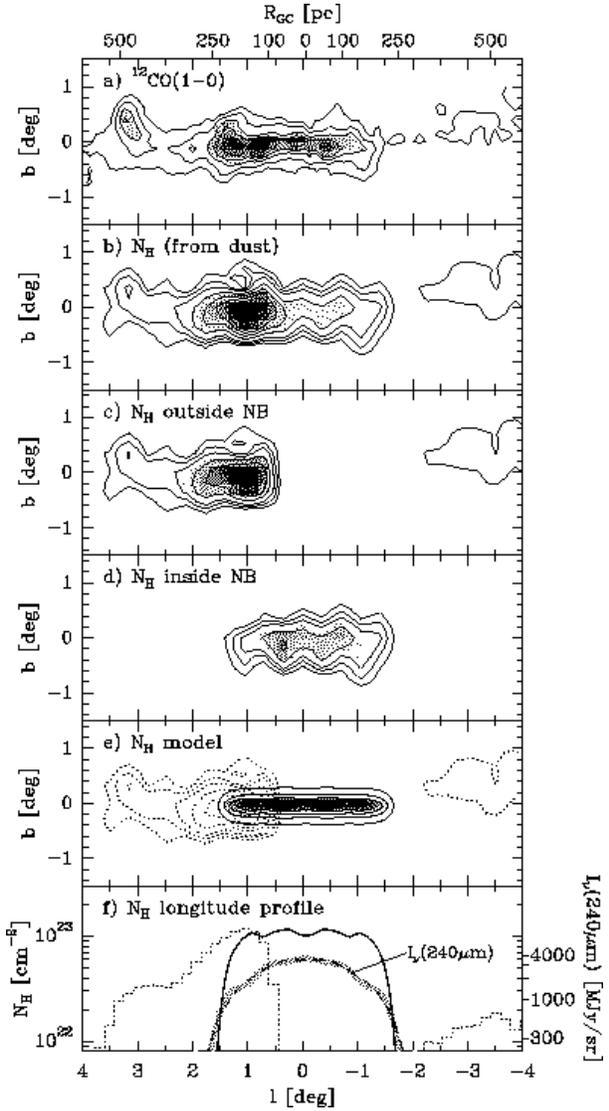}
% {gcnb_nh_in_out_new.ps} (gcnb_nh_in_out_paper_new.graphic)
\caption{\label{nb_nh_inout}
 {\bf a)} Distribution of $^{12}$CO(1--0) (HPBW\,=\,0\grdp15; Bitran et al. 1997; 
          see also Mauersberger \& Bronfman 1998).~~
 {\bf b)} Hydrogen column density map of the central 8\degr\ derived from the 
          COBE 240\,\mim\ dust emission map (HPBW\,=\,0\grdp7). 
          Contributions from the GD are subtracted. 
          Contour lines are at 1.5, 2.5, 3.5 to 15.5 by 2\,$\times$\,10$^{22}$\,\scm.~~
 {\bf c)} Column density map of ISM outside the NB.~~ 
 {\bf d)} Column density map of ISM in the NB.~~ 
 {\bf e)} Hydrogen column density map of the NB model (see Sect. \ref{disismmorph}).
          Contours are in steps of 10\% of the maximum. Dotted contours show the 
          distribution of ISM outside the NB.~~
 {\bf f)} Longitude profile of hydrogen column density inside (black solid line; cf. map e) and outside 
          the NB (dotted curve, not deconvolved; cf. map c). 
          Light grey curve: observed 240\,\mim\ surface brightness profile of dust inside the NB.
          Dashed line: modeled 240\,\mim\ surface brightness profile of dust inside the NB 
          (convolved with the COBE beam).}
\end{figure}

\begin{figure}[htb]
\includegraphics[width=0.45\textwidth]{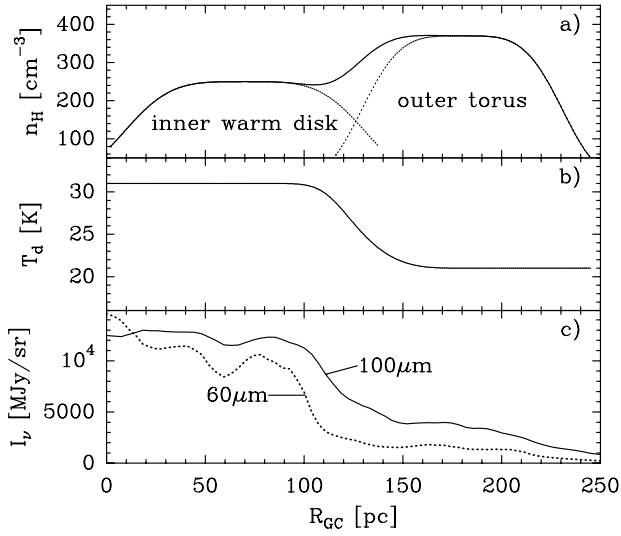}
% {gcnb_nh_td_profiles.ps} (gcnb_nh_td_profiles.graphic)
\caption{\label{nb_nh_td}
 {\bf a)} Modeled radial mid-plane profile of the average hydrogen number density 
          in the NB (assuming homogeneous matter distribution; see discussion in 
          Sect. \ref{disismmorph}). Note that the central density depression 
          is not well-constrained. ~~
 {\bf b)} Mass-averaged dust temperature.~~
 {\bf c)} IRAS 60 and 100\,\mim\ surface brightness profiles, representing warm dust 
          (positive and negative offsets from the GC averaged).}
\end{figure}

\begin{figure}[htb]
\includegraphics[width=0.40\textwidth]{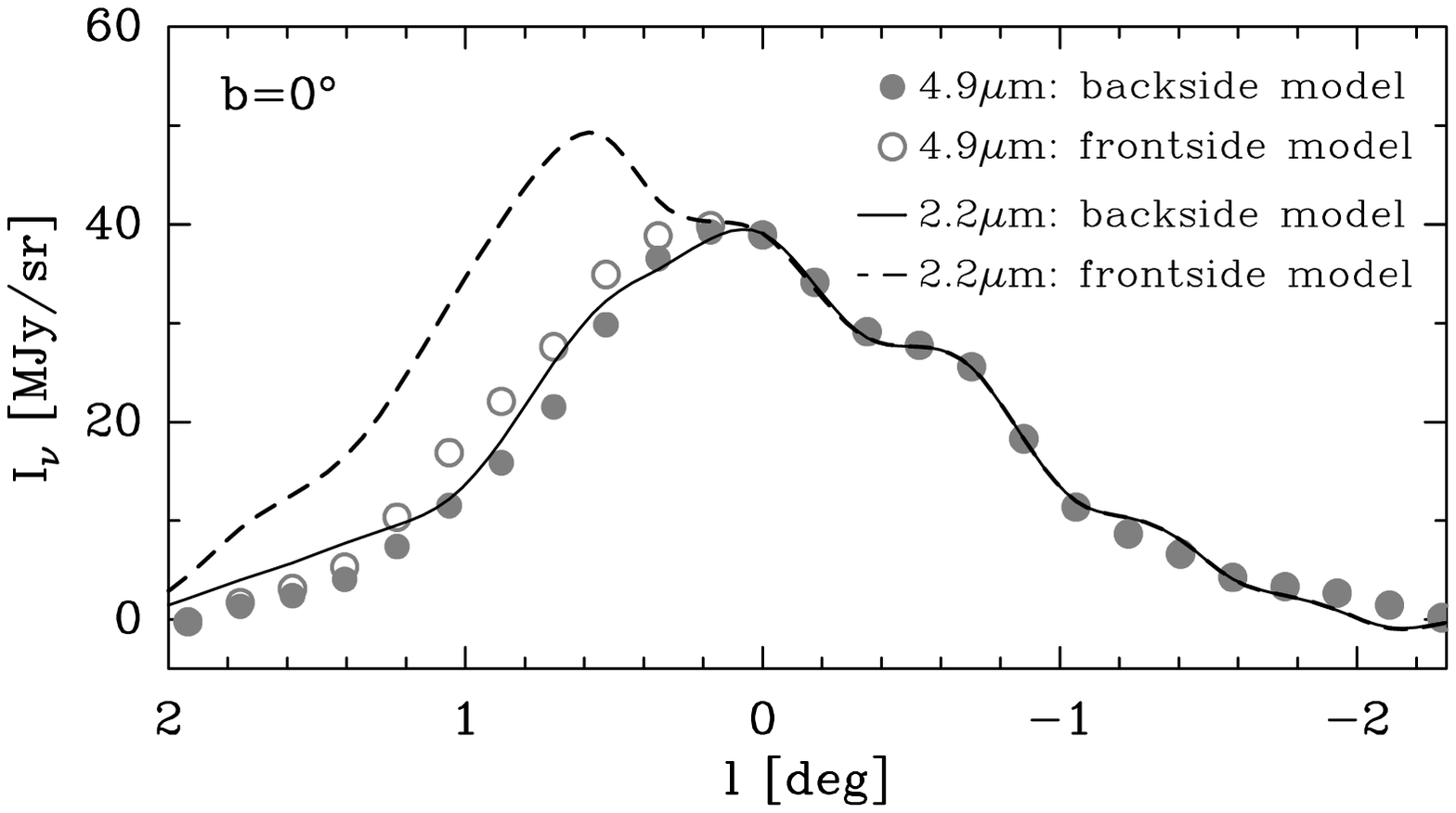}
% {gcnb_comp_2_4_profiles.ps} (gcnb_comp_2_4_profiles.graphic)
\caption{\label{nb_comp24}
 Longitude profiles of the 2.2\,\mim\ and 4.9\,\mim\ 
 emission from the NB (COBE). The observed emission is dereddened 
 with two different extinction models, assuming the ``+1\degr\ cloud complex'' 
 being located in front (dashed curve and empty circles) and behind 
 the stellar NB (solid curve and filled circles).}
\end{figure}

In Sect. \ref{resnbnh} we derived a total hydrogen mass of 
$6.8\times 10^7$\,\msun\ for the central $\sim$\,1\,kpc
and confirmed the well-known asymmetry in the mass distribution of 
ISM in the CMZ. 
A comparison of the distribution of stars 
(Figs. \ref{nirsizemap}a and \ref{nb_lm}c)
and that of ISM (Figs. \ref{nb_lm}a and \ref{nb_nh_inout}b)
indicates that not all of the ISM in the CMZ is 
associated with the stellar NB. 
There are several indicators that allow to distinguish between dust inside and outside 
the stellar NB:\\
1. The stellar radiation density inside the NB is considerably higher than outside. 
Therefore, dust located inside the NB must have a higher 
temperature and specific luminosity than dust outside the NB. 
Figures \ref{nb_lm}b and d show that both quantities are 
strongly enhanced in the central $|l| \le 0$\grdp6 
($\sim 25$\,K and 30\,\lsun/\msun, respectively), 
indicating the presence of a warm and luminous inner region 
($R \le 100$\,pc) in the NB. 
At $-0$\grdp6\,$\ge l \ge -1$\grdp9, both quantities decrease to 
$\sim 21$\,K and 13-15\,\lsun/\msun, respectively, indicating the presence of 
colder dust in the outer region of the NB. At $l \le -1$\grdp9, the 
approximate outer ``edge'' of the NB, both quantities 
drop sharply and approach 16-17\,K and 2-3\,\lsun/\msun, respectively, values 
typical for dust located well outside the stellar NB, i.e., in the GB.\\
2. At negative $l$, the longitude profiles of the total dust luminosity 
and the 4.9\,\mim\ surface brightness agree well (Fig. \ref{nb_lm}c), 
indicating that the main dust-heating sources have the same spatial 
distribution as evolved stars, which dominate the NIR luminosity. 
Dust-heating mechanisms will be discussed in Sect. \ref{nbmod}.
In contrast, there is a relative dust luminosity excess at positive $l$.\\
3. At $l > 0$\grdp3 both the dust temperature and the 
specific dust luminosity drop sharply from values typical for the inner NB 
to values typical for ISM outside the NB (Figs. \ref{nb_lm}b and d). 
Together with the large peak at $l\sim 0$\grdp9 
of the hydrogen column density (see Fig. \ref{nb_lm}a), this indicates 
the presence of large amounts of cold dust along the line of sight at 
positive longitudes.
Hence we can attribute the dust luminosity excess at positive $l$\ to cold dust 
located outside the stellar NB.  
By subtracting this excess from the total dust luminosity map, we obtain the luminosity 
map of dust associated with the stellar NB (longitude profile shown in Fig. \ref{nb_lm}c). 

Since we showed in Sect. \ref{disstarsmorph} that the distribution of stars 
(i.e., the density of the stellar NIR radiation field) is approximately 
symmetric with respect to the GC, 
we make the reasonable assumption that the specific luminosity ($L_{\rm d}/M_{\rm H}$) 
of dust inside the NB 
is the same at positive $l$\ as it is at negative $l$, where no cold dust intersects 
the line of sight, and 'symmetrize' the map of the specific dust luminosity.
The mass distribution of warm luminous dust inside the stellar NB is then derived 
by dividing the corrected dust luminosity map (profile shown in Fig. \ref{nb_lm}c, dotted curve) 
by the corrected map of the specific dust luminosity (profile shown as dotted curve in Fig. \ref{nb_lm}d). 
Low-luminosity dust outside the NB was filtered out from this map by setting 
a cut-off level for the specific dust luminosity at 5\,\lsun/\msun. 
The resulting hydrogen column density map of the NB is shown in Fig. \ref{nb_nh_inout}d.
This map is almost symmetric with respect to the GC. 

However, because the column density map was calculated with the mass-averaged 
dust temperature, it is only a first-order estimate to the actual distribution of ISM 
in the NB. Figure \ref{nb_lm}b shows that the 
dust temperature in the inner part of the NB is considerably higher than in the outer 
parts. While we do not expect line-of-sight averaging effects at $|l|>1$\degr, 
the value of $\sim25$\,K at $|l|<0$\grdp4 represents certainly an average 
of cold dust in the outer NB and warm dust in the inner NB.
To correct the $N_{\rm H}$\ map of the NB (Fig. \ref{nb_nh_inout}d) for this dust 
temperature gradient and to derive the three-dimensional density distribution, 
we transformed it back to a 240\,\mim\ surface brightness map using an 
effective dust temperature map (profile in Fig. \ref{nb_lm}b) symmetrized in the 
same way as the specific dust luminosity (Fig. \ref{nb_lm}d). 
We then constructed and optimized a volume density model of ISM in the NB to fit, 
after convolution with the COBE beam, both the `observed' 240\,\mim\ surface brightness 
and line-of-sight integrated effective dust temperature.
Although the sparsely resolved COBE maps do not allow to put strong constraints on the 
shape of the radial density profile, the flatness of the longitude FIR emission profile
is not consistent with a homogeneously filled, optically thin disk. 
It rather suggests a torus-like configuration. 
Since molecular line data indicate the presence of a 
{\it ``180-pc Molecular Ring''} (see Sect. \ref{nbmod} for discussion and references), 
the input configuration for our model was a 180-pc torus with $T_{\rm d}=21$\,K 
(see Fig. \ref{nb_lm}b) and an inner disk with a central density depression and 
$T_{\rm d}>25$\,K. The free parameters were the (radial) thickness of the torus, 
its density profile, and the radial density profile and temperature of the inner disk.
The best fit to the data was obtained with:\\
{\bf 1.} A torus with $R_0=180$\,pc, $T_{\rm d}=21$\,K, a radial density profile of the 
   form given given by Eq. (\ref{eqnbnirmod}) with $n=4$, $\sigma_{\rm FWHM}=100$\,pc 
   (i.e., $R_{\rm in}=130$\,pc and $R_{\rm out}=230$\,pc), 
   and $n_{\rm H}^0=370\pm50$\,\ccm, and\\
{\bf 2.} an inner disk with $R_{\rm out}=120$\,pc, $T_{\rm d}=(31\pm3)$\,K, and 
   $n_{\rm H}^0=250\pm50$\,ccm. The size and shape of the inner density depression 
   could, of course, not be constrained.
 
The presence of a sharp transition in the physical characteristics of ISM and/or 
a change of the interstellar radiation field at $R_{\rm GC}\sim 100-120$\,pc 
is also confirmed by the extremely sharp drop-off of the 60 and 100\,\mim\ emission 
at $|l|\sim 0$\grdp7, which indicates a nearly complete absence 
of warm dust ($T_{\rm d}>30$\,K) in the outer torus at $R> 120$\,pc (see Figs. 
\ref{nb_lm}e and \ref{nb_nh_td}c). 
The total hydrogen masses of the inner disk and outer torus are 
$\sim 4\times 10^6$\,\msun\ and $\sim 1.6\times 10^7$\,\msun, respectively. 
These numbers are compiled in Table \ref{nbsumtab}.
The notations used for the different 
stellar and ISM components of the NB and their interrelation are summarized in 
Table \ref{nbcomptab}. 

Since the NB is basically unresolved in latitude in the COBE maps, only an upper limit 
for its full scale height of 60\,pc (0\grdp40) could be derived. 
The IRAS maps provide a much higher angular resolution. The 
100\,\mim\ emission is already representative for the bulk of the dust mass, although 
it may be somewhat biased towards warmer dust. A comparison between the COBE 
60\,\mim, 100\,\mim, and 240\,\mim\ maps shows that cold dust extends further 
than warm dust in longitude, but not in latitude.  
Since the dust ridge is very prominent in the 100\,\mim\ IRAS map and its scale height 
is considerably larger than the IRAS beam, saturation 
of the 100\,\mim\ IRAS map in the very centre does not falsify the result much. 
The vertical surface brightness profile could be best fitted by a 
volume density distribution of the form of Eq. (\ref{eqnbnirmod}) 
with $n=1.4$\  and $\sigma= 45\pm 5$\,pc (0\grdp30$\pm$0\grdp03). 
The distribution is centered at $b_0 \sim -0.05$\degr, the position of SgrA$^*$. 
This is in good agreement with the FWHM scale height of $^{12}$CO at 
$l = 0$\degr\ of 50\,pc (Bitran et al. 1997; see Fig. \ref{nb_nh_inout}a). 
Since the $^{12}$CO emission is optically thick, the actual scale height is 
expected to be somewhat overestimated by this tracer.

In Fig. \ref{nb_nh_inout}e, the resulting $N_{\rm H}$\ map of the NB 
is displayed, and Fig. \ref{nb_nh_inout}f shows the corresponding longitude 
profile as well as a comparison between the modeled (beam-convolved) 
and observed 240\,\mim\ surface brightness profiles. 
The corresponding radial profiles of the mid-plane hydrogen density and 
mass-averaged dust temperature are shown in Fig. \ref{nb_nh_td}.
We call this configuration, i.e., the inner warm disk and the outer cold torus, 
the ``{\it Nuclear Molecular Disk}''. 
It has the same size as the {\it Nuclear Stellar Disk}, 
i.e., $R_{\rm out}=230$\,pc and $\Delta h_{\rm FWHM}=45\pm5$\,pc. 

Note that the continuum data used in this paper 
do not allow to distinguish between a disk/torus-like and a bar-like 
morphology. For example, the CMZ in IC\,342,  
which resembles in many aspects our own GC region, 
seems to resemble a bent bar rather then a disk/torus (e.g., Schulz et al. 2001).
However, the main parameters of the NB (compiled in Table \ref{nbsumtab}) would not change 
dramatically if a bar-like morphology is assumed rather than a disk or torus-like
morphology. 

We attribute all other cold, low-specific-luminosity dust in the central kpc 
to ISM not heated by stars in the NB, and therefore located outside the stellar NB. 
Figure 
\ref{nb_nh_inout}b shows the column density distribution of this material. 
The hydrogen masses at positive and negative $l$\ are 
$M_{\rm H}^{l^+} \sim 2.9\times 10^7$\,\msun\ and 
$M_{\rm H}^{l^-} \sim 1.1\times 10^7$\,\msun, respectively. 
The average dust temperature and specific dust luminosity of this material 
are $T_{\rm d}\sim 15-17$\,K and $L_{\rm d}/M_{\rm H}\sim 2-3$\,\lsun/\msun, 
respectively, compared 
to $\sim 20-30$\,K and $10-50$\,\lsun/\msun\ for dust inside the NB.
The total hydrogen mass in the central 1\,kpc of our Galaxy (the CMZ) 
amounts then to 
$M_{\rm H}^{\rm CMZ} \sim 6\times 10^7$\,\msun. 
Of this, one third is located in the {\it Nuclear Molecular Disk}, 
which is associated with the {\it Nuclear Stellar Disk} (i.e., the NB).
Our estimate of the total hydrogen mass in the CMZ is intermediate between the 
values of $2\times 10^7$\,\msun\ and $2\times 10^8$\,\msun\ 
derived by Mauersberger \& Bronfman (1998) from the optically thin C$^{18}$O(1--0) 
and optically thick $^{12}$CO(2--1) lines, respectively, and agrees well with 
$M_{\rm H} \sim $2--6$\times 10^7$\,\msun\ derived by Sodroski et al. (1995) 
from COBE DIRBE data.

The total mass of the cold high-column density complex at $l \sim 0$\grdp9 
is $M_{\rm H}\sim 10^7$\,\msun, 
of which the Sgr\,B2 GMC at $l \sim 0$\grdp7 
contains $\sim$10\%\ (Gordon et al. 1993). 
If this complex would be located in front of the NB, it would cause high 
extinction.
We assumed that the same low fraction of the total 
mass in this complex as inside the NB contributes to diffuse extinction 
while most of the mass is concentrated in small, ultra-opaque clouds which block 10\% 
of the light (see Sect. \ref{disismprop}), and tested the effect of this cloud complex 
on the final (extinction-corrected) stellar NIR maps of the NB. 
Figure \ref{nb_comp24} shows clearly that the model with the cloud complex located in 
front of the NB yields a strong, implausible 2.2\,\mim\ excess, while 
the model with the cloud complex on the back side yields a good 
agreement of the dereddened 2.2\,\mim\ and 4.9\,\mim\ surface brightness profiles. 
There is no reason to believe that the 
average effective stellar temperature in the NB is much higher at positive 
longitudes than at negative longitudes,
nor does the specific dust luminosity indicate 
an excess of hot stars at $l > 0.5$\degr\ (see Fig. \ref{nb_lm}d). 
Since this test may be affected by the low angular resolution of the COBE data, 
we cannot ultimately conclude that the entire ``+1\degr\ cloud complex'' 
is located behind the NB. Indeed, the new 2MASS and 
MSX images\footnote{2MASS Image mosaic and 2MASS-MSX image combination by E. Kopan (IPAC)\\  
(http://antwrp.gsfc.nasa.gov/apod/ap000705.html)} 
of the NB
indicate the presence of some very opaque dust lanes in front of the NB 
at $l>0.4$\degr. However, on a large scale, we can exclude that this material 
causes strong extinction towards the NB (see Sect. \ref{datanextnb}). 

%_________________________________________________________________

\subsection{Physical characteristics of Interstellar Matter in the Nuclear Bulge} 
     \label{disismprop}

From the model described in the previous section, we derive a total 
hydrogen mass of the {\it Nuclear Molecular Disk} of 
$\sim 2\times 10^7$\,\msun, of which 
$4\times 10^6$\,\msun\ are located in the inner warm disk at $R\le 120$\,pc 
and $1.6\times 10^7$\,\msun\ ($\sim$80\%) are distributed in 
the outer cold torus.  
The corresponding dust luminosities amount to $\sim 2\times10^8$\,\lsun\ 
in each of the two components.
Average densities in the inner disk and outer torus are 
$\langle n_{\rm H}\rangle\sim 100$\,\ccm\ and $\sim 150$\,\ccm, respectively.
This high average hydrogen density supports the observation that most of the gas 
in the NB is in molecular form. 
The mid-plane hydrogen column density through the entire NB is 
$\sim 1\times 10^{23}$\,\scm.
Assuming homogeneous matter distribution, this corresponds to a 
total visual extinction of $\sim100$\,mag between the front side of the 
NB and the GC, of which $\sim$80\,mag are due to the cold outer dust torus. 
This clearly contradicts both the value of $\sim$\,10-15\,mag derived from the 
NIR colours of the NB (Sect. \ref{resgbext}and \ref{resnbsed}) 
and the total visual extinction of 30\,mag between the Sun and the GC 
(e.g., Whitelock \& Glass 1990).

This apparent contradiction can be solved when we assume that only $\sim$10\% 
of ISM (by mass) in the NB is homogeneously distributed, while 90\% 
of the mass is trapped in small, discrete clouds which cover only a small fraction 
of the total area. 
A high degree of clumpiness is indeed confirmed by high-resolution submillimetre 
continuum maps of the NB at $\lambda$\,800\,\mim\ (Lis \& Carlstrom 1994) and 
$\lambda$\,1.2\,mm\ (Zylka et al. 2001, in prep.). 
These maps show only very little extended emission between and around the 
dense clumps/clouds. 
We also find that the average column density of the central 120\,pc of 
$N_{\rm H} \sim 7\times 10^{22}$\,\scm\ (not deconvolved; see Sect. \ref{resnbnh}) 
agrees well with the value of $\sim 8\times 10^{22}$\,\scm\ 
derived from the $\lambda$\,1.2\,mm map (when smoothed to the same angular 
resolution and assuming $T_{\rm d} = 25$\,K). 
It also agrees well with the value $N_{\rm H} \sim 8\times 10^{22}$\,\scm\ which we 
derive from the C$^{18}$O(1--0) map by Dahmen et al. (1997) using Eq.\,(7) 
of Mauersberger \& Bronfman (1998) with ${\rm [C^{18}O]/[H_2]} \sim 4\times 10^{-7}$\ 
and adopting $T_{\rm kin} \sim 20\ldots30$\,K.
The fact that the chopped submm continuum maps, the total power low-resolution 
COBE FIR maps as well as the C$^{18}$O data see all the 
same hydrogen mass in the central 120\,pc rules out that the high-resolution submm 
measurements have resolved out a considerable amount of extended emission. 
This confirms the assumption that most of the ISM in the NB is concentrated in very compact 
molecular clouds. A first inspection of the 1.2\,mm map suggests an area filling factor of 
dense clouds in the NB of $\sim 15$\% 
or less. This translates into a volume filling factor of only a few per cent. 
The average density of these compact molecular clouds, which contain $\sim$\,90\% 
of the total hydrogen mass in the NB, becomes then $n_{\rm H} \sim 10^4$\,\ccm. 
Note that this is a very rough estimate and that the actual densities in the cloud 
cores may be much higher. 
The average visual extinction of these clouds would be 
of order 1000\,mag and they become optically thin only at $\lambda\ge 100$\,\mim.
Thus, our mass estimate, which is based on the assumption that the 240\,\mim\ 
emission is optically thin, should still be valid. 
We call these clouds ``{\it ultra-opaque}'' and assume that $\sim$\,10\% 
of the stellar emission from the NB and the back side of the GB is completely 
blocked out by them (see Sect. \ref{datanextnb}). 
Further characteristics of these clouds are being investigated in Paper\,IV 
(Zylka et al., in prep.).

The physical reason for this clumpiness is most probably the tidal stability limit. 
We find that in the gravitational potential of the mass distribution shown in 
Fig. \ref{nb_photmass} only clouds with densities higher than a few times 
$10^3 - 10^4$\,\ccm\ 
are stabilized against tidal forces by their own gravitation (cf. G\"usten \& Downes 1980). 
In contrast to G\"usten \& Downes (1980), who use the mass distribution 
derived by Sanders \& Lowinger (1972), we find that this limit holds only for the 
outer NB ($R\ge 120$\,pc) and that the shallow mass distribution in the inner 
{\it Nuclear Stellar Disk} causes a stability window between $R_{\rm GC}\sim 50$\ 
and 110\,pc. Inside 50\,pc only superdense clouds are stable against tidal disruption.

For the central region of IC\,342, which resembles in many aspects our own GC 
region, but is seen face-on, Schulz et al. (2001) also find that the molecular gas 
is extremely clumpy and that the bulk of the gas mass in the inner $\sim$50\,pc of this 
galaxy is concentrated in compact clouds with $M_{\rm H} \sim 10^6$\,\msun\ and 
$n_{\rm H} \sim 10^4$\,\ccm. Quantitatively, this is a striking similarity to the
conditions in our own GC region.

The total mass and average density of the homogeneously distributed intercloud medium 
in the NB are then $M_{\rm H}= 2.0\pm0.5\times 10^6$\,\msun\ and 
$n_{\rm H}\sim 10$\,\ccm, respectively. 
The presence of a diffuse, hot, ionized intercloud medium in the NB is supported by the presence 
of extended thermal (free-free) radio emission (see Fig \ref{radiomapsplo}, 6\,cm map).
Mezger \& Pauls (1979) derived for the extended, low-density ionized gas in the NB an average 
electron temperature of $T_{\rm e} \sim 5000$\,K and an electron density of 
$n_{\rm e} \sim 10$\,\scm. The total mass of this component is 
$M_{\rm H_{II}} \sim 2\times 10^6$\,\msun. Both values are in good agreement with our estimate. 
The detection of extended X-ray emission in the inner $\sim$\,1\grdp8 of the NB 
indicates the presence of second diffuse component, i.e. a very hot 
Plasma (e.g., Koyama et al. 1996).

%%%%%%%%%%%%%%%%%%%%%%%%%%%%%%%%%%%%%%%%%%%%%%%%%%%%%%%%%%%%%%%%%%%%%%%%%%%%%%%%%%%%%%

\section{Towards an empirical model of the Nuclear Bulge}                    \label{nbmod}

\begin{figure}[htb]
\includegraphics[width=0.47\textwidth]{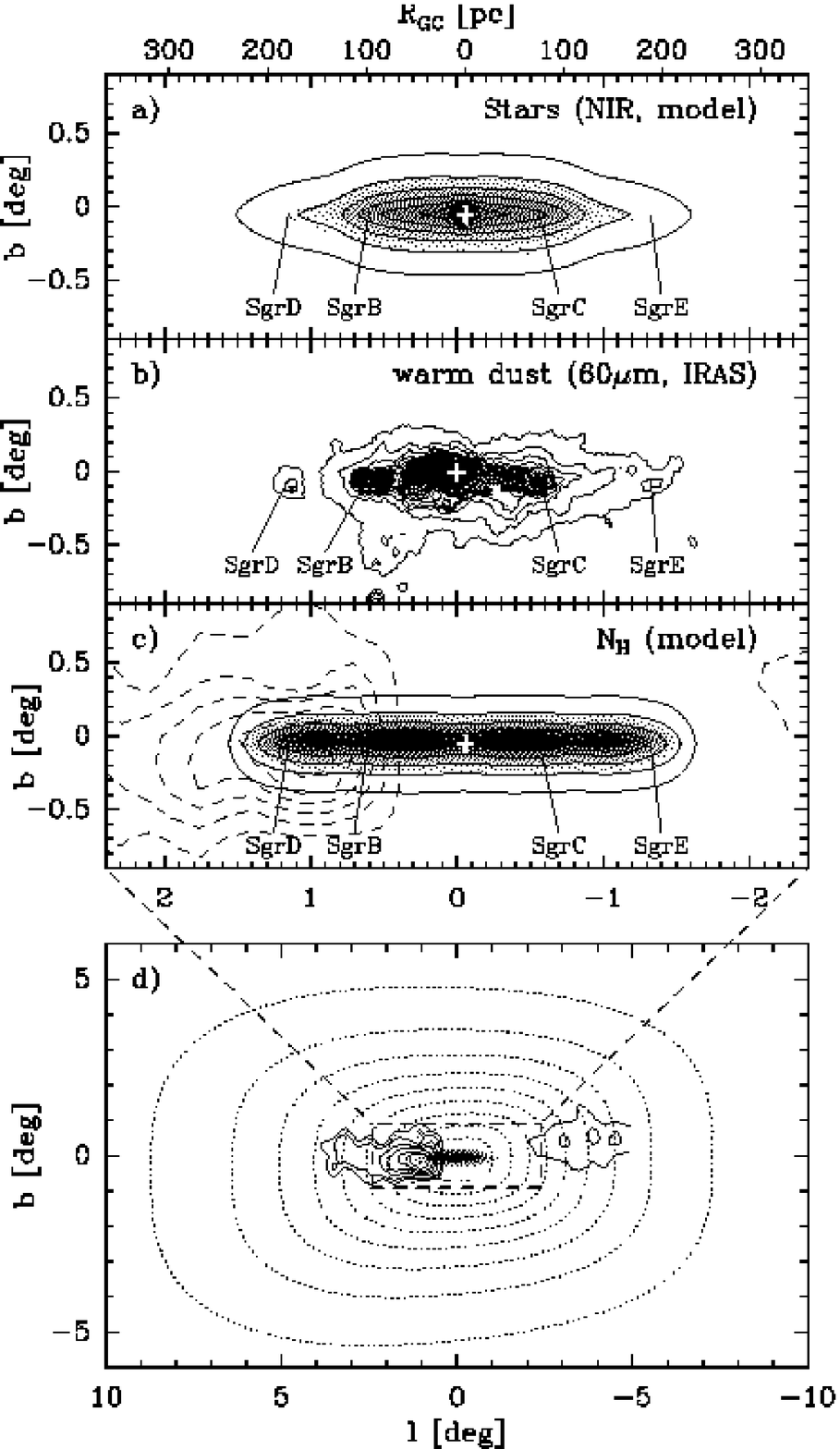}
% {gcnb_nirmod_nhmod_new.ps} (gcnb_nirmod_nhmod_plo.graphic)
\caption{\label{nb_nir_nh}
 {\bf a)} Model of the completely dereddened 4.9\,\mim\ surface brightness of the NB.~~
 {\bf b)} IRAS 60\,\mim\ emission from warm dust.~~
 {\bf c)} Model of the hydrogen column density in the NB, representing mainly 
          the distribution of cold dust. Dashed lines: $N_{\rm H}$\ contours 
          of cold ISM outside the NB (as observed with the large COBE beam).~~
 {\bf d)} Schematic overview of the Galactic Centre Region viewed from the Sun. 
          Dotted lines: stellar NIR emission from the bar-like Galactic Bulge. 
          Solid lines: cold ISM in the Central Molecular Zone, located outside the NB. 
          Grey-scale plot: the Nuclear Bulge.}
\end{figure}

\begin{table*}[htb]                  
 \caption[]{Notations used for different physical components and spatial regions 
   of the NB \label{nbcomptab}}
%  \begin{flushleft}
   \begin{tabular}[t]{ll} 
\hline \noalign{\smallskip}
{\bf Stellar Components:}     & Explanation \\
\noalign{\smallskip} \hline \noalign{\smallskip}
Nuclear Stellar Cluster (NSC):& Central $\rho\propto r^{-2}$\ stellar cluster, as first described by Becklin \& Neugebauer (1968). Its\\
                              & central part is identical with the star cluster discussed in Paper\,I (Fig.6).        \\
Nuclear Stellar Disk (NSD):   & Large-scale disk ($R_{\rm out}\sim 230$\,pc) of mainly cool stars with a possible density drop at\\
                              & $R\sim 120$\,pc. Contains most of the mass in the NB.                                  \\
Nuclear Bulge (NB):           & The sum of NSC and NSD.                                                                  \\
\noalign{\smallskip} \hline \noalign{\smallskip}
{\bf ISM Components:}         &  \\
\noalign{\smallskip} \hline \noalign{\smallskip}
Inner warm disk:              & Warm dust and molecular gas in the NB which is restricted to $R<120$\,pc.                \\
Outer cold torus:             & Colder dust and molecular gas in the outer NB (still heated by these stars), possibly    \\
                              & related to the ``180\,pc Molecular Ring''. The reason for the sharp transition in ISM    \\ 
                              & characteristics at $R\sim 120$\,pc is not yet understood.                                \\
Nuclear Molecular Disk (NMD): & The sum of the inner warm disk and the outer cold torus. Represents all molecular gas    \\ 
                              & which is physically associated with the stellar NB and heated by these stars.            \\ 
Central Molecular Zone (CMZ): & All molecular gas in the central kpc of the Galaxy. Includes the Nuclear Molecular Disk  \\ 
                              & {\it and} cold molecular gas outside the NB in the GB.                                   \\
\noalign{\smallskip} \hline \noalign{\smallskip}
{\bf Spatial regions:} (see Table \ref{nbsumtab}) &  \\
\noalign{\smallskip} \hline \noalign{\smallskip}
Central 1.25\,pc:             & Sphere of $R=0.625$\,pc around the GC (Volume\,$\sim 1$\,pc$^3$). Dominated by the NSC.  \\
Inner NB:                     & Stars and ISM at $R<120$\,pc, defined by the sharp transition of ISM characteristics     \\
                              & and a possible drop of the stellar density at $R\sim 120$\,pc. Includes the central 1.25\,pc \\
Outer NB:                     & Outer part of the NB and NMD (the outer cold ISM torus), mainly defined by the           \\
                              & complete lack of warm and hot dust at $R> 120$\,pc (the outer cold ISM torus).           \\
                              & Stellar population dominated by the NSD. \\
\noalign{\smallskip} \hline \noalign{\smallskip} 
  \end{tabular}
% \end{flushleft} 
\end{table*}

\begin{table*}[htb]                  
 \caption[]{Adopted characteristics of the Nuclear Bulge \label{nbsumtab}}
  \begin{flushleft}
   \begin{tabular}[t]{lccccl} 
\hline \noalign{\smallskip}
Component: & Total & Central 1.25\,pc & Inner NB & Outer NB & Ref./Sect.$^{(e)}$\\
\noalign{\smallskip} \hline \noalign{\smallskip}
{\bf Size:} & & & & & \\ 
$R$/pc                                                        &
 230$\pm$20                & 0.625            & 120$\pm$20                & 120--230              &    \ref{disstarsmorph}, \ref{disismmorph} \\
$h_{\rm FWHM}$/pc                                             &
  45$\pm$5                 & --               &  45$\pm$5                 &  45$\pm$5             & 
   \ref{disstarsmorph}, \ref{disismmorph} \\[1mm]
{\bf Stars:} & & & & & \\
$M_{\star}$/\msun                                             &
 $1.4\pm 0.6\times 10^9$      & $7\times 10^5$   & $8\pm 3\times 10^8$     & $6\pm 2\times 10^8$       & \ref{disstarsmass} \\[1mm] 
$\langle\rho_{M_{\star}}\rangle$/(\msun\,pc$^{-3}$)           &
 $200\pm 50$                  & $7\times 10^5$   & $400\pm 100$            & $120\pm 30$               & \ref{disstarsmass} \\[1mm] 
$L_{\rm \star,NIR}$/\lsun$^{(a)}$                             &
 $7\pm 2\times 10^8$          & $1.4\times 10^6$ & $4\pm 1\times 10^8$     & $3\pm 1\times 10^8$       & \ref{disstarslum}  \\[1mm] 
$L_{\rm \star,tot}$/\lsun$^{(b)}$                             &
 $2.5\pm 1\times 10^9$        & $4\times 10^7$   & $2\pm 1\times 10^9$     & $5\pm 2\times 10^8$       & \ref{disstarslum}  \\[1mm] 
$L_{\rm \star,NIR}/L_{\rm \star,tot}$                         & 
 0.28                         & 0.035            & 0.20                    & 0.60                      & \ref{disstarslum}  \\
$\langle\rho_{L_{\rm \star,tot}}\rangle$/(\lsun\,pc$^{-3}$)   &
 $100\pm 50$                  & $4\times 10^7$   & $200\pm 50$             & $60\pm 20$                & \ref{disstarslum}  \\[1mm]
$T_{\star,{\rm eff}}$/K$^{(a)}$                               &
 $4400\pm 400$                & $\ge$4500        & $4400\pm 400$           & $4400\pm 400$             & \ref{resnbsed}     \\[1mm]
$(M_{\star}/L_{\rm \star,NIR})$/(\msun/\lsun)$^{(a)}$         &
 $2\pm 0.3$                   & $0.5\pm 0.2$     & $2\pm 0.3$              & $2\pm 0.3$                & \ref{disstarslum}, \ref{disstarsmass} \\[1mm]
$(M_{\star}/L_{\rm \star,tot})$/(\msun/\lsun)$^{(b)}$         &
 $0.6\pm 0.3$                 & $0.02\pm 0.01$   & $0.4\pm 0.2$            & $1.2\pm 0.5$              & \ref{disstarslum}, \ref{disstarsmass} \\[1mm]
{\bf ISM:} & & & & & \\
$M_{\rm H}$/\msun                                             & 
 $2.0\pm 0.3\times 10^7$      & --$^{(d)}$       & $4\pm 1\times 10^6$     & $1.6\pm 0.3\times 10^7$   & \ref{disismmorph}  \\ 
$\langle n_{\rm H}\rangle$/cm$^{-3}$\,$^{(c)}$                &
 $140\pm 30$                  & --$^{(d)}$       & $100\pm 30$             & $150\pm 30$               & \ref{disismprop}   \\
$L_{\rm dust}$/\lsun                                          &
 $4.0\pm 0.2\times 10^8$      & --$^{(d)}$       & $2\pm 0.3\times 10^8$   & $2\pm 0.3\times 10^8$     & \ref{disismmorph}  \\ 
$T_{\rm dust}$/K                                              &
 20--30                       & --$^{(d)}$       & $31\pm3$                & $21\pm1$                  & \ref{disismmorph}  \\
$(L_{\rm dust}/M_{\rm H})$/(\lsun/\msun)                      &
 $20\pm 3$                    & --$^{(d)}$       & $50\pm10$               & $12.5\pm2$                & \ref{disismmorph}  \\[1mm]
{\bf ISM/Stars:} & & & & & \\
$(M_{\rm H}/M_{\star})$                                       &
 $1.4\pm 0.5\times 10^{-2}$   & --$^{(d)}$       & $5\pm 1.5\times 10^{-3}$& $2.7\pm0.5\times 10^{-2}$ &          \\[1mm]
$(L_{\rm dust}/L_{\rm \star,NIR})$$^{(a)}$                    &
 $0.6\pm 0.2$                 & --$^{(d)}$       & $0.5\pm 0.2$            & $0.7\pm 0.2$              &          \\
$(L_{\rm dust}/L_{\rm \star,tot})$$^{(b)}$                    &
 $0.16\pm 0.08$               & --$^{(d)}$       & $0.10\pm 0.05$          & $0.4\pm 0.2$              &          \\
\noalign{\smallskip} \hline \noalign{\smallskip} 
  \end{tabular}
 \end{flushleft} 
\begin{list}{}{}
\item[$^{(a)}$] $T_{\rm NIR}$\ derived from NIR SEDs. Represents $\sim T_{\rm eff}$\ of cool stars and does not include 
      contribution from hot massive stars
\item[$^{(b)}$] Includes contribution from hot massive stars
\item[$^{(c)}$] Average $n_{\rm H}$, assuming homogeneous matter distribution. 
      The ISM is actually composed of discrete dense molecular clouds  
      with $\langle n_{\rm H}\rangle \sim 10^4$\,\ccm\ and 
      a thin intercloud medium with $n_{\rm H}\sim 10$\,\ccm\
      (see discussion in Sect. \ref{disismprop}).
\item[$^{(d)}$] Values could not be derived
\item[$^{(e)}$] Note that some values for the three sub-regions of the NB represent average 
       characteristics of different physical components  which overlap 
       in these regions. Some of these values are not individually derived in the 
       corresponding sections, but arise\\ \hspace*{3.5mm}  from the combined model of the NB. 
\end{list}
\end{table*}

%%% Short summary 
Using COBE and IRAS NIR to FIR data, we derived the morphology and physical characteristics 
of stars and ISM in the central kpc of the Galaxy.  
Figure \ref{nb_nir_nh} summarizes the projected large-scale distribution 
of stars and ISM in the Galactic Centre Region. 
We find that the central $D\sim 500$\,pc are dominated by the {\it Nuclear Bulge}, 
a dense and massive disk-like complex of stars and ISM, which is clearly 
distinguished from the much larger {\it Galactic Bulge} in which it is embedded. 
Similar distinct nuclear star clusters have been found by {\it HST} observations 
in the centers of many exponential bulges in intermediate and late-type 
spiral galaxies (e.g., Carollo et al. 1998). 
Here, we derived density and luminosity distributions of the various 
stellar and gaseous components populating the {\it Nuclear Bulge} 
of our own Galaxy. Notations and spatial extents of these components 
are described in Table \ref{nbcomptab} and their physical characteristics are 
compiled in Table \ref{nbsumtab}.

%%% General morphology:
The {\it Nuclear Bulge} appears, in projection, as a flat bar with an outer 
radius of 230$\pm$20\,pc, FWHM scale height 45$\pm$5\,pc, and 
diameter-to-thickness ratio 5\,:\,1. 
Its overall morphology appears disk-like and symmetric with respect to the 
Galactic Centre. 
This disk may be elliptical, in which case it would rather resemble a bar. 
The {\it Nuclear Bulge} consists of the large {\it Nuclear Stellar Disk} with the 
{\it $R^{-2}$\ Nuclear Stellar Cluster} at its centre, and the {\it Nuclear Molecular Disk}.
In our analysis, we neglected the small-scale asymmetry seen in the 
COBE NIR images (see Fig. \ref{cobemapsplo} and Sect. \ref{disstarsmorph}).
On small scales, the {\it Nuclear Bulge} may well be not as symmetric as 
suggested by our results.
This remains, however, subject to further studies which have to combine 
NIR, MIR, and submm data with much higher angular resolution than provided 
by the COBE data.

%%% Total stellar mass and mass distribution:
The total mass of the {\it Nuclear Bulge} amounts to 
$1.4\pm 0.6\times 10^9$\,\msun, 
of which $\sim 99$\%\ is contained in stars and 
1\%\ in ISM, mainly in the form of dense molecular clouds. 
The {\it Nuclear Bulge} accounts thus for 
$\sim$1\%\ of the total stellar mass in the Galactic Disk and Bulge.
In contrast to the usual assumption that the mass distribution in the central 
300\,pc is dominated by an $R^{-2}$\ cluster, we find that the density profile 
of the {\it Nuclear Stellar Cluster} steepens outside $R_{\rm GC}\sim 5-10$\,pc 
and that it only dominates the inner $R_{\rm GC}\sim 20-30$\,pc.  
At larger radii, the mass distribution is dominated by the 
{\it Nuclear Stellar Disk} which has a rather flat radial density profile
inside $R_{\rm GC}\sim 120$\,pc and a well-defined outer boundary at $\sim$230\,pc.
The {\it Galactic Bulge} begins to dominate the mass distribution 
only for $R_{\rm GC} > 400$\,pc. 
The photometric mass distribution in the inner 500\,pc derived from our models 
of the {\it Nuclear} and {\it Galactic Bulge} (with the central Black Hole included) 
agrees well with the dynamical mass distribution derived by other authors 
and with the total stellar mass in the central 30\,pc as deduced in Paper\,II 
from the KLF of this region (see Fig. \ref{nb_photmass}). 
%
%%% Total stellar luminosity:
The total stellar luminosity of the {\it Nuclear Bulge} amounts to 
$L_{\ast} \sim 2.5\pm 1\times 10^9$\,\lsun, of which $7\pm 2\times 10^8$\,\lsun\ 
($\sim$30\%) arise from cool low-mass MS and evolved stars. 
The contribution by hot and massive stars ($\sim$70\%) is highly uncertain. 
The {\it Nuclear Bulge} thus contributes $\sim$5\%\ to the total luminosity of the 
Galactic Disk and Bulge.

%%% Stellar population:
The stellar population of the {\it Nuclear Bulge} appears to resemble 
that of the  {\it Galactic Bulge}. 
Unique for the {\it Nuclear Bulge}, however, is a population of 
young massive stars and an over-abundance of luminous giant and supergiant stars, 
which are both strongly 
concentrated towards the centre and are mainly associated with the 
{\it Nuclear Stellar Cluster}. However, the distribution of hot dust, 
H{\small II} regions, and supernova remnants (Fig. \ref{nb_ms}; 
see LaRosa et al. 2000) indicates that star formation 
occurs throughout the entire {\it Nuclear Bulge} and not exclusively at 
its centre. Ongoing star formation is one of the 
key characteristics which distinguish the {\it Nuclear Bulge} from the 
{\it Galactic Bulge}. The star formation history of the {\it Nuclear Bulge}, 
based on the analysis of the KLF of the central 30\,pc, has been discussed in Paper\,II. 

%%% Gas mass and morphology:
The total hydrogen mass in the central kpc, referred to as the 
{\it Central Molecular Zone}, amounts to $M_{\rm H}\sim 6\times 10^7$\,\msun\ 
or, if corrected for He and metals, to $M_{\rm ISM}\sim 1\times 10^8$\,\msun. 
We find that $\sim$\,2/3 of this material is located outside the 
stellar {\it Nuclear Bulge} and not heated by these stars. 
About $3\times 10^7$\,\msun\ are at positive $l$\ (0\grdp5\,--\,4\degr) 
and $1\times 10^7$\,\msun\ at negative $l$ ($-2$\degr\,--\,$-4$\degr), thus causing 
the observed asymmetry in the distribution of ISM in the 
{\it Central Molecular Zone}. The physical reason for this asymmetry is not known.
Only $\sim 2\times 10^7$\,\msun\ are directly associated with the stellar 
{\it Nuclear Bulge}. 
Gas and dust in this {\it Nuclear Molecular Disk}, which is approximately 
symmetric with respect to the Galactic Centre, are distributed in a thin and warm 
inner disk ($T_{\rm d}\sim 30$\,K) and a colder, dense outer torus ($\sim$20\,K) 
between $R_{\rm GC}\sim 120$\ and 220\,pc, which contains $\sim 80$\%\ of the 
total gas mass in the {\it Nuclear Bulge}. 

%%% Interpretation of morphology in terms of dynamical models:
The presence of such a molecular torus in a large-scale bar potential 
(due to the barred {\it Galactic Bulge}) is predicted 
by the same dynamical models as described in Sect. \ref{intro}. 
Due to shocks and angular momentum loss on the inner self-intersecting 
$X_1$\ orbits in the {\it H{\small I} Central Disk}, the gas 
is driven further inward and being compressed to densities at which molecular 
clouds form, and finally settles on stable $X_2$\ orbits 
(e.g.,  Binney et al. 1991; Englmaier \& Gerhard 1999; Englmaier \& Shlosman 2000). 
The outer {\it Central Molecular Zone} would then represent the region between 
the innermost stable $X_1$\ orbit and the more circular $X_2$\ orbits. 
The massive outer torus of the {\it Nuclear Molecular Disk} 
can be associated with the region where this gas has accumulated on stable 
$X_2$\ orbits and becomes dense enough to form stars. 
One implication of this interpretation is that star formation in the 
{\it Nuclear Bulge} should occur -- apart from the very centre -- 
mainly in the outer molecular torus. Indeed, the distribution of giant 
H\,{\small II} region complexes and supernova remnants across the entire 
{\it Nuclear Bulge}, with the outermost ones being at $R_{\rm GC}=170-190$\,pc 
(Sgr\,D and E; see LaRosa et al. 2000 and Fig. \ref{nb_ms}) is consistent with 
the presence of a modest star-forming ring with a mean radius of 180\,pc 
around the galactic centre.
The relation between this massive torus and the narrow 
{\it 180-pc Molecular Ring} (Bally et al. 1987) 
is not fully understood (see Sect. \ref{intro}). 

%%% Clumpiness of ISM:
Extinction arguments lead to the conclusion that 
ISM in the {\it Nuclear Molecular Disk} is extremely clumpy. 
This is confirmed by high-resolution submm continuum maps of the inner 
{\it Nuclear Bulge} (Lis \& Carlstrom 1994; Zylka et al., in prep.). 
We find that more than 90\% 
of the total gas mass in the {\it Nuclear Molecular Disk} is concentrated 
in very compact, 
{\it ultra-opaque} molecular clouds with average densities of 
$\langle n_{\rm H}\rangle > 10^4$\,\ccm\ 
and a volume filling factor of only a few per cent. 
These clouds are completely optically thick for UV, optical, and NIR 
radiation, but they become optically thin for the thermal FIR/submm 
radiation at wavelengths longer than 100\,\mim.
The physical reason for this clumpiness is most probably the tidal stability limit, 
since only clouds with densities higher than a few times $10^3 - 10^4$\,\ccm\ 
are stabilized against tidal forces by their own gravitation in the gravitational 
potential of the mass distribution in the {\it Nuclear Bulge}.
The remaining $<$\,10\%\ of interstellar mass in the {\it Nuclear Bulge} 
are contained in a diffuse, thin and warm intercloud medium. 

%%% Dust heating sources:
The dominant heating source of the thin intercloud medium in the {\it Nuclear Bulge} 
is the strong UV radiation field due to young massive stars in the 
{\it Nuclear Stellar Cluster} and other regions of high-mass star-formation 
in the outer {\it Nuclear Bulge}, as well as a widely distributed population 
of intermediate-mass MS stars. 
Due to the extreme clumpiness of the ISM, 
at least the soft UV radiation can 
penetrate the entire {\it Nuclear Bulge}, ionize the thin intercloud gas, 
and heat the dust in this component to high temperatures. 
This soft UV radiation together with the strong stellar NIR radiation field also 
dominates the heating of the dense molecular gas 
(e.g., Poglitsch et al. 1991; H\"uttemeister et al. 1998; Schulz et al. 2001).
The relative lack of hot dust (200 -- 400\,K) and the abrupt decrease of 
specific dust luminosity beyond $R_{\rm GC}\sim 100-120$\,pc supports our 
finding that most of the gas and dust in the {\it Nuclear Molecular Disk} 
is contained in very compact and massive molecular clouds distributed mainly 
in a torus between $R_{\rm GC}\sim 120$\ and 220\,pc. Most of their mass then 
is effectively shielded from the strong interstellar radiation field in the 
{\it Nuclear Bulge}.
Gas in the inner {\it Nuclear Molecular Disk} must also be clumpy so that 
the radiation field can penetrate and maintain the high mass-averaged dust 
temperature of $>$\,30\,K. However, the absence of a stringent tidal stability 
limit between $R_{\rm GC}\sim 50$\ and 110\,pc allows smaller clouds to exist 
in this region, which have a larger surface-to-mass ratio and can be more 
effectively heated, thus explaining the large excess of hot dust compared to 
the outer torus. 

%%% Final comparison with starbursts:
With its flat disk and ring-like morphology, high stellar density, high degree of 
clumpiness of ISM, and ongoing star formation, the {\it Nuclear Bulge} may be 
tentatively considered as counterpart of the central complex in IC\,342 
(e.g., Wright et al. 1993; B\"oker et al. 1997; Mauersberger \& Bronfman 1999; 
Schulz et al. 2001) or even of the massive gaseous tori seen
in starburst galaxies (e.g., Downes \& Solomon 1998).

%%% Distance remark:
Since we used a distance of 8.5\,kpc to the Galactic Centre, 
all sizes have to be scaled by (8/8.5)\,$\sim 0.94$, and masses and 
luminosities by (8/8.5)$^2\sim 0.89$, when the numbers are 
being compared to other studies which use  
a distance of 8\,kpc. These corrections are, however, smaller than 
the uncertainties of our results.

%%%%%%%%%%%%%%%%%%%%%%%%%%%%%%%%%%%%%%%%%%%%%%%%%%%%%%%%%%%%%%%%%%%%%%%%%%%%%%%%%%%%%%

\begin{acknowledgements}

   We wish to thank W. Reich for assisting with radio continuum data retrieval 
   from the MPIfR archive. L. Bronfman kindly provided the $^{12}$CO(1--0) map 
   of the inner Milky Way and  
   D. Lis provided the 800\,\mim\ continuum map of the inner NB. 
   We wish to thank A. Schulz, S. Jogee, E. Schinnerer, S. Stolovy, and 
   N. Scoville for helpful comments and stimulating discussions. 
   Our special thanks go to W. Duschl who elucidated the results given here 
   in terms of Active Galactic Nuclei 
   and to the referee, Ian Glass, for useful suggestions.
   RL acknowledges financial support by a research grant of the Max Planck 
   Society and by NSF grant AST 99-81546. RZ acknowledges finacial
   support by the Deutsche Forschungsgemeinschaft.
   The $COBE$\ datasets were developed by the NASA Goddard Space Flight Center
   under the guidance of the COBE Science Working Group and were
   provided by the National Space Science Data Center.
   $SkyView$\ was developed and is maintained under 
   NASA ADP Grant NAS 5-32068 with P.I. Thomas A. McGlynn under the
   auspices of the High Energy Astrophysics Science Archive Research
   Center (HEASARC) at the Goddard Space Flight Center Laboratory for 
   High Energy Astrophysics.
 
\end{acknowledgements}

%%%%%%%%%%%%%%%%%%%%%%%%%%%%%%%%%%%%%%%%%%%%%%%%%%%%%%%%%%%%%%%%%%%%%%%%%%%%%%%%%%%%%%

\begin{appendix}

\section{The interstellar dust extinction curve} \label{apdust}

\begin{figure}[htb]
\includegraphics[width=0.50\textwidth]{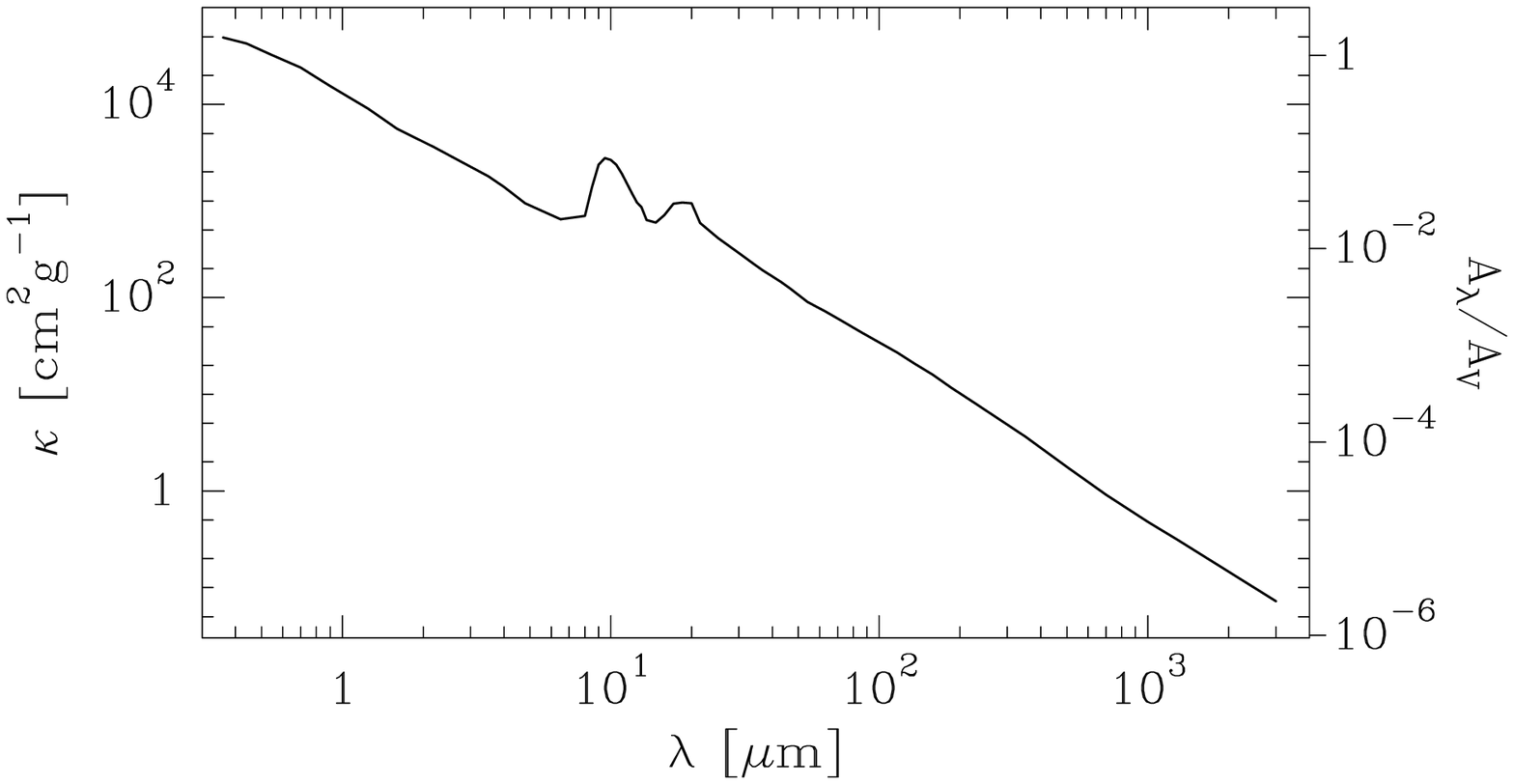}
% {extcurve.ps} (ext/extcurve_paper.greg)
\caption{\label{extcurve}
 Opacity spectrum of the dust model used in this paper (cm$^2$\ per gram of dust).}
\end{figure}

\begin{table}[htb]                  
 \caption[]{Dust opacities (per gram of dust)\label{optab}}\vspace{-0mm}
  \begin{flushleft}
   \begin{tabular}[t]{rrrrrr} 
\hline \noalign{\smallskip}
$\lambda$~~ & $\kappa(\lambda)$~~~ & $\lambda$~~  & $\kappa(\lambda)$~~~ & $\lambda$~~  & $\kappa(\lambda)$~~~\\
 $[$\mim$]$        & [cm$^2$\,g$^{-1}$] & $[$\mim$]$        & [cm$^2$\,g$^{-1}$] & $[$\mim$]$        & [cm$^2$\,g$^{-1}$]  \\
\noalign{\smallskip} \hline \noalign{\smallskip}
0.55 & 32100~~ & 4.9  &   926~~~ & 140  &    20~~~ \\
1.25 &  9030~~ & 12   &  1190~~~ & 240  &   7.3~~~ \\
1.65 &  5300~~ & 25   &   415~~~ & 800  &  0.71~~~ \\
2.2  &  3590~~ & 60   &    77~~~ & 1300 &  0.31~~~ \\
3.5  &  1800~~ & 100  &    34~~~ & 3000 &  0.072~~~ \\
\noalign{\smallskip} \hline\noalign{\smallskip} 
\end{tabular} 
\end{flushleft}
\end{table}

A ``standard'' extinction curve was adopted for
the entire GD and the NB since extinction along the line of
sight toward the GC is believed to be dominated by dust
in the diffuse ISM which accounts for the intercloud gas as well as
for low- and intermediate-density molecular clouds (Tielens et al. 1996).
The dust opacity spectrum used here is based on data from 
Rieke \& Lebofsky (1985) for $\lambda = 0.36\ldots 13$\,\mim\ and 
Draine \& Lee (1984) for $\lambda = 13$\,\mim$\ldots 1.3$\,mm. 
Tabulated values for dust grains without ice mantles and no coagulation 
were taken from Ossenkopf \& Henning 1994, who assumed 
a MRN grain size distribution (Mathis et al. 1977). 
Figure \ref{extcurve} shows the dust opacity spectrum used in 
this paper and Table \ref{optab} lists opacities per gram of dust 
for selected wavelengths. 
A ``standard'' H-to-dust mass ratio in the solar neighbourhood of
$(M_{\rm H}/M_{\rm d})_{\odot}$\,=\,110 was adopted (Draine \& Lee
1984; Mathis \& Wallenhorst 1981; Kr\"ugel et al. 1990), which scales
with the reciprocal of the relative metallicity $Z/Z_{\odot}$\ 
(see Fig. \ref{diskmodpar}).  The
resulting $A_{\rm V}/N_{\rm H}$\ ratio amounts to 
$(A_{\rm V}/N_{\rm H})_{\odot} = 5.3\,10^{-22}$\ (for $Z/Z_{\odot}$\,=\,1; Bohlin et
al. 1978). The FIR/submm spectral index of the dust opacity is
$\beta = -1.8$.  

%________________________________________________________________________

\section{Distribution of Interstellar Matter in the Galactic Disk}  \label{apgalmod}

\begin{figure}[htb]
\includegraphics[width=0.50\textwidth]{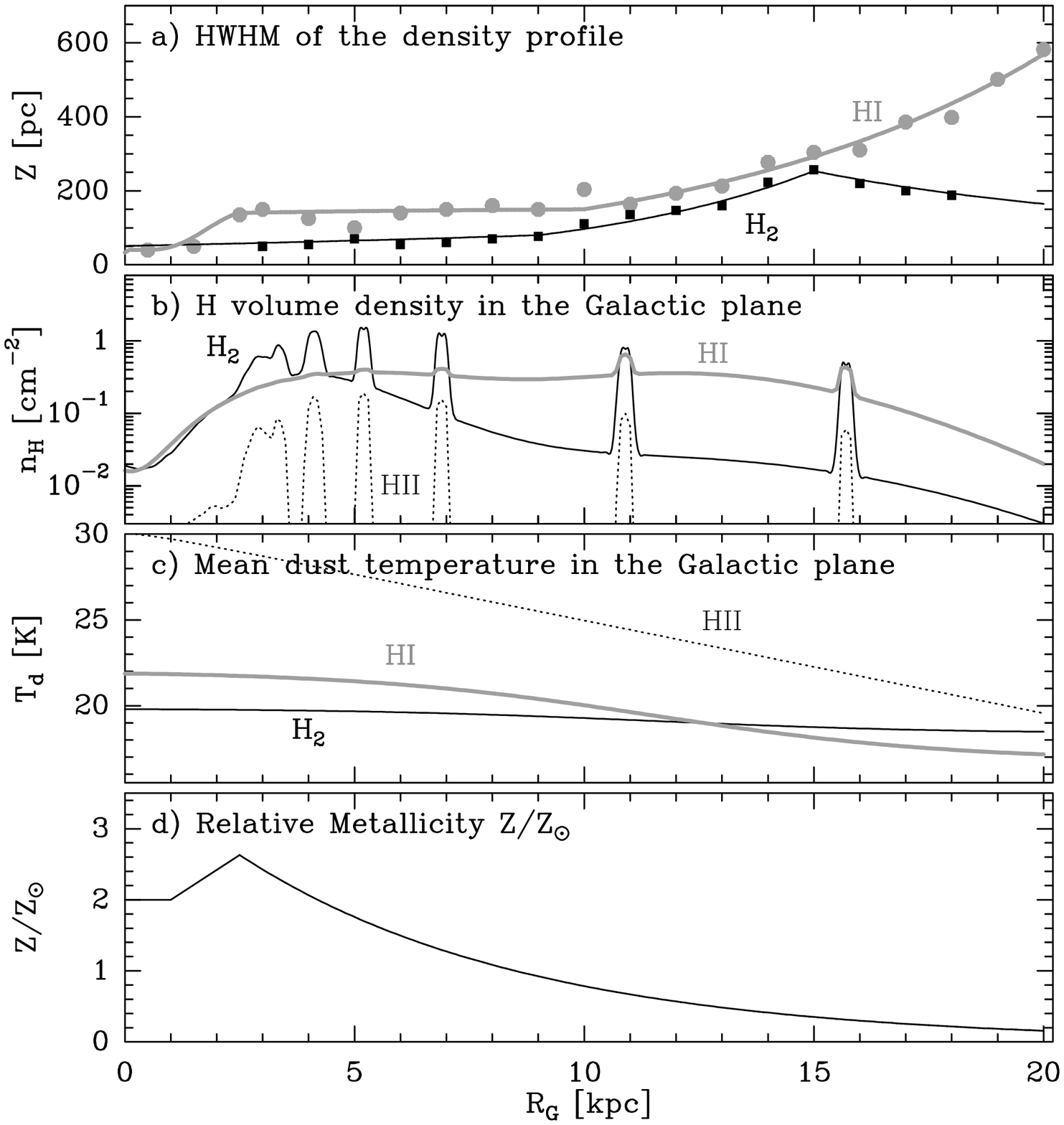}
% {disk_mod_par_paper.ps} (disk_mod_par_paper.graphic)
\caption{\label{diskmodpar} Main parameters of the Galactic Disk model.\newline
{\bf a)} HWHM height of the density profiles of H\,I and H$_2$\ in the GD. 
   Data points (black squares for H$_2$\ and grey circles for H\,I) were taken 
   from Wouterloot et al. (1990). The solid lines show the scale heights  
   of H\,I and H$_2$\ adopted for the model.~~
{\bf b)} Ring-averaged volume density profiles (\,H\,cm$^{-3}$\,) at $b = 0$\degr\
   The ripples are due to the spiral structure.~~
{\bf c)} Mean dust temperature profiles of the H\,I and H$_2$\ components 
   of the Galactic Disk.~~
{\bf d)} Relative abundance of heavy elements as function of galactic radius 
   (after Afflerbach et al. 1997).}
\end{figure}

\begin{figure}[htb]
\includegraphics[width=0.50\textwidth]{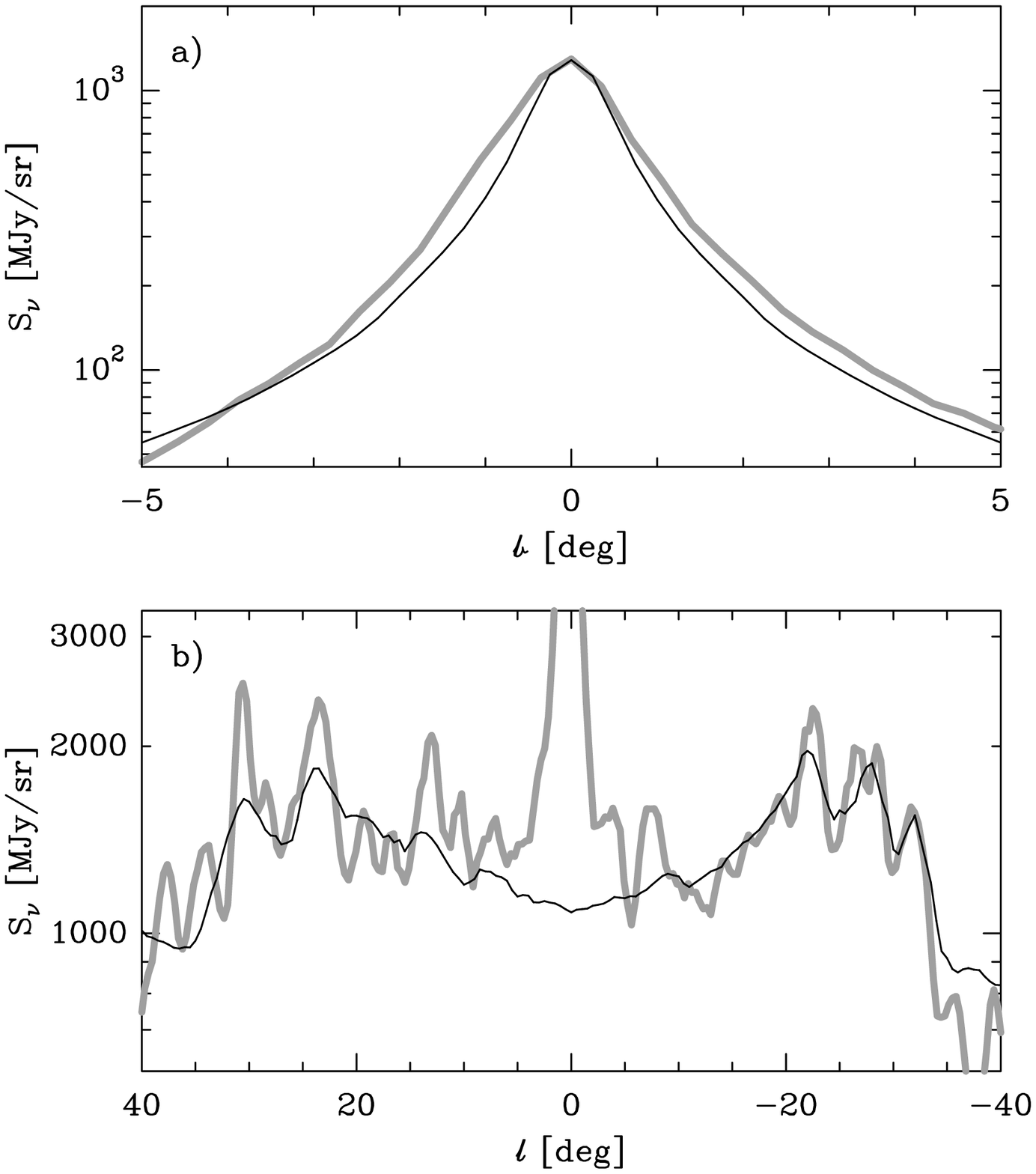}
% {disk_mod_profiles_paper_new.ps} (disk_mod_l_profile_paper.graphic)
\caption{\label{diskmodprof} 
    Observed and modeled 240\,\mim\ emission profiles. 
    Grey squares and thick grey curve: COBE data, Black solid lines: model.~~
{\bf a)} Average latitude profile of the Galactic Disk between
   $-$20\degr$< l <$\,20\degr, excluding spiral arms and the NB region.~~
{\bf b)} Longitude profile at $b = 0$\degr.}
\end{figure}

To treat the emission and extinction by dust in the GD in an appropriate way 
and to correctly derive NIR surface brightness maps of the NB,  
a three-dimensional model of the distribution of ISM in the GD was developed. 
The primary goal of this model was to derive extinction maps towards the 
GB and NB and to separate the MIR to FIR emission from the 
GD into contributions from the front and back sides with respect to the GC region. 
Proper modeling of the NIR surface brightness distribution of the GB, 
which is impossible without applying correct extinction corrections, 
is particularly important since due to its exponential profile the separation of 
GB and NB is very sensitive to how the surface brightness of the GB is 
interpolated into the region occupied by the NB. 

Here, the GD was modeled by an axisymmetric disk with a central hole and 
an overlayed spiral arm pattern.
Three gas components mixed with dust were considered: 
$i)$\ atomic hydrogen (H{\small I}), $ii)$\ molecular Hydrogen (H$_2$), and $iii)$\ H{\small II}.  
The dust model is described in Appendix \ref{apdust}.
Radial density profiles at $b = 0$\degr\ were adopted from 
Liszt (1992) for H\,I and from 
Bronfman (1992) for H$_2$, and were adjusted to fit the FIR data.
A sech$^2$\ vertical density profile was adopted for the H$_2$\ disk 
and a double-Gaussian profile for the H\,I disk.
The scale heights of the H\,I and H$_2$\ disks as function of galactocentric 
radius - including the exponential flare and the warp of the outer disk - 
were adapted to various measurements summarized by Wouterloot et al. (1990). 
The dust temperatures were, to first order, adopted from Sodroski et al. (1994) and 
then varied to fit the COBE FIR data with the above model. 
The metal abundance gradient was taken from Afflerbach et al. (1997). 
For the GC region we use a relative metallicity of $Z/Z_{\odot} = 2$\ 
as derived by Mezger et al. (1979).
A four-arm, modified logarithmic spiral density pattern was overlayed on the axisymmetric disk. 
The general spiral geometry was adopted from Georgelin \& Georgelin (1976) 
and Englmaier \& Gerhard (1999), and adjusted to fit the 
COBE FIR data. It resembles the pattern derived by 
Drimmel \& Spergel (2001), but is not identical. 
The details of this spiral arm model may not necessarily reflect the true 
spiral pattern of the Galaxy. However, it fits the COBE FIR data sufficiently well 
for our purpose and provides a reasonable estimate of how the extinction towards the 
GB rises with increasing galactocentric radius towards the GD molecular ring.
The spiral arms have a Gaussian density cross section. The density contrast between spiral 
arms and inter-arm regions and its radial gradient were adopted from 
Heyer \& Terebey (1998). 
The local Orion arm was not included in our model.
The main parameters of the GD model are summarized in Fig. \ref{diskmodpar}. 

With these parameters, the GD was modeled out to $R = 20$\,kpc with a grid cell 
size of (100\,pc)$^3$.
Figure \ref{diskmodprof} compares the observed 240\,\mim\ emission profiles 
with the profiles predicted by the model (see Fig. \ref{avlat} for the latitude 
extinction profile). 
The GD model contains $M_{\rm H_2} \sim 1\times 10^9$\,\msun\ molecular hydrogen  
and $M_{\rm H\,I} \sim 3\times 10^9$\,\msun\ atomic hydrogen of which 
$\sim 2\times 10^9$\,\msun\ are located at $R < 15$\,kpc. These values are in good 
agreement with other estimates.

This model allows to derive the thermal emission and 
extinction due to dust along any line of sight in the inner Galaxy. 
Although our model is not intended to represent the exact morphology of the entire 
GD and does not include the stellar disk, 
its strength is that it is based on a complete set of 
parameters derived from observations without artificial assumptions and 
that it reproduces the observed FIR emission and colour temperatures 
{\it and} extinction in the inner Galaxy. 
A much more sophisticated three-dimensional model of the GD, 
which is also based on COBE/DIRBE data, 
has recently been developed by Drimmel \& Spergel (2001).

\end{appendix}

%______________________________________________________________


\begin{thebibliography}{}

\bibitem{}
   Afflerbach, A., Churchwell, E., \& Werner, M.W. 1997, ApJ, 478, 190
\bibitem{}
   Athanassoula, E., Morin, S., Wozniak, H., et al. 1990, MNRAS, 245, 130
\bibitem{} 
   Bally, J., Stark, A., Wilson, R.W., \& Henkel, C. 1987, ApJS, 65, 13
\bibitem{} 
   Becklin, E.E., \& Neugebauer, G. 1968, ApJ, 151, 145
\bibitem{} 
   Binney, J., Gerhard, O.E., Stark, A.A., Bally, J., \& Uchida, K.I. 1991, MNRAS, 252, 210
\bibitem{} 
   Bitran, M., Alvarez, H., Bronfman, L., May, J., \& Thaddeus, P. 1997, A\&AS, 125, 99
\bibitem{} 
   Blitz, L., \& Spergel, D.N. 1991, ApJ, 379, 631
\bibitem{} 
   Blitz, L., Binney, J., Lo, K.Y., Bally, J., \& Ho, P.T.P. 1993, Nature, 361, 417
\bibitem{} 
   Bohlin, R.C., Savage, B.D., \& Drake, J.F. 1978, ApJ, 224, 132
\bibitem{} 
   B\"oker, T., F\"orster-Scheiber, N.M., \& Genzel, R. 1997, AJ, 114, 1883
\bibitem{} 
   Bronfman, L. 1992, ASSL Vol.~180, 
   The center, bulge, and disk of the Milky Way, 
   ed. L. Blitz (Kluwer, Dordrecht), 131
\bibitem{} 
   Burton, W.B., \& Liszt, H.S. 1978, ApJ, 225, 815
\bibitem{} 
   Carollo, C.M., Stiavelli, M., \& Mack, J. 1998, AJ, 116, 68
\bibitem{} 
   Catchpole, R.M., Whitlock, P.A., \& Glass, I.S. 1990, MNRAS, 247, 479 
\bibitem{} 
   $COBE$\ Diffuse Infrared Background Experiment (DIRBE) Explanatory Supplement,
   Version 2.1. 1997, ed. M.G. Hauser, T. Kelsall, D. Leisawitz, J. Weiland, 
   ($COBE$\ Ref. Pub. 97-A; Greenbelt: NASA/GSFC), available in electronic form 
    from the NSSDC at http://www.gsfc.nasa.gov/astro/cobe/cobe\_home.html
\bibitem{} 
   Cox, P., \& Laurejis, R. 1989, in IAU Symp. 136, The Center of the Galaxy, 
   ed. M. Morris (Kluwer, Dordrecht), 121
\bibitem{} 
   Cox, P., \& Mezger, P.G. 1989, AAR, 1, 49
\bibitem{} 
   Dahmen, G., H\"uttemeister, S., Wilson, T.L., et al. 1997, A\&AS, 126, 197
\bibitem{} 
   Downes, D., \& Solomon, P.M. 1998, ApJ, 507, 615
\bibitem{} 
   Draine, B.T., \& Lee, H.M. 1984, ApJ, 285, 89
\bibitem{} 
   Drimmel, R., \& Spergel, D.N. 2001, ApJ, 556, 181
\bibitem{} 
   Dwek, E., Arendt, R.G., Hauser, M.G., et al. 1995, ApJ, 445, 716
\bibitem{} 
   Eckart, A., \& Genzel, R. 1998, in 
   Black Holes: Theory and Observation, ed. 
   F. W. Hehl, C. Kiefer, \& R.J.K. Metzler 
   (Springer), 60
\bibitem{} 
   Englmaier, P., \& Gerhard, O.E. 1999, MNRAS, 304, 512
\bibitem{} 
   Englmaier, P., \& Shlosman, I. 2000, ApJ, 528, 677
\bibitem{} 
   Figer, D.F., McLean, I.S., \& Morris, M. 1999, ApJ, 415, 202
\bibitem{} 
   Freudenreich, H.T. 1998, ApJ, 492, 495
\bibitem{} 
   Genzel, R., \& Townes, C.H. 1987, ARA\&A, 25, 377
\bibitem{} 
   Genzel, R., Hollenbach, D., \& Townes, C.H. 1994, Rep. Prog. Phys., 57, 417
\bibitem{} 
   Genzel, R., Thatte, N., Krabbe, A., Kroker, H., \& Tacconi-Garman, L.E. 1996, 
   ApJ, 472, 153
\bibitem{} 
   Genzel, R., Eckart, A., Ott, T., \& Eisenhauer, F. 1997, MNRAS, 291, 219
\bibitem{} 
   Georgelin, Y.M., \& Georgelin, Y.P. 1976, A\&A, 49, 57
\bibitem{} 
   Gordon, M.A., Berkermann, U., Mezger, P.G., et al. 1993, A\&A, 280, 208
\bibitem{} 
   G\"usten, R., \& Downes, D. 1980, A\&A, 87, 6
\bibitem{} 
   Haller, J.W., Rieke, M.J., Rieke, G.H., et al. 
   1996, ApJ, 456, 194
\bibitem{} 
   Handa, T., Sofue, Y., Nakai, N., Hirabayashi, H., \& Inoue, M. 
   1987, Publ. Astron. Soc. Japan., 39, 709
\bibitem{} 
   Hauser, M.G., Kelsall, T., Moseley, S.H. Jr., et al. 1991, 
   in AIP Conf. Proc. 222, After the First Three Minutes,
   ed. S.S. Holt, C.L. Bennett, \& V. Trimble, 161
\bibitem{} 
   Hauser, M.G., Arendt, R.G., Kelsall, T., et al. 1998, ApJ, 508, 25
\bibitem{} 
   Haynes, R.F., Caswell, J.L., \& Simons, L.W.J. 1978, Austr. J. of Phys.  
   Suppl., 45, 1
\bibitem{} 
   Heyer, M.H., \& Terebey, S. 1998, ApJ, 502, 265
\bibitem{} 
   H\"uttemeister, S., Dahmen, G., \& Mauersberger, R., et al. 1998, A\&A, 334, 646
\bibitem{} 
   Kelsall, T., Weiland, J.L., Franz, B.A., et al. 1998, Ap,J 508, 44
\bibitem{} 
   Koyama, K., Maeda, Y., Sonobe, T., et al. 
   1996, PASJ, 48, 249
\bibitem{} 
   Kr\"ugel, E., Chini, R., \& Steppe, H. 1990, A\&A, 229, 17
\bibitem{} 
   LaRosa, T.N., Kassim, N.E., Lazio, T.J.W., \& Hyman, S.D. 2000, ApJ, 119, 207.
\bibitem{} 
   Lebofsky, M.J., \& Rieke, G.H. 1987, in AIP Conf. Proc. 155, 
   The Galactic Center, ed. D.C. Backer, 79 
\bibitem{} 
   Lindqvist, M., Habing, H.J., \& Winnberg, A. 1992, A\&A, 259, 118
\bibitem{} 
   Lis, D.C., \& Carlstrom, J.E. 1994, ApJ, 424, 189
\bibitem{} 
   Liszt, H.S. 1992, ASSL Vol.~180, 
   The center, bulge, and disk of the Milky Way, 
   ed. L. Blitz (Kluwer, Dordrecht), 111
\bibitem{} 
   Liszt, H.S., \& Burton, W.B. 1978, ApJ, 226, 790
\bibitem{} 
   Maihara, T., Oda, N., Sugiyama, T., \& Okuda, H. 1978, PASJ, 30, 1
\bibitem{} 
   Mathis, J.S., Rumpl, W., \& Nordsieck, K.H. 1977, ApJ, 217, 425 (MRN)
\bibitem{} 
   Mathis, J.S., \& Wallenhorst, S.G. 1981, ApJ, 244, 483
\bibitem{} 
   Mauersberger, R., \& Bronfman, L. 1998, RvMA., 11, 209
\bibitem{} 
   McGinn, M.T., Sellgren, K., Becklin, E.E., \& Hall, D.N.B. 1989, ApJ, 338, 824
\bibitem{} 
   McNamara, D.H., Madsen, J.B., Barnes, J., \& Ericksen, B.F. 2000, PASP, 112, 202
\bibitem{} 
   Mezger, P.G., \& Pauls, T. 1978, IAU-Symp. 84, 357
\bibitem{} 
   Mezger, P.G., Pankonin, V., Schmid-Burgk, J., Thum, C., \& Wink, J. 1979, A\&A, 80, L3
\bibitem{} 
   Mezger, P.G., Duschl, W.J., \& Zylka, R. 1996, AAR, 7, 289 (MDZ96)
\bibitem[Paper\,II]{p2} 
   Mezger, P.G., Zylka, R., Phillip, S., \& Launhardt, R. 1999, A\&A, 348, 457 (Paper II)
\bibitem{} 
   Morris, M., \& Serabyn, E. 1996, ARA\&A, 34, 645
\bibitem{} 
   Ossenkopf, V., \& Henning, Th. 1994, A\&A, 291, 943
\bibitem{} 
   Phillip, S., Zylka, R., Mezger, P.G., et al. 
   1999a, A\&A, 348, 768 (Paper I)
\bibitem{} 
   Phillip, S., Tuffs, R.J., Mezger, P.G., \& Zylka, R. 1999b, A\&A, 350, 582
\bibitem{} 
   Poglitsch, A., Stacey, G.J., Geis, N., et al. 1991, ApJ, 374, L33
\bibitem{} 
   Reich, W. 1994, in 
   The Nuclei of Normal Galaxies - Lessons from the Galactic Center,
   ed. R. Genzel, \& A.I. Harris (Kluwer, Dordrecht), 55
\bibitem{} 
   Reich, W., Reich, P., \& F\"urst, E. 1990a, A\&AS, 83, 539
\bibitem{} 
   Reich, W., F\"urst, E., Reich, P., \& Reif, K. 1990b, A\&AS, 85, 633
\bibitem{} 
   Reid, M.J. 1993, ARA\&A, 31, 348
\bibitem{} 
   Rieke, G.H., \& Lebofsky, M.J. 1985, ApJ, 288, 618
\bibitem{} 
   Rieke, G.H., \& Rieke, M.J. 1988, ApJ, 330, L33
\bibitem{} 
   Sanders, R.H., \& Lowinger, T. 1972, AJ, 77, 292 
\bibitem{} 
   Schulz, A., G\"usten, R., K\"oster, B., \& Krause, D. 2001, A\&A, 371, 25 
\bibitem{} 
   Scoville, N.Z. 1972, ApJ, 175, L127
\bibitem{} 
   Scoville, N.Z., Solomon, P.M., \& Jefferts, K.B. 1974, ApJ, 187, L63
\bibitem{} 
   Serabyn, E., \& Morris, M. 1996, Nature, 382, 602
\bibitem{} 
   Sodroski, T.J., Bennett, C., Boggess, N., et al. 1994, ApJ, 428, 638
\bibitem{} 
   Sodroski, T.J., Odegard, N., Dwek, E., et al. 1995, ApJ, 452, 262
\bibitem{} 
   Sodroski, T.J., Odegard, N., Arendt, R.G., et al. 1997, ApJ, 480, 173
\bibitem{} 
   Sofue Y. 1994,  in 
   The Nuclei of Normal Galaxies - Lessons from the Galactic Center,
   ed. R. Genzel, \& A.I. Harris (Kluwer, Dordrecht), 43
\bibitem{} 
   Sofue, Y. 1995, PASJ 47, 551
\bibitem{} 
   Stanek, K.Z., Udalski, A., Szymanski, M., et al. 1997, ApJ, 477, 163
\bibitem{} 
   Tielens, A.G.G.M., Wooden, D.H., Allamandola, L.J., Bregman, J., \& Witteborn, F.C. 
   1996, ApJ, 461, 210
\bibitem{} 
   Weiland, J.L, Arendt, R.G., Berriman, G.B., et al. 1994, ApJ, 425, L81
\bibitem{} 
   Wheelock, S.L., Gautier, T.N., Chillemi, J., et al. 1994, IRAS Sky Survey Atlas, 
   Explanatory Supplement
\bibitem{} 
   Wouterloot, J.G.A., Brand, J., Burton, W.B., \& Kwee, K.K. 1990, A\&A, 230, 21
\bibitem{} 
   Wright, M.C.H., Ishizuki, S., Turner, J.L., Ho, P.T.P., \& Lo, K.Y. 1993, ApJ, 406, 470 

\end{thebibliography}
\end{document}